\documentclass[
  aps,
  pre,
  onecolumn,
  superscriptaddress,
  longbibliography
]{revtex4-2}

\usepackage[T1]{fontenc}
\usepackage[utf8]{inputenc}
\usepackage[english]{babel}

\usepackage{amsmath,amssymb,bm,mathtools}

\usepackage{graphicx}

\usepackage{booktabs}
\usepackage{dcolumn}
\usepackage{array}
\usepackage{tabularx}

\usepackage{xcolor}
\usepackage{url}

\usepackage[colorlinks=true,linkcolor=blue,citecolor=blue,urlcolor=blue]{hyperref}

\newcommand{\bea}{\begin{eqnarray}}
\newcommand{\eea}{\end{eqnarray}}

\let\oldref\ref
\renewcommand{\ref}[1]{(\oldref{#1})}

\begin{document}
\raggedbottom

\title{Effects of confinement in a Brownian gas with simultaneous stochastic resetting and dynamically emergent correlations}

\author{Gabriele de Mauro}
\affiliation{LPTMS, CNRS, Univ.  Paris-Sud,  Universit\'e Paris-Saclay,  91405 Orsay,  France}
\author{Satya N. Majumdar}
\affiliation{LPTMS, CNRS, Univ.  Paris-Sud,  Universit\'e Paris-Saclay,  91405 Orsay,  France}
\author{Gr\'egory Schehr}
\affiliation{Sorbonne Universit\'e, Laboratoire de Physique Th\'eorique et Hautes Energies, CNRS UMR 7589, 4 Place Jussieu, 75252 Paris Cedex 05, France}

\begin{abstract}
We study $N$ non-interacting Brownian particles in an external potential under simultaneous stochastic resetting to the origin. Although they do not interact directly, common resets generate strong dynamically emergent correlations (DEC). We analyze how confinement modifies these correlations and the nonequilibrium stationary state for $V(x)=\kappa |x|^\alpha$, $\alpha\geq0$, focusing mainly on two analytically tractable cases: harmonic confinement (HC), $\alpha=2$, and box confinement (BC), $\alpha\to\infty$. In both cases the stationary state is controlled by the competition between confinement and resetting lengths. We derive exact results for the stationary joint distribution, density, correlations, extreme value statistics (EVS), and gap statistics. While the density behaves similarly in HC and BC, the normalized correlation coefficient differs sharply. In BC it is non-monotonic and overshoots the unconfined value, as hard walls suppress decorrelating trajectories. In HC it instead increases monotonically toward the unconfined limit. For general $\alpha$, the behavior is monotonic for $0<\alpha<\alpha_c=1+\sqrt{5}$ and non-monotonic for $\alpha>\alpha_c$. The difference between HC and BC is also visible in edge observables. In HC, the maximum scales as $M_1=O(\sqrt{\ln N})$ and has a limiting distribution with bounded support and a shape transition controlled by the ratio of the two length scales. In BC, the maximum is at distance $O(1/N)$ from the boundary, as in equilibrium, but its fluctuations have a broad power-law tail with logarithmic corrections. The first gap shows a similar contrast: BC gives a smaller typical gap but stronger anomalous fluctuations than HC. Finally, we extend the EVS analysis to general $\alpha$ and identify, via simulations and scaling arguments, three universality classes: $0\leq\alpha\leq1$, $1<\alpha<\infty$, and the singular limit $\alpha\to\infty$.
\end{abstract}

\maketitle

\section{Introduction}

Brownian motion is one of the most fundamental stochastic processes, originally introduced to describe the erratic motion of particles suspended in a fluid \cite{Brown1828,Einstein1905}. For a one-dimensional free Brownian particle, the propagator is Gaussian, with variance $2Dt$. Hence the typical distance explored by the particle grows as $\sqrt{t}$, and the probability distribution keeps broadening indefinitely \cite{Gardiner2009,Risken1996}. As a consequence, in the absence of confinement or any other localizing mechanism, no stationary state is reached.
In recent years, a simple yet profound modification of Brownian motion has attracted considerable attention, namely stochastic resetting \cite{EM2011,EM2011JPA,EM2014,Reuveni2016,BhatDeBaccoRedner2016,NagarGupta2016,PalReuveni2017,ChechkinSokolov2018,PalKusmierzReuveni2019PRE,DurangLeeLizanaJeon2019,GuptaPlataPal2020,EMSch2020,Bressloff2020,Bressloff2020Queueing,Bressloff2021Accumulation,PKR2022,TsVdLkRmAc2022,GuptaJ2022,NagarG2023,HM2025}. In this process, the diffusive dynamics is intermittently interrupted at random times, and the particle is instantaneously brought back to a fixed position, typically the origin. This mechanism fundamentally alters the long-time behavior of the system. Indeed, the competition between diffusion, which tends to spread the particle, and resetting, which localizes it, leads to the emergence of a nonequilibrium stationary state (NESS), i.e. a time-independent state in which a nonzero probability current persists \cite{EM2011,EM2011JPA,EMSch2020,PKR2022,GuptaJ2022,NagarG2023,EM2014,HM2025,GMS2014,BKP2019,MMS2020,PatelAS2026,Santra2026,Besga2020,Tal-Friedman2020,Faisant2021,GinotBechinger2025}. Another important direction concerns stochastic resetting with memory, in which the resetting position depends on the past history of the process~\cite{BSS2014,BRC2014,BEM2017}. 
Such memory effects have also been studied for single particles in confining potentials~\cite{BM2024b,BEM2025}, and more recently in many-particle systems, in the absence of confinement~\cite{BM2026a}.

The effect of stochastic resetting has also been explored in many-body systems \cite{NagarG2023,GMS2014,BKP2019,MMS2020,MS2026_DEC,Biroli2023,BLMS2024,deMauro2026_NonPoisson,SabM2024,MMS2025,BM2026a,BMS2026_fpg,BMS2025,deMauro2026,MCPL2022,KMS2025,SLP2025,Galla2026,BKMS2024,Olsen2026,VilkAssafMeerson2022,MeersonVilk2026,Vilk2026,Majumder2024,Acharya2025,Majumder2026}. 
In particular, a gas of $N$ non-interacting Brownian particles subject to simultaneous resetting was introduced in Ref.~\cite{Biroli2023}, where it was shown to exhibit a rich and non-trivial phenomenology. 
In this case, the stationary state again results from the competition between diffusion, which tends to spread the particles, and resetting, which localizes them. 
However, despite the absence of direct interactions between particles, the simultaneity of the resets generates strong correlations in the system.
In this work, we refer to these correlations as dynamically emergent correlations (DEC) \cite{MS2026_DEC}. 
By this, we mean correlations that are not produced by direct interactions between particles, but emerge dynamically from the stochastic evolution itself. 
More generally, DEC arise when several non-interacting particles are coupled to the same fluctuating environment, or to the same stochastic dynamical variable. 
Since all particles experience the same realization of this randomness, they become strongly correlated. 
In the present model, the common source of randomness is the simultaneous resetting protocol. 
Its fluctuations are encoded in the random variable $\tau$, which represents the time elapsed since the last reset. 
Conditioned on this time, the particles are independent; after averaging over it, they become correlated. 
Systems with DEC have recently attracted significant research interest~\cite{MS2026_DEC,Biroli2023,BLMS2024,deMauro2026_NonPoisson,SabM2024,MMS2025,BM2026a,BMS2026_fpg,BMS2025,deMauro2026,MCPL2022,KMS2025,SLP2025,Galla2026,BKMS2024,Olsen2026}. 
One of the main reasons is that strongly correlated systems are typically hard to analyze, and only a few examples are exactly solvable. 
Systems with DEC sometimes provide a rare setting in which non-trivial correlations emerge while still allowing for analytical progress.

In Ref.~\cite{Biroli2023}, between two successive resetting events the particles are free to diffuse over the whole real line. This naturally raises a fundamental question: how is this picture modified when the particles are no longer freely diffusing, but instead evolve in a confining environment? For instance, consider $N$ independent Brownian particles diffusing in a one-dimensional box with hard walls. In the absence of resetting, this system relaxes at long times to the \emph{equilibrium} state of an ideal gas, with a uniform spatial distribution. What happens when stochastic resetting is introduced in a system that already possesses an equilibrium stationary state?
In this case, the stationary state is no longer generated by the competition between diffusion and resetting alone. Instead, it results from the interplay between two distinct mechanisms: relaxation toward equilibrium, induced by confinement, and nonequilibrium localization, induced by resetting. Understanding the structure of this stationary state, and in particular how confinement modifies the DEC generated by simultaneous resetting, is the central goal of this work.

The effect of external potentials on a single particle under stochastic resetting has already been investigated extensively~\cite{BM2024b,BEM2025,Pal2015conf,CS2015conf,RG2017conf,PP2019conf,ANBND2019conf,SMS2020conf,GPKP2021conf1,GPK2021conf2}. The focus of the present work is different: we study a system of $N$ particles subject to the same resetting events, and ask how confinement modifies the collective structure of the gas, in particular the dynamically emergent correlations induced by the common reset clock.

Our work is also relevant from another perspective. Consider an ideal gas composed of $N$ independent Brownian particles confined in a one-dimensional box of size $L$ with hard walls. This system is a fundamental model in statistical physics: at equilibrium it is simply an ideal gas and it is fully understood.  
As pointed out in \cite{MS2026_DEC}, if we let the box size $L(t)$ fluctuate stochastically in time, the system is driven out of equilibrium. Also in this case, even though the particles do not interact, they all share the same fluctuating environment and therefore become strongly correlated. This is a quite general problem which could provide a benchmark to study how a fluctuating environment can turn an otherwise ideal gas into a strongly correlated nonequilibrium system. 
However, considering a general stochastic process $L(t)$ is analytically intractable, and one must resort to simplified systems. A natural first step is to consider a dichotomous process, in which the box instantaneously switches between two sizes $L_1$ and $L_2$, with rates $r_1$ and $r_2$. Even in this case, the problem remains difficult, as discussed in \cite{MS2026_DEC}.  
To make further analytical progress, we introduce additional approximations. Specifically, we can take the limit $L_1 \to 0$, $r_1 \to \infty$, $L_2 \to \infty$, and set $r_2 = r$. In this regime, contraction events force all particles to collapse to the origin. Because $r_1 \to \infty$, the particles are immediately released following a contraction, after which they diffuse freely along the real line. 
The problem therefore reduces to simultaneous stochastic resetting introduced above.  
In the present work, we take a step further and relax one of these conditions. Starting from the dichotomous noise picture, we consider the limit $L_1 \to 0$ and $r_1 \to \infty$, while keeping $L_2 = L$ finite and $r_2 = r$. The dynamics then consists of global contraction events that reset all particles to the origin, followed by diffusion inside a finite box.
It therefore provides a new analytically tractable realization of DEC, and allows us to study how correlations are modified by the presence of confinement.

We also mention here that another approach consists of replacing hard walls
with a harmonic trap whose stiffness switches between two values, $\mu_1$
and $\mu_2$, at rates $r_1$ and $r_2$. This setting allows for exact results
and was solved in \cite{BKMS2024}. In the same spirit as the dichotomous-box
construction discussed above, the present HC resetting model can be obtained
by taking $\mu_1\to\infty$ and $r_1\to\infty$, with $r_1/\mu_1\to0$, while
keeping $\mu_2=\mu$ finite and setting $r_2=r$. In this work, however, we rederive all results directly, as this particular limit was neither explicitly extracted nor analyzed in Ref.~\cite{BKMS2024}. We also show that harmonic confinement and hard-wall confinement lead to very different stationary states.

\begin{figure}[htbp]
\centering

\begin{minipage}{0.4\textwidth}
    \centering
    \includegraphics[width=\linewidth]{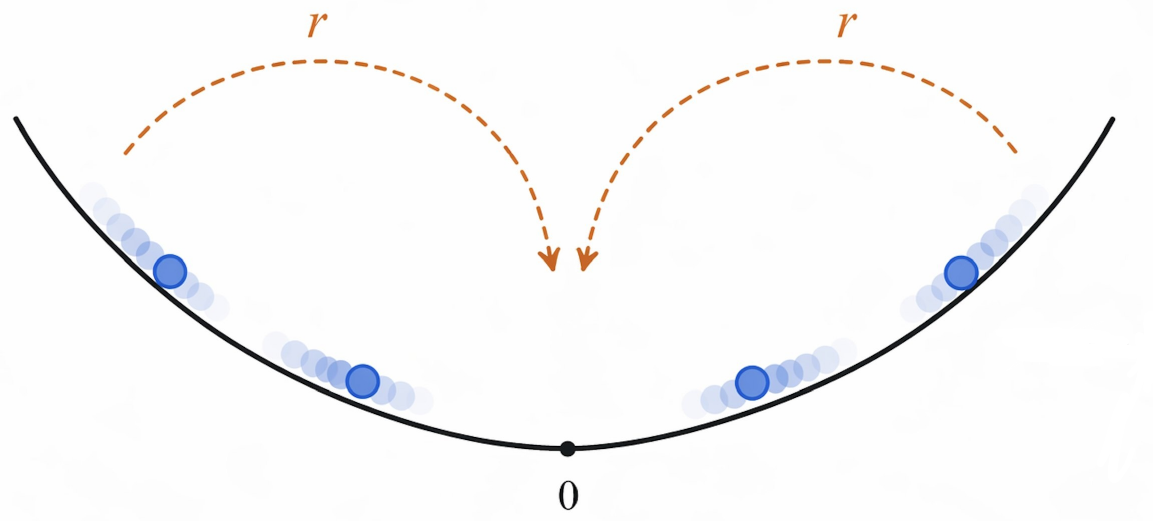}
\end{minipage}%
\hspace{0.1\textwidth}
\begin{minipage}{0.4\textwidth}
    \centering
    \includegraphics[width=\linewidth]{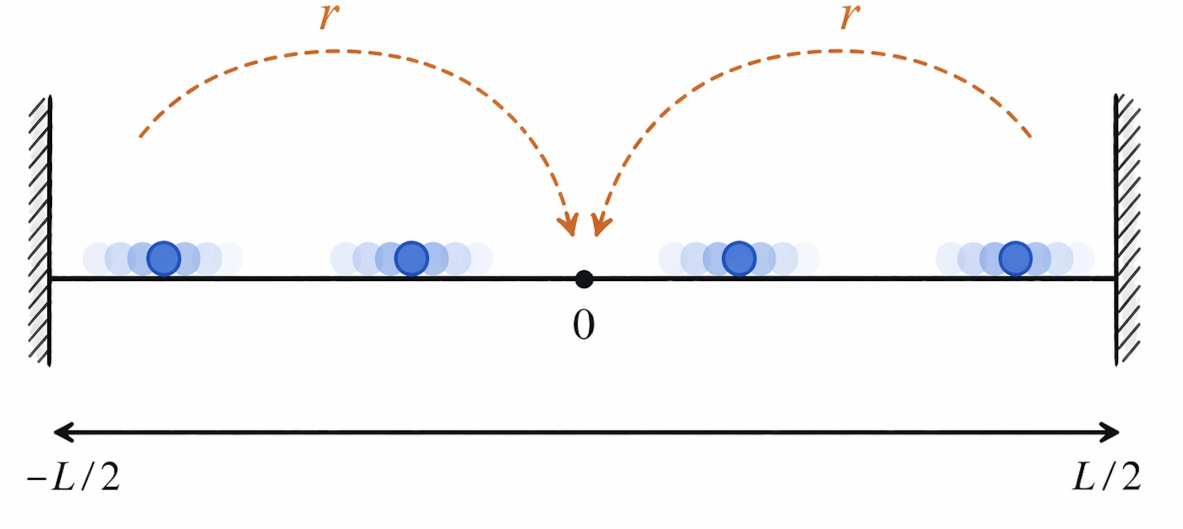}
\end{minipage}

\vspace{0.05cm}

\begin{minipage}{0.4\textwidth}
    \centering
    \includegraphics[width=\linewidth]{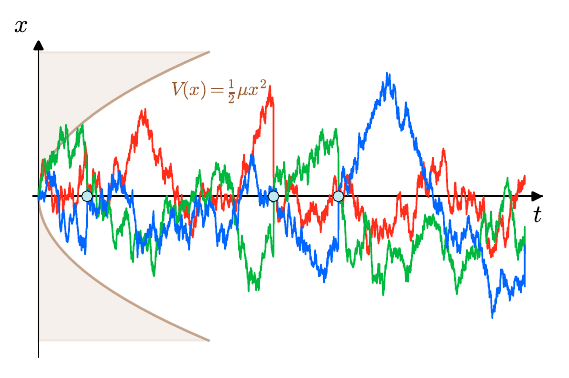}
\end{minipage}%
\hspace{0.1\textwidth}%
\begin{minipage}{0.4\textwidth}
    \centering
    \includegraphics[width=\linewidth]{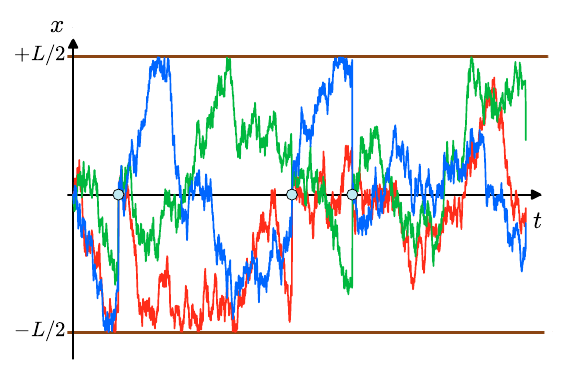}
\end{minipage}
\caption{
Schematic representation of the confined resetting gas.
The left column shows harmonic confinement (HC), where particles evolve in the potential $V(x)=\frac{1}{2}\mu x^2$, while the right column shows box confinement (BC), where particles diffuse in the interval $[-L/2,L/2]$ with reflecting boundaries.
In the upper panels, the dashed arrows indicate simultaneous reset events to the origin, occurring with rate $r$.
The lower panels show representative trajectories $x(t)$ of $N=3$ particles in the two geometries.
Between reset events, the particles evolve independently under the corresponding confined dynamics, whereas at each reset they are all brought back to $x=0$.
Reset events are marked by light blue circles.
}
\label{fig:initialimage}
\end{figure}

To address the questions raised above, we study a system of $N$ non-interacting particles diffusing in a confining potential $V(x)$ under simultaneous stochastic resetting. We focus mainly on two analytically tractable geometries: harmonic confinement (HC), corresponding to $V(x)=\frac{1}{2}\mu x^2$, and box confinement (BC), obtained by taking $V(x)=0$ inside the symmetric box $[-L/2,L/2]$ and $V(x)=+\infty$ outside it. These two settings are shown schematically in Fig.~\ref{fig:initialimage}. 
Our first goal is to understand how confinement modifies the DEC. We quantify this through exact computations of suitable correlation measures and find a rich phenomenology. Beyond these two analytically tractable cases, we also extend part of the analysis to the general family of confining potentials $V(x)=\kappa |x|^\alpha$, with $\alpha\geq0$, using numerical simulations. This class interpolates between the previously known unconfined resetting gas for $\alpha=0$, the HC case for $\alpha=2$ and setting $\kappa=\mu/2$, and the BC case in the limit $\alpha\to\infty$.

We then characterize the stationary state through several observables. We first analyze the particle density, which interpolates between the equilibrium profile imposed by confinement and the NESS of the free resetting gas~\cite{Biroli2023}.
We further investigate the effect of confinement by probing the edges of the gas, where confinement is expected to have its strongest impact. To this end, we study the extreme value statistics (EVS) \cite{Gumbel1958,Leadbetter1983,FortinClusel2015,Arnold1992,DavidNagaraja2003,MPSch2019,MajumdarSchehr2024} and the gap statistics \cite{Gumbel1958,Leadbetter1983,FortinClusel2015,Arnold1992,DavidNagaraja2003,MPSch2019,MajumdarSchehr2024}. The EVS describe the distribution of the rightmost particle, denoted by $M_1$, while the gap statistics characterize the distance $d_1=M_1-M_2$ between the first and the second rightmost particles. 
These observables are particularly useful also because they have been extensively studied both in uncorrelated equilibrium gases, and in unconfined gases with simultaneous resetting \cite{Biroli2023,BLMS2024,deMauro2026_NonPoisson,BM2026a,BMS2026_fpg,BMS2025,deMauro2026}. Consequently, they serve as natural benchmarks for analyzing two distinct effects: the impact of DEC relative to uncorrelated equilibrium systems, and the specific role of confinement compared to unconfined resetting gases.

Beyond their analytical tractability, the HC and BC geometries are also natural from an experimental perspective. Indeed, simultaneous stochastic resetting has already been implemented with colloidal particles in liquids, where harmonic traps generated by lasers are commonly used to perform resetting events \cite{Vatash2025,biroli2025exp}. This suggests that the HC case, and possibly also the BC case, could be experimentally tested with only minor modifications of existing experimental setups.

The rest of the paper is organized as follows. In Sec.~\ref{sec:TheModel}, we introduce the model and derive the exact expression for the full joint probability density function (JPDF), valid for a resetting gas in an arbitrary confining potential $V(x)$.
We summarize all the main results in Sec.~\ref{sec:MainResults}.
In Sec.~\ref{sec:AverageDensity}, we analyze the average particle density, which already captures several important features of the underlying physics in both models. We then turn to the analysis of the DEC in Sec.~\ref{sec:Correlations}, where we treat the HC and BC analytically and extend the study numerically to the broader family of potentials $V(x)=\kappa |x|^\alpha$ for $\alpha>0$.
In Sec.~\ref{sec:observables}, we investigate further the effects of confinement by probing the edges of the gas through two observables: the EVS and the gap statistics.
In Sec.~\ref{sec:GeneralPotential}, we generalize the EVS to the broader class of confining potentials $V(x)=\kappa |x|^{\alpha}$ for any $\alpha\geq0$. Finally, we conclude in Sec.~\ref{sec:conclusions}. Some details are relegated to appendices~\ref{sec:criticalalpha_APP}--\ref{sec:potenzsqrt_APP}.

\section{The model}\label{sec:TheModel}

In this section, we introduce a general framework to study a Brownian gas confined by an external potential $V(x)$ and subject to simultaneous stochastic resetting. We consider $N$ one-dimensional, non-interacting Brownian particles, with positions $\mathbf{x}=(x_1,\dots,x_N)$, diffusing with diffusion constant $D$ in the potential $V(x)$. The dynamics proceeds in two steps
\begin{enumerate}
    \item[(i)] We choose a random time $\tau_1$, drawn from the probability distribution $p(\tau)=r e^{-r\tau}$.
    During this time interval, the position of each particle, starting from the
    origin, evolves via the Langevin equation
    \begin{equation}\label{eq:general_lang_eq}
        \frac{dx_i(t)}{dt}=-V'(x_i(t))+\sqrt{2D}\,\eta_i(t),
        \qquad i=1,2,\dots,N,
    \end{equation}
    where $V(x)$ is the external confining potential, $V'(x)=\frac{dV(x)}{dx}$ and the $\eta_i(t)$'s are independent Gaussian white noises with
    \begin{equation}
        \langle \eta_i(t)\rangle =0,
        \qquad
        \langle \eta_i(t)\eta_j(t')\rangle=
        \delta_{ij}\delta(t-t') .
    \end{equation}

    \item[(ii)] At the end of the time interval $\tau_1$, the positions of all
    particles are reset to $0$ simultaneously and instantaneously.
\end{enumerate}
Following step (ii), one chooses again a random time interval $\tau_2$, drawn
from the same distribution $p(\tau)=r e^{-r\tau}$, and the particles evolve
independently via Eq. \eqref{eq:general_lang_eq} during the duration $\tau_2$, followed by an instantaneous simultaneous resetting of step (ii). One then keeps repeating the two steps (i) and (ii).

In the absence of resetting, i.e. when $r=0$, the particles evolve
independently forever according to Eq. \eqref{eq:general_lang_eq}. In that case, the joint distribution of the particle positions remains
factorized at all times $t$, namely
\begin{equation}\label{eq:propag_Npart_noreset}
    P_0(\mathbf{x},t)
    =
    \prod_{i=1}^N p_0(x_i,t),
\end{equation}
where the subscript $0$ denotes the absence of resetting, $r=0$.
Here $p_0(x,t)$ is the single-particle propagator, i.e. the probability
density of a single particle at time $t$. It evolves according to the
Fokker--Planck equation associated with the Langevin equation
\eqref{eq:general_lang_eq},
\begin{equation}\label{eq:FPeq_1part_noreset}
    \frac{\partial p_0(x,t)}{\partial t}
    =
    \frac{\partial}{\partial x}
    \left[
        V'(x)p_0(x,t)
    \right]
    +
    D\frac{\partial^2 p_0(x,t)}{\partial x^2},
\end{equation}
with initial condition $p_0(x,0)=\delta(x)$.
At long times, if $V(x)$ is sufficiently confining, the position distribution
approaches the equilibrium Gibbs--Boltzmann distribution
\begin{equation}\label{eq:equildistrib_1part_generalV}
    p_0^{\rm eq}(x)
    =
    \frac{1}{Z_0}
    \exp\left[-\frac{1}{D}V(x)\right],
\end{equation}
where
\begin{equation}
    Z_0
    =
    \int_{-\infty}^{\infty}
    dx\,
    \exp\left[-\frac{1}{D}V(x)\right]
\end{equation}
is the normalizing partition function. For example, if
\begin{equation}\label{eq:harmpotential_initialdef}
    V(x)=\frac{\mu}{2}x^2,
\end{equation}
the equilibrium distribution is Gaussian:
\begin{equation}\label{eq:gaussian_equil_1part}
    p_0^{\rm eq}(x)\big|_{\rm Harmonic}
    =
    \frac{1}{\sqrt{2\pi D/\mu}}
    \exp\left[
        -\frac{\mu x^2}{2D}
    \right].
\end{equation}

In contrast, if the confining potential is a box, i.e.
\begin{equation}\label{eq:potentiaL_BOX}
V(x)=
\begin{cases}
0, & |x|\leq \frac{L}{2},\\
+\infty, & |x|>\frac{L}{2},
\end{cases}
\end{equation}
then the equilibrium distribution is simply uniform:
\begin{equation}\label{eq:uniform_equil_1part}
    p_0^{\rm eq}(x)\big|_{\rm Box}
    =
    \frac{1}{L},
    \qquad
    -\frac{L}{2}\le x\le \frac{L}{2}.
\end{equation}

If we now switch on the resetting rate $r>0$, the joint probability
distribution function of the particle positions $P_r(\mathbf{x},t)$ no longer
factorizes as in Eq. \eqref{eq:propag_Npart_noreset}. The simultaneous resetting introduces correlations between particles. 
The finite resetting rate $r>0$ also drives the system into
a NESS at long times:
\begin{equation}
    P_r^{\rm st}(\mathbf{x})
    =
    P_r(\mathbf{x},t\to\infty),
\end{equation}
which still remains unfactorized.
The exact distribution $P_r(\mathbf{x},t)$ can be written down using the renewal property of the process \cite{EMSch2020}:
\begin{equation}\label{eq:renewaleqgenerlaV}
    P_r(\mathbf{x},t)
    =
    e^{-rt}
    \prod_{i=1}^N p_0(x_i,t)
    +
    r\int_0^t d\tau\,
    e^{-r\tau}
    \prod_{i=1}^N p_0(x_i,\tau).
\end{equation}
Its interpretation is straightforward. We divide all trajectories contributing to $P_r(\mathbf{x},t)$ into two mutually exclusive classes.
The first term corresponds to trajectories for which no reset has occurred up to time $t$. The probability of this event is $e^{-rt}$ and, conditioned on it, the system evolves exactly as in the reset-free dynamics for a duration $t$. This gives the contribution $e^{-rt}P_0(\mathbf{x},t)$.
The second term accounts for trajectories in which at least one reset has occurred before time $t$. In this scenario, only the last reset is relevant, as each resetting event acts as a renewal process that completely erases the system's prior history.
We denote by $t-\tau$ the time of the last reset, with $0<\tau<t$, so that $\tau$ is the time elapsed since that event. The probability density associated with this history is $r e^{-r\tau}d\tau$: a reset occurs at time $t-\tau$, and no further reset occurs during the interval $[t-\tau,t]$. After this last reset, all particles start again from the origin and evolve without further resetting for a duration $\tau$, so their configuration is distributed according to $P_0(\mathbf{x},\tau)$. Integrating over all possible values of $\tau$ gives the second contribution.

Taking the limit $t\to\infty$ in \eqref{eq:renewaleqgenerlaV}, the no-reset
contribution vanishes and we obtain the NESS
\begin{equation}\label{eq:NESS}
    P_r^{\rm st}(\mathbf{x})
    =
    P_r(\mathbf{x},t\to\infty)
    =
    r\int_0^\infty d\tau\,
    e^{-r\tau}
    \prod_{i=1}^N p_0(x_i,\tau),
\end{equation}
where $p_0(x,t)$ is the solution of Eq. \eqref{eq:FPeq_1part_noreset}. This has to be compared with
the $r=0$ equilibrium stationary state, where $p_0(x,t)$ relaxes to
$p_0^{\rm eq}(x)$ given by Eq. \eqref{eq:equildistrib_1part_generalV}. In that case the fully factorized stationary distribution is
\begin{equation}\label{eq:JPDFequilibrium}
    P_0^{\rm st}(\mathbf{x})=
    \prod_{i=1}^N p_0^{\rm eq}(x_i),
\end{equation}
with $p_0^{\rm eq}(x)$ given by \eqref{eq:equildistrib_1part_generalV}.
Thus, as $r$ increases from $0$, the joint probability distribution evolves
from the fully factorized equilibrium distribution in Eq. \eqref{eq:JPDFequilibrium} to the NESS
described by Eq. \eqref{eq:NESS}. This NESS is not factorized, indicating nonzero correlations between particles generated by simultaneous resetting with rate $r$.

In the rest of the paper, we mainly focus on two specific confinement geometries:
harmonic confinement (HC), corresponding to the potential in Eq. \eqref{eq:harmpotential_initialdef}, and box confinement (BC), corresponding to the hard-wall potential in Eq. \eqref{eq:potentiaL_BOX}. In some cases, we also consider the more general family of external potentials
\begin{equation}\label{eq:familypotent}
V(x)=\kappa |x|^{\alpha},
\qquad \alpha\geq0.
\end{equation}
For $\alpha>0$, this family provides a class of confining potentials, which includes the HC for $\alpha=2$ with $\kappa=\mu/2$. In the limit $\alpha\to\infty$, after an appropriate rescaling of lengths, it approaches the BC. The case $\alpha=0$ instead corresponds to the unconfined resetting gas studied in \cite{Biroli2023}. In this latter case, the reset-free dynamics, obtained by setting $r=0$, is not confined and therefore does not admit a stationary state.

\subsection{Single-particle propagator in the HC without resetting}

In this subsection we derive the single-particle propagator $p_0^{(H)}(x,t)$ for a reset-free one-dimensional particle diffusing in a potential $V(x)=\frac{1}{2}\mu x^2$.  Here the process reduces to the so-called Ornstein--Uhlenbeck (OU) process \cite{Gardiner2009,Risken1996}. The propagator is then the solution of the Fokker--Planck equation
\begin{equation}\label{eq:OUprocessFPeq}
\partial_t p_0^{(H)}(x,t)
= \mu\,\partial_x\!\big[x\,p_0^{(H)}(x,t)\big]
+ D\,\partial_x^2 p_0^{(H)}(x,t)\, .
\end{equation}
with the initial condition $p_0^{(H)}(x,0)=\delta(x)$. Instead of solving directly Eq. \eqref{eq:OUprocessFPeq}, it is actually easier to notice that the dynamics of an OU particle can be described by the following stochastic differential equation
\begin{equation}\label{eq:SDE_OUprocess}
\dot{x}(t) = -\mu x(t) + \sqrt{2D}\,\eta(t),
\end{equation}
where $x(t)$ is the position of the particle at time $t$ and $\eta(t)$ is a Gaussian white noise with unit variance.
Since the noise is Gaussian and \eqref{eq:SDE_OUprocess} is linear in $x$, the distribution $p_0^{(H)}(x,t)$ will also be Gaussian at all times $t$. It is therefore sufficient to determine the mean and the variance of this Gaussian in order to fully characterize $ p_0^{(H)}(x,t)$. This can be easily done using \eqref{eq:SDE_OUprocess}. The mean is obviously zero because of the initial condition, while the variance reads
\begin{equation}\label{eq:OU_variance}
\sigma_H^{2}(t)=\frac{D}{\mu}\left(1-\mathrm{e}^{-2\mu t}\right).
\end{equation}
Then we simply have 
\begin{equation}\label{eq:OU_propagator}
p^{(H)}_0(x,t)
=\frac{1}{\sqrt{2\pi\,\sigma_H^{2}(t)}}
\exp\!\left[-\frac{x^{2}}{2\sigma_H^{2}(t)}\right],
\end{equation}
In the stationary state, the OU particle is effectively confined within a region of typical size 
\begin{equation}
    L_H=\sigma_H(t\to\infty)=\sqrt{\frac{D}{\mu}},
\end{equation}
This length-scale $L_H$ plays a role analogous to the half-size $L_B=L/2$ in the BC.

\subsection{Single-particle propagator in the BC without resetting}

In this section we derive the single-particle probability density $p_0^{(B)}(x,t)$ for a Brownian particle confined in the box potential given in Eq.~\eqref{eq:potentiaL_BOX}, in the absence of resetting. Since in our problem we will introduce a resetting protocol that brings all particles back to the origin, it is sufficient here to determine the propagator starting from the initial condition $x_0=0$. It satisfies
\begin{equation}
\label{eq:FP_Box}
\partial_t p_0^{(B)}(x,t)=
D\,\partial_x^2 p_0^{(B)}(x,t),
\qquad
x\in\left[-\frac{L}{2},\frac{L}{2}\right],
\end{equation}
with initial condition $p_0^{(B)}(x,0)=\delta(x)$,
and reflecting boundary conditions at the walls $x=\pm L/2$. The probability current is $J(x,t)=-D\,\partial_x p_0^{(B)}(x,t)$, 
so that reflecting boundaries correspond to
\begin{equation}
\label{eq:Neumann_BC}
J\!\left(\pm\frac{L}{2},t\right)=0
\qquad
\Longleftrightarrow
\qquad
\partial_x p_0^{(B)}\!\left(\pm\frac{L}{2},t\right)=0.
\end{equation}
Because both the box and the initial condition are symmetric with respect to $x=0$, the solution is even in $x$. Therefore, we can expand it directly over the even eigenfunctions of the Laplacian satisfying the boundary condition in Eq. \eqref{eq:Neumann_BC}. We therefore look for a solution of the form
\begin{equation}
\label{eq:spectral_ansatz_box}
p_0^{(B)}(x,t)=
\sum_{n=0}^{\infty}
c_n\,\phi_n(x)\,e^{-Dk_n^2 t},
\end{equation}
where the eigenfunctions $\phi_n(x)$ read
\begin{equation}
\label{eq:eigs_even_box_n}
\phi_n(x)
=
\sqrt{\frac{2}{L}}
\cos\left(\frac{2\pi n x}{L}\right),
\qquad
k_n=\frac{2\pi n}{L}.
\end{equation}
for $n\geq1$, while for $n=0$ we have
\begin{equation}
\label{eq:eigs_even_box}
\phi_0(x)=\frac{1}{\sqrt{L}},
\qquad
k_0=0,
\end{equation}
Indeed, these functions satisfy
\begin{equation}
-\partial_x^2\phi_n(x)=k_n^2\phi_n(x),
\qquad
\phi_n'\!\left(\pm\frac{L}{2}\right)=0.
\end{equation}
The coefficients $c_n$ are fixed by the initial condition. Since $\delta(x)=\sum_{n=0}^{\infty}
\phi_n(0)\phi_n(x)$, one has $c_n=\phi_n(0)$. Therefore,
\begin{equation}
\label{eq:propagator_spectral_box}
p_0^{(B)}(x,t)
=
\sum_{n=0}^{\infty}
\phi_n(x)\phi_n(0)\,
e^{-Dk_n^2 t}.
\end{equation}
Using $\phi_0(0)=1/\sqrt{L}$ and $\phi_n(0)=\sqrt{2/L}$ for $n\geq 1$, we finally obtain
\begin{equation}
\label{eq:propag_1}
p_0^{(B)}(x,t)
=
\frac{1}{L}
+
\frac{2}{L}
\sum_{n=1}^{\infty}
\cos\left(\frac{2\pi n x}{L}\right)
\exp\!\left[-D\left(\frac{2\pi n}{L}\right)^2 t\right].
\end{equation}
We can now use the identity $\cos\theta=(e^{i\theta}+e^{-i\theta})/2$ in \eqref{eq:propag_1}, to rewrite it as
\begin{equation}
\label{eq:propag_2}
p_0^{(B)}(x,t)=\frac{1}{L}\sum_{n=-\infty}^{\infty}
\exp\!\left[-D\left(\frac{2\pi n}{L}\right)^2 t\right]\,
e^{\,i\,2\pi n x/L}.
\end{equation}
Finally, applying the Poisson summation formula 
\begin{equation}
    \sum_{n=-\infty}^{\infty} e^{-a n^2}\, e^{i b n}
=\sqrt{\frac{\pi}{a}}
\sum_{m=-\infty}^{\infty}
\exp\!\left[-\frac{(b-2\pi m)^2}{4a}\right], 
\qquad
a=\frac{4\pi^2Dt}{L^2}>0,
\qquad b=\frac{2\pi x}{L},
\end{equation}
to \eqref{eq:propag_2} we obtain the equivalent representation
\begin{equation}
\label{eq:propag_3}
p_0^{(B)}(x,t)=\frac{1}{\sqrt{4\pi Dt}}
\sum_{m=-\infty}^{\infty}
\exp\!\left[-\frac{(x-mL)^2}{4Dt}\right].
\end{equation}
In the following, we will repeatedly use both Eq.~\eqref{eq:propag_1} and Eq.~\eqref{eq:propag_3}, since the first one is useful in the large-$t$ regime, while the second one is more suited to the small-$t$ regime.

\subsection{Stationary state and control parameters}\label{sec:statstateandcontrolparam}

We now go back to the NESS of the $N$ particles system with resetting \eqref{eq:NESS}.
This stationary state is governed by the competition between two mechanisms. On the one hand, confinement alone would drive the system toward an equilibrium steady state. On the other hand, stochastic resetting in infinite space generates a NESS. The competition between these two tendencies is controlled by the ratio between the confinement length and the typical distance traveled by a particle between two consecutive reset events.
In the HC, the relevant confinement length is $L_H=\sqrt{D/\mu}$, while in the BC it is the half-size of the box, $L_B=L/2$.
The resetting length scale is instead $\ell_r=\sqrt{D/r}$. This naturally leads to the dimensionless control parameters
\begin{equation}\label{eq:deiofbothnu}
\nu_H=\frac{L_H}{\ell_r}=\sqrt{\frac{r}{\mu}}
\qquad (\text{HC}),
\qquad
\nu_B=\frac{L_B}{\ell_r}=\frac{1}{2}\sqrt{\frac{rL^2}{D}}
\qquad (\text{BC}).
\end{equation}
These parameters control the behavior of the system throughout the paper. Small values of $\nu_H$ or $\nu_B$ correspond to the confinement-dominated regime, where reset events are rare on the relaxation time scale of the confined dynamics and the system has enough time to equilibrate between two consecutive resets. 
One therefore recovers the equilibrium stationary state imposed by confinement:
a Gaussian density with variance $D/\mu$ in the HC and a uniform density on $[-L/2,L/2]$ in the BC.
In the opposite limit, $\nu_H,\nu_B\gg1$, the resetting length is much smaller than the confinement length. Resetting then occurs so frequently that particles typically remain close to the origin and do not feel either the hard walls or the harmonic trapping force. The system therefore approaches the unconfined simultaneous-resetting NESS of Ref.~\cite{Biroli2023}, whose average density is
\begin{equation}
\frac{1}{2\ell_r}e^{-|x|/\ell_r},
\qquad x\in\mathbb{R}.
\end{equation}
Thus, the same two regimes can be reached either by changing the reset rate $r$ at fixed confinement, or by changing the confinement lengths $L_H$ and $L_B$ at fixed $r$.

It is also useful to associate with each confinement geometry a characteristic relaxation time. In the HC, this time scale is \begin{equation}\label{eq:HCrelaxtime} 
t_H^\ast=\frac{1}{2\mu}, 
\end{equation} 
as can be seen from Eq.~\eqref{eq:OU_variance}. In the BC, the diffusive time needed to explore the box is instead of order \begin{equation}\label{eq:BCrelaxtime}
t^\ast_B=\frac{L^2}{4\pi^2 D}=\frac{L_B^2}{\pi^2 D}, 
\end{equation} 
as can be directly seen from the slowest decaying mode in Eq.~\eqref{eq:propag_1}.
These relaxation times quantify the time scales over which, after each reset, the reset-free confined dynamics tends to relax towards its equilibrium stationary state. They therefore compete with the typical resetting time $t^\ast_r=1/r$, which interrupts this relaxation process and drives the system away from equilibrium. Using the definitions of $\nu_H$ and $\nu_B$ given above we can also notice
\begin{equation}\label{eq:relaxandnus}
    \frac{t^\ast_H}{t^\ast_r}=r t_H^\ast=\frac{\nu_H^2}{2},
    \qquad
    \frac{t^\ast_B}{t^\ast_r}=r t^\ast_B=\frac{\nu_B^2}{\pi^2}.
\end{equation}
Thus, small values of $\nu_H$ or $\nu_B$ correspond to the regime where the system typically has enough time to equilibrate between two consecutive resets, whereas large values correspond to the regime where resetting occurs before the particles can explore the confining length.

We also mention here that a key ingredient of our analysis is the conditionally independent and identically distributed (CIID) structure of the stationary state~\cite{Biroli2023,BLMS2024}. Conditioned on the time interval $\tau$ since the last reset, all particles evolve independently according to the reset-free propagator associated with the chosen confinement. As a consequence, the stationary $N$-particle distribution derived in Eq.~\eqref{eq:NESS} can be written as
\begin{equation}\label{eq:NESS_CIIDform}
P_r^{\rm st}(\mathbf x)
= r\int_0^\infty \!d\tau\, e^{-r\tau}
\prod_{i=1}^N p_0(x_i,\tau).
\end{equation}
Although the particles are independent at fixed $\tau$, the same time $\tau$ is shared by all of them. Therefore, after averaging over this common random variable, the JPDF $P_r^{\rm st}(\mathbf{x})$ no longer factorizes. This is the origin of the DEC studied in this work, and it is also the structure that makes many observables exactly computable.

\section{Main results}\label{sec:MainResults}

In this section we summarize the main results of this paper and highlight how the two confinement geometries, HC and BC, compare. The stationary state is controlled by the dimensionless parameters $\nu_H$ and $\nu_B$, introduced in Eq.~\eqref{eq:deiofbothnu}, which measure the ratio between the confinement length and the resetting length in the two cases. We first discuss the average density, then the correlation measures quantifying the DEC, and finally the edge observables, namely EVS and gap statistics.

We begin with the average particle density. We compute it exactly in both geometries, see Eq.~\eqref{eq:density_HARM_final} for the HC and Eq.~\eqref{eq:density_BOX_final} for the BC, together with Fig.~\ref{fig:density_comparison}. In both cases, the density admits a scaling form controlled by $\nu_H$ or $\nu_B$, and interpolates between the equilibrium profile at small $\nu_H$ or $\nu_B$ and the exponential profile of the unconfined resetting gas at large $\nu_H$ or $\nu_B$.

We then turn to fluctuations and DEC in the stationary state. Simultaneous resetting induces strong correlations between particles, even though they do not interact directly. To quantify these correlations, we introduce several measures. The quantities $\mathcal{C}_1$ and $\mathcal{C}_2$, introduced in Eq.~\eqref{eq:C2C1_def_together}, measure the total single-particle fluctuations and the connected correlations between two particles, respectively. The normalized coefficient $A=\mathcal{C}_2/\mathcal{C}_1\in[0,1]$, defined in Eq.~\eqref{eq:def_of_a}, quantifies the fraction of the total fluctuations that is collective. For clarity, we denote the corresponding functions by $A_H(\nu_H)$ in the HC and by $A_B(\nu_B)$ in the BC, see Fig.~\ref{fig:a_Harmonic_vs_Box}. We study these quantities as functions of the control parameter $\nu_H$ in the HC, or $\nu_B$ in the BC.

For both HC and BC, the connected inter-particle correlator $\mathcal{C}_2$ exhibits a similar qualitative dependence on the control parameter $\nu_H$ or $\nu_B$ (see Figs.~\ref{fig:F1_F2_Harmonic} and \ref{fig:F1_F2_BOX}). In particular, $\mathcal{C}_2$ vanishes in both limits of weak and strong resetting, i.e. for small and large values of $\nu_H$ or $\nu_B$, with the same leading-order asymptotic behavior in HC and BC, and reaches a maximum at intermediate values.

A markedly different behavior emerges when one considers the normalized correlation coefficient. In the HC, $A_H(\nu_H)$ increases monotonically with increasing $\nu_H$ and approaches the unconfined value from below for large $\nu_H$. In the BC, instead, $A_B(\nu_B)$ is non-monotonic: it exceeds the unconfined value at intermediate $\nu_B$ before decreasing back to it at large values of $\nu_B$ (see Fig.~\ref{fig:a_Harmonic_vs_Box}). This overshoot is a consequence of the fact that the hard walls cut off rare trajectories in which some particles would otherwise wander far from the bulk of the gas between two consecutive reset events, while leaving the typical bulk trajectories almost unaffected. In Sec.~\ref{subsec:normalizedcorrel_dim1} we explain this phenomenon in more detail.
In the HC, by contrast, rare trajectories that wander far from the origin are less strongly suppressed than in the BC, and therefore not enough to produce the same overshoot. As a result, the normalized correlation coefficient approaches the unconfined value monotonically, without developing an overshoot.

In order to test this interpretation, we extend the analysis in two ways. First, in Sec.~\ref{sec:correl_generalalpha} we consider a generic confining potential of the form given in Eq. \eqref{eq:familypotent}.
This family continuously connects several important limits. It recovers the HC case for $\alpha=2$ and $\kappa=\mu/2$, and it approaches the BC, with $L_B=1$, in the limit $\alpha\to\infty$. We show that the stationary correlations remain monotonic as a function of the ratio $\nu_{\alpha}=L_{\alpha}/\ell_r$ between the confining length scale $L_{\alpha}=\left(\frac{D}{\kappa}\right)^{1/\alpha}$ and the resetting length scale $\ell_r=\sqrt{D/r}$ for $0<\alpha<\alpha_c$, with $\alpha_c=1+\sqrt{5}=3.236\dots$, whereas they become non-monotonic for $\alpha>\alpha_c$ (see Fig.~\ref{fig:Anualpha}). This is because, as $\alpha$ increases, the confining force suppresses increasingly efficiently the rare excursions far away from the origin, while affecting the typical "bulk" trajectories much less. Since these rare excursions are precisely the ones that contribute most to decorrelating the particles, their suppression enhances the normalized correlation coefficient and ultimately leads to the overshoot.

Second, we consider the BC geometry and generalize it to higher dimensions, considering a disk for $d=2$ and a sphere for $d=3$, both with reflecting boundaries. Since rare large excursions carry a larger statistical weight in higher dimensions \cite{Gardiner2009}, the overshoot is expected to become more pronounced. This is confirmed in Fig.~\ref{fig:aind1d2d3}, which shows that the size of the overshoot increases with increasing $d$, at least up to $d=3$.

We then investigate other observables in the NESS in order to better understand the structure of the gas and the effect of different confinement geometries on it. In particular, we study the EVS and gap statistics, namely the statistics of the rightmost particle $M_1$ and of the spacing between the first and second rightmost particles, $d_1=M_1-M_2$.
Interestingly, the different correlation structures arising in the two confinement geometries (HC and BC) are also reflected in these edge observables. In Sec.~\ref{sec:EVS}, we study the probability density of the rightmost particle $M_1$, denoted by $\operatorname{Prob}\{M_1=m\}$.
For the HC, we find that for large $N$, the rightmost particle is typically at a distance of order $M_1=O(\sqrt{\ln N})$, which is the same order as in the unconfined resetting gas \cite{Biroli2023}. However, the shape of the distribution is drastically different from the unconfined case. In the unconfined geometry, the rescaled distribution has an unbounded support and a tail that decays faster than exponentially. By contrast, in HC the limiting distribution has a bounded support, even though particles are in principle free to explore the whole real line.
Remarkably, the distribution in HC also exhibits a shape transition at its upper edge as the control parameter $\nu_H$ is varied. There exists a critical value $\nu_H^c=\sqrt{2}$, such that for $\nu_H>\nu_H^c$, the scaling function vanishes as $z\to1$, whereas for $\nu_H<\nu_H^c$ it diverges. Exactly at $\nu_H=\nu_H^c$, the probability density remains finite at the edge (see Eq. \eqref{eq:SHforHC} and Fig. \ref{fig:EVS}).
In the BC, the EVS are completely different. Typically, the maximum $M_1$
is located very close to the right boundary $L/2$. For this reason, we define
the dimensionless distance from the boundary as
\begin{eqnarray}
    \Delta=\frac{1}{2}-\frac{M_1}{L}.
\end{eqnarray}
This random variable can in principle take values in $[0,1]$, since $M_1$
can in principle lie anywhere in the box, $[-L/2,L/2]$. However, in the
large-$N$ EVS regimes considered below, the relevant sector is
$0\leq M_1\leq L/2$, or equivalently $0\leq \Delta\leq 1/2$. 
In the typical regime, $\Delta=O(1/N)$. Without resetting, particles in the box would simply equilibrate to a uniform distribution, and the corresponding EVS would be described by the Weibull law. In that case, the scaled distance from the boundary, $N\Delta$, has an exponential tail, so large deviations from the boundary are strongly suppressed. Here, instead, simultaneous resetting changes the scaling function completely: it develops a very broad algebraic tail with logarithmic corrections. In particular, all moments of $N\Delta$, including the mean, diverge. This means that the gas is "fluffy": although $M_1$ is typically extremely close to $L/2$, anomalously large distances from the boundary are much more likely than in the equilibrium case. For this reason, we also investigate the large-deviation properties of the variable $N\Delta$ (see Eq. \eqref{eq:FULLEVSBCwithLD_M1} and Fig. \ref{fig:EVSBC_FULLL}).

We then turn to the gap statistics in Sec.~\ref{sec:GAPstats}, which again reveal strong differences between the two geometries. In the HC, the first gap $d_1$ is typically of order $O(1/\sqrt{\ln N})$, and its distribution decays exponentially for large values.
In contrast, in the BC the typical gap is much smaller, of order $O(1/N)$. However, its distribution possesses a much broader tail, which again leads to a divergent first moment. This shows that, although the typical gap is significantly smaller in the BC than in the HC, atypically large gaps are much more likely to occur in the BC. For this reason, as in the EVS problem, we also analyze the large-deviation properties of the first gap in the BC (see Eq. \eqref{eq:recapGAPBC_begin} and Fig. \ref{fig:GAPBC_FULLL}).

Finally, in Sec.~\ref{sec:GeneralPotential} we consider again the potential $V(x)=\kappa |x|^{\alpha}$.
Using extensive numerical simulations, together with some physical arguments, we provide evidence for the existence of three universality classes for the EVS within the family of potentials defined in \eqref{eq:familypotent}. One class corresponds to $0\leq \alpha \leq 1$, while another one applies to $1<\alpha<\infty$. The BC limit, namely $\alpha\to\infty$, is instead singular and defines a third distinct case. The reason is that the BC limit is the only hard confinement, namely the only case with a strictly bounded accessible region, whereas every finite $\alpha$ corresponds to soft confinement, where particles remain confined but can still, in principle, explore arbitrarily large distances.
In particular, for $0\leq \alpha \leq 1$, we argue that the maximum always scales as $M_1=O(\sqrt{\ln N}),$ and that the associated probability distribution is also universal, coinciding with the one found in \cite{Biroli2023} for the unconfined resetting gas.
For $\alpha>1$, by contrast, we expect $M_1$ to be of order $M_1=O\!\left((\ln N)^{1/\alpha}\right)$, while the limiting distribution, once $M_1$ is rescaled by a factor $O\!\left((\ln N)^{1/\alpha}\right)$, becomes bounded and exhibits a shape transition at the upper boundary. Therefore, this regime shares the same qualitative features as the HC case corresponding to $\alpha=2$.
The origin of the change of behavior at $\alpha=1$ can be understood from the large distance properties of the confining force $-V'(x)=-\alpha\kappa\,x|x|^{\alpha-2}$.
For $\alpha<1$, the magnitude of the force vanishes as $|x|\to\infty$, whereas for $\alpha>1$ it increases with distance. Therefore, for $\alpha<1$ it is natural that the statistics of the maximum remain the same as in the $\alpha=0$ case, since extreme events probe particles located at the largest distances, where the confining force becomes negligible.
By contrast, for $\alpha>1$, the restoring force grows with the distance from the origin. As a consequence, the EVS behavior is expected to be fundamentally different. These results are summarized in Table \ref{tab:EVS_general_alpha} in Sec.~\ref{sec:GeneralPotential}.

Related problems involving EVS for Brownian particles in confining potentials have recently been studied in the context of Brownian reshuffling~\cite{BK2025,BKM2026}, while the distribution of the time at which the maximum is reached in stationary stochastic processes was analyzed in Ref.~\cite{MMS2022}.

\section{Average density of particles}\label{sec:AverageDensity}

We begin by analyzing the average particle density, which by symmetry coincides with the one-point marginal of the stationary JPDF $P_r^{\rm st}(\mathbf x)$.
As discussed in Sec.~\ref{sec:MainResults}, the stationary behavior is controlled by the competition between confinement and resetting, governed by the dimensionless parameters $\nu_H$ (HC) and $\nu_B$ (BC), defined in Eq. \eqref{eq:deiofbothnu}. The density provides the simplest observable where this competition is directly visible.
For $\nu_H, \nu_B \ll 1$, resetting is ineffective and the system relaxes close to the equilibrium stationary state imposed by the confinement (Gaussian in the harmonic well, uniform in the box).
For $\nu_H, \nu_B \gg 1$, resetting dominates and the density approaches the unconfined resetting NESS, characterized by an exponential decay away from the origin.
More precisely, we define the average particle density $\rho_N(x)$ as 
\begin{equation}\label{eq:defdensitygeneral}
    \rho_N(x)
    \equiv \frac{1}{N}
    \left\langle \sum_{i=1}^{N} \delta(x - x_i) \right\rangle
    = \int\!\!\int dx_2 \cdots dx_N \,
    P_r^{\rm st}(x,x_2,\dots,x_N).
\end{equation}
With this convention, $\rho_N(x)$ is normalized to unity.
The average is taken with respect to the stationary JPDF $P_r^{\rm st}(\mathbf{x})$ in Eq.~\eqref{eq:NESS}, and the integral runs over the allowed domain of the remaining coordinates, namely $\mathbb{R}$ in the HC and $[-L/2,L/2]$ in the BC. The last equality follows from permutation symmetry. Using Eq.~\eqref{eq:NESS}, we then obtain
\begin{equation}\label{eq:densitydef_HC}
    \rho_H(x)=r \int_{0}^{\infty} d\tau \, e^{-r\tau}\, p_0^{(H)}(x,\tau),
\end{equation}
for the HC, and, equivalently,
\begin{equation}\label{eq:densitydef_BC}
    \rho_B(x)=r\int_0^{\infty} \, d\tau \, e^{-r\tau}\, p_0^{(B)}(x,\tau),
\end{equation}
for the BC.
Here we have dropped the $N$ dependence since both $\rho_H$ and $\rho_B$ are independent of $N$.

\subsection{Harmonic confinement}

\begin{figure}[h!]
    \centering

    \begin{minipage}{0.48\textwidth}
        \centering
        \includegraphics[width=\textwidth]{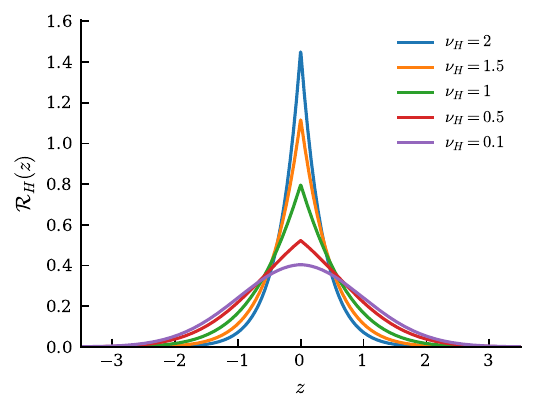}
    \end{minipage}
    \hfill
    \begin{minipage}{0.48\textwidth}
        \centering
        \includegraphics[width=\textwidth]{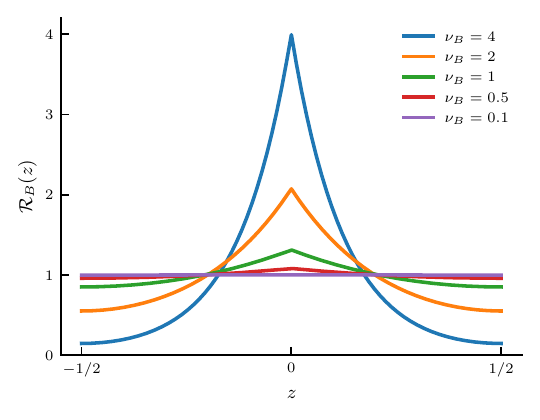}
    \end{minipage}

    \vspace{-0.3cm}
    \caption{Comparison between the scaling functions of the stationary density in the HC (left), $\mathcal{R}_H(z)$, and in the BC (right), $\mathcal{R}_B(z)$, for different values of the control parameters $\nu_H$ and $\nu_B$. The explicit expressions of $\mathcal{R}_H(z)$ and $\mathcal{R}_B(z)$ are given in Eqs. \eqref{eq:density_HARM_final} and \eqref{eq:density_BOX_final}, respectively.}
    \label{fig:density_comparison}
\end{figure}

By substituting \eqref{eq:OU_propagator} in \eqref{eq:densitydef_HC} and performing the change of variable $y=e^{-2\mu\tau}$ we get
\begin{equation}
\rho_H(x)=\frac{r}{2\mu}\sqrt{\frac{\mu}{2\pi D}}
\int_{0}^{1}dy\; y^{\frac{r}{2\mu}-1}(1-y)^{-1/2}\,
\exp\!\left(-\frac{\mu x^{2}}{2D}\,\frac{1}{1-y}\right).
\end{equation}
Setting $t=\dfrac{y}{1-y}$, the above expression becomes
\begin{equation}
\rho_H(x)=\frac{r}{2\mu}\sqrt{\frac{\mu}{2\pi D}}\;e^{-c}
\int_{0}^{\infty}dt\; e^{-ct}\,t^{\,b-1}(1+t)^{-b-\frac12}, \qquad b=\frac{r}{2\mu},\qquad c=\frac{\mu x^{2}}{2D}.
\end{equation}
We can now use the identity
\begin{equation}
\int_{0}^{\infty}dt\; e^{-ct}\,t^{\,b-1}(1+t)^{-b-\frac12}
=\Gamma(b)\,U\!\left(b,\frac12,c\right),
\end{equation}
where $U(\alpha,\beta,\gamma)$ is Tricomi's confluent hypergeometric function, defined by the integral representation
\begin{equation}
U(\alpha,\beta,\gamma)=\frac{1}{\Gamma(\alpha)}
\int_{0}^{\infty} dt\; e^{-\gamma t}\,t^{\alpha-1}(1+t)^{\beta-\alpha-1},
\end{equation}
and $\Gamma(z)$ is the Euler's Gamma function defined as $\Gamma(z) = \int_{0}^{\infty} t^{\,z-1} e^{-t}\, dt$.
This leads to 
\begin{equation}
\rho_H(x)=\frac{r}{2\mu}\sqrt{\frac{\mu}{2\pi D}}\;
e^{-\frac{\mu x^{2}}{2D}}\,
\Gamma\!\left(\frac{r}{2\mu}\right)\,
U\!\left(\frac{r}{2\mu},\frac12,\frac{\mu x^{2}}{2D}\right),
\end{equation}
or, using the property $x\Gamma\!\left(x\right)=\Gamma\!\left(1+x\right)$ of the Gamma function, we get
\begin{equation}\label{eq:density_HARM_final}
\rho_H(x)=\frac{1}{L_H}\,\mathcal{R}_H\!\left(\frac{x}{L_H}\right),
\qquad 
\mathcal{R}_H(z)=\frac{\Gamma\!\left(1+\frac{\nu_H^{2}}{2}\right)}{\sqrt{2\pi}}\,
\exp\!\left(-\frac{z^{2}}{2}\right)\,
U\!\left(\frac{\nu_H^{2}}{2},\frac12,\frac{z^{2}}{2}\right),
\end{equation}
where we recall $\nu_H=\sqrt{\frac{r}{\mu}}$.
The asymptotic behavior of the scaling function $\mathcal{R}_H(z)$ reads
\begin{equation}
\mathcal{R}_H(z) \approx
\begin{cases}
\displaystyle
\frac{\Gamma\!\left(1+\frac{\nu_H^2}{2}\right)}
{\sqrt{2}\,\Gamma\!\left(\frac{1+\nu_H^2}{2}\right)}
-\frac{\nu_H^2}{2}|z|,
& z\to 0, \\[12pt]
\displaystyle
\frac{\Gamma\!\left(1+\frac{\nu_H^2}{2}\right)}{\sqrt{2\pi}}\,
\left(\frac{2}{z^2}\right)^{\frac{\nu_H^2}{2}}
\exp\!\left(-\frac{z^2}{2}\right),
& |z|\to \infty.
\end{cases}
\end{equation}
On the other hand, the limiting density profiles obtained by varying the dimensionless parameter $\nu_H$ are
\begin{equation}
\mathcal{R}_H(z)\approx
\begin{cases}
\dfrac{1}{\sqrt{2\pi}}\exp\!\left(-\dfrac{z^{2}}{2}\right),
& \nu_H \ll 1,\\[8pt]
\dfrac{\nu_H}{2}\exp\!\left(-\nu_H|z|\right),
& \nu_H \gg 1.
\end{cases}
\end{equation}
as shown in Fig. \ref{fig:density_comparison}.
For $\nu_H=L_H/\ell_r\ll1$, the resetting length is much larger than the harmonic confinement length. Resetting is then rare on the relaxation time scale of the trap, and the system relaxes to the equilibrium Gaussian distribution. In the opposite limit, $\nu_H\gg1$, particles are reset before they can explore the scale $L_H$. The trap is therefore effectively irrelevant, and the density reduces to the exponential profile of the unconfined resetting NESS.

We note here that Eq.~\eqref{eq:density_HARM_final} coincides with the single particle stationary distribution obtained in Ref.~\cite{Pal2015conf}, which studies the steady state of a single particle in a confining harmonic potential under stochastic resetting. Although we consider an $N$ particle system, $\rho_H(x)$ is a single particle marginal. Indeed, using the CIID form in Eq.~\eqref{eq:NESS_CIIDform}, integration over the remaining $N-1$ coordinates simply leaves the single particle propagator averaged over the reset age. Therefore, the average density of our $N$ particle system is the same as the steady state density of a single particle.
Ref.~\cite{Pal2015conf} treats the more general case in which the reset position $x_0$ does not necessarily coincide with the center of the harmonic potential. Our result is obtained by setting $x_0=0$, so that the reset point and the trap minimum coincide. The expression in Ref.~\cite{Pal2015conf} is written in terms of Hermite polynomials of negative order, denoted by $H_{\lambda}$ here, and can be rewritten in terms of Tricomi's function by using the identity
\begin{equation}
U\!\left(\frac{r}{2\mu},\frac{1}{2},\frac{\mu x^2}{2D}\right)
=
2^{\frac{r}{\mu}}\,
H_{-\frac{r}{\mu}}\!\left(\sqrt{\frac{\mu}{2D}}\,|x|\right).
\end{equation}
This gives the compact form in Eq.~\eqref{eq:density_HARM_final}.

\subsection{Box confinement}

By substituting Eq. \eqref{eq:propag_1} into Eq. \eqref{eq:densitydef_BC} we get
\begin{equation}
    \rho_B(x)=\frac{1}{L} \left( 1+2\sum_{n=1}^{\infty} \frac{r}{r+k_n^2D} \cos(k_nx) \right)
\end{equation}
with $k_n=\frac{2\pi n}{L}$.
We can now use the identity
\begin{equation}
    \sum_{n=1}^{\infty} \frac{\cos(n\theta)}{n^{2}+\zeta^{2}}
    = \frac{\pi}{2\zeta}\,
    \frac{\cosh\!\big(\zeta(\pi-\theta)\big)}{\sinh(\zeta\pi)}
    \;-\; \frac{1}{2\zeta^{2}} \, .
\end{equation}
to rewrite $\rho_B(x)$ as
\begin{equation}\label{eq:density_BOX_final}
    \rho_B(x)= \frac{1}{L} \mathcal{R}_B\left(\frac{x}{L}\right), \qquad 
    \mathcal{R}_B(z)=\frac{\nu_B\,\cosh\!\left(\nu_B(1 - 2|z|)\right)}{\sinh\!\left(\nu_B\right)},
   \qquad |z|\le \frac{1}{2},
\end{equation}
where we recall $\nu_B=\frac{1}{2}\sqrt{\frac{rL^2}{D}}.$ The asymptotic behavior of the scaling function $\mathcal{R}_B(z)$ reads
\begin{equation}
\mathcal{R}_B(z) \approx
\begin{cases}
\nu_B \coth(\nu_B)
-2\nu_B^2 |z|,
& z\to 0, \\[6pt]
\displaystyle
\frac{\nu_B}{\sinh(\nu_B)}
\left[
1+2\nu_B^2\left(\frac{1}{2}-|z|\right)^2
\right],
& |z|\to \frac{1}{2}.
\end{cases}
\end{equation}
On the other hand, the limiting density profiles obtained by varying the dimensionless parameter $\nu_B$ are
\begin{equation}
\mathcal{R}_B(z)\approx
\begin{cases}
1 
+ \nu_B^2\!\left[\frac{\left(1-2|z|\right)^{2}}{2}
- \dfrac{1}{6}\right],
& \nu_B \ll 1,\\[6pt]
\nu_B\,e^{-2\nu_B|z|}, & \nu_B \gg 1.
\end{cases}
\end{equation}
Thus, the density profile interpolates smoothly between the two regimes as the parameter $\nu_B$ is varied (see Fig.~\ref{fig:density_comparison}). For small $\nu_B$, the resetting length scale $\ell_r$ is much larger than the confinement length $L/2$. In this regime, confinement dominates: between two consecutive resets, particles have enough time to relax to equilibrium inside the box. The stationary density therefore approaches the uniform equilibrium distribution, $\rho_B(x)=1/L$.
In contrast, for $\nu_B \gg 1$, one has $\ell_r \ll L/2$. Resetting events are then frequent and localize the particles near the resetting position before they can explore the boundaries. The walls become effectively irrelevant, and the stationary density coincides with that of the unconfined resetting NESS \cite{EM2011,Biroli2023}.

\section{Correlations}\label{sec:Correlations}

We now turn to the study of correlations in the stationary state. We show how nontrivial effects emerge as the corresponding control parameter, $\nu_H$ or $\nu_B$, is varied. In addition, we uncover profound differences in the structure of the DEC in the two geometries.
As a first step, we introduce suitable observables to characterize the structure of DEC in our system. We need a measure of single-particle fluctuations and a measure of correlations between pairs of particles. The simplest natural choice would be
\begin{align}
    &\langle x_i^{2}\rangle - \langle x_i\rangle^{2},\\[2mm]
    &\langle x_i x_j\rangle - \langle x_i\rangle\,\langle x_j\rangle,
    \qquad i \neq j,
\end{align}
which correspond respectively to the variance of the position of particle $i$ and the covariance between the positions of two distinct particles $i$ and $j$. The notation $\langle \cdot \rangle$ denotes averages with respect to the JPDF in the NESS, given in Eq.~\eqref{eq:NESS}.
However, our system is invariant under the flip symmetry $x_i \to -x_i$. As a consequence, the first-order covariance vanishes identically. We therefore move to second-order quantities and consider
\begin{align}\label{eq:C2C1_def_together}
    &\mathcal{C}_1 = \langle x_i^{4}\rangle - \langle x_i^{2}\rangle^{2},\\[2mm]
    &\mathcal{C}_2 = \langle x_i^{2} x_j^{2}\rangle - \langle x_i^{2}\rangle\,\langle x_j^{2}\rangle,
    \qquad i \neq j.
\end{align}
Since all particles are statistically equivalent, these observables do not depend on the indices $i$ and $j$, and we omit them in the following. They play a role analogous to the previous ones: $\mathcal{C}_1$ quantifies the total fluctuations of a single particle, while $\mathcal{C}_2$ measures the correlations between distinct particles.
It is also useful to introduce the normalized correlation coefficient
\begin{equation}\label{eq:def_of_a}
    A = \frac{\mathcal{C}_2}{\mathcal{C}_1}.
\end{equation}
It can be shown, see \cite{BM2026a,deMauro2026}, that $A \in [0,1]$: it vanishes when $x_i^2$ and $x_j^2$ are statistically independent, and approaches one in the limit of perfect correlation. Therefore, this observable quantifies the fraction of the total single-particle fluctuations that is shared between different particles. 
This normalized coefficient in \eqref{eq:def_of_a} has proved particularly useful in previous studies on DEC \cite{deMauro2026,BM2026a}, as it isolates the collective part of the fluctuations from the overall fluctuation scale.
We now introduce the notation $\langle \cdot \rangle_0$ to denote averages over the $N$-particles reset-free propagator at a fixed time $t = \tau$, defined in Eq.~\eqref{eq:propag_Npart_noreset}. For instance, the $q$-th moment of the single-particle propagator in the absence of resetting reads
\begin{equation}\label{eq:sigmasquared_def_clean}
    \langle x_i^q(\tau) \rangle_0 = \int\!\! \int d\mathbf{x} \,x_i^q\,P_0(\mathbf{x},\tau) = \int_{-\Lambda^*}^{\Lambda^*} dx_i \, x_i^q \, p_0(x_i,\tau),
\end{equation}
where $p_0(x,\tau)$ is the reset-free single-particle propagator. It is given by Eq.~\eqref{eq:OU_propagator} for the HC, with $\Lambda^* = \infty$, and by Eqs.~\eqref{eq:propag_1} or \eqref{eq:propag_3} for the BC, with $\Lambda^* = L/2$.
With this definition, it follows directly from Eq.~\eqref{eq:NESS} that the $\mathcal{C}_1$ and $\mathcal{C}_2$ can be rewritten as
\begin{equation}\label{eq:C1statasvar}
    \mathcal{C}_1 =
    r\int_0^{\infty}  \, d\tau \, e^{-r\tau} \, \langle x^4(\tau) \rangle_0-\left(r
    \int_0^{\infty}  \, d\tau \, e^{-r\tau} \, \langle x^2(\tau) \rangle_0
    \right)^2,
\end{equation}
and
\begin{equation}\label{eq:C2asvarianceofsigma2}
    \mathcal{C}_2 =
    r\int_0^{\infty}  \, d\tau \, e^{-r\tau} \, \langle x^2(\tau) \rangle_0^2-\left(r\int_0^{\infty} \, d\tau \, e^{-r\tau} \, \langle x^2(\tau) \rangle_0\right)^2,
\end{equation}
where we have dropped the $i$-dependence of $\langle x_i^q(\tau) \rangle_0$ since it is independent of $i$ for all $q$.
Equation $\eqref{eq:C2asvarianceofsigma2}$ can be interpreted as the variance, with respect to the random reset age $\tau$, of the observable $\langle x^2(\tau) \rangle_0$, i.e. the typical squared displacement after a time interval $\tau$ elapsed since the last reset. This shows that correlations arise from the fluctuations of the reset age $\tau$. Since this random time is common to all particles, its fluctuations are shared across the entire system, thereby inducing correlations.
Both $\mathcal{C}_1$ and $\mathcal{C}_2$ depend on the resetting length $\ell_r$ and on the relevant confinement length, $L_H$ in the HC or $L_B$ in the BC, but we keep this dependence implicit in the notation. Their ratio is instead dimensionless and it depends only on the corresponding ratio of length scales. We will then denote in the following $A_H(\nu_H)$ in the HC and $A_B(\nu_B)$ in the BC, with $\nu_H=L_H/\ell_r$ and $\nu_B=L_B/\ell_r$.

\subsection{Stationary correlations}

We start by analyzing $\mathcal{C}_1$ and $\mathcal{C}_2$ from Eqs.~\eqref{eq:C1statasvar} and \eqref{eq:C2asvarianceofsigma2}. This requires the reset-free moments $\langle x^2(\tau)\rangle_0$ and $\langle x^4(\tau)\rangle_0$ at fixed reset age $\tau$. By direct integration, for the HC we obtain
\begin{equation}
\langle x_i^2 (\tau) \rangle_0^{(H)}=
\frac{D}{\mu}\left(1-e^{-2\mu\tau}\right),
\qquad 
\langle x_i^4 (\tau) \rangle_0^{(H)}=
3 \left( \frac{D}{\mu} \right)^2
\left(1-e^{-2\mu\tau}\right)^2.
\end{equation}
\begin{figure}[h!] 
    \centering
    \includegraphics[width=0.75\textwidth]{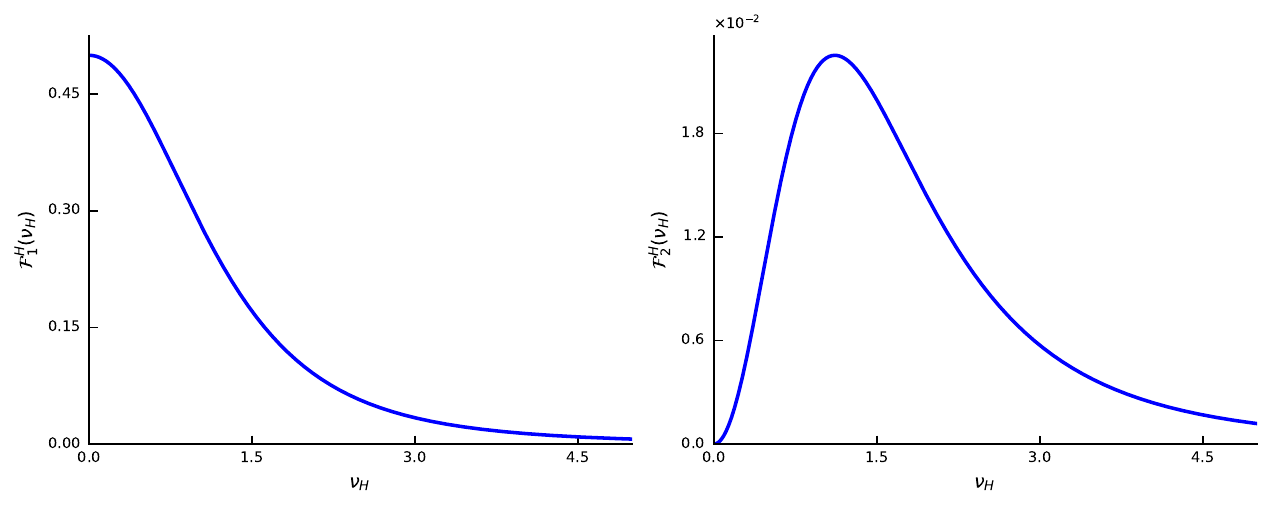}
    \caption{Plot of the two functions $\mathcal{F}^H_1(\nu_H)$ (left) and $\mathcal{F}^H_2(\nu_H)$ (right) given respectively in \eqref{eq:C1OUprocess} and \eqref{eq:C2OUprocess}. The function $\mathcal{F}_2^H(\nu_H)$ has a maximum at a value $\nu_H^{*}=\sqrt{\sqrt{5}-1}=1.1118\dots .$}
    \label{fig:F1_F2_Harmonic}
\end{figure}
For the BC instead, the averages are taken over the propagator \eqref{eq:propag_1}, which leads to
\begin{equation}\label{eq:x2_moment}
\langle x_i^2(\tau) \rangle_0^{(B)}=
\frac{L^2}{12}+
\frac{L^2}{\pi^2}
\sum_{n=1}^{\infty}
\frac{(-1)^n}{n^2}
\exp\!\left[
-\frac{4\pi^2 D n^2}{L^2}\,\tau
\right].
\end{equation}
and
\begin{equation}\label{eq:4momet_Box}
\langle x_i^4(\tau) \rangle_0^{(B)}
=
\frac{L^4}{80}
+ \frac{L^4}{2\pi^2}
\sum_{n=1}^{\infty}\frac{(-1)^n}{n^2}
\exp\!\left(-\frac{4\pi^2 D n^2}{L^2}\tau\right)
-
\frac{3L^4}{\pi^4}
\sum_{n=1}^{\infty}\frac{(-1)^n}{n^4}
\exp\!\left(-\frac{4\pi^2 D n^2}{L^2}\tau\right).
\end{equation}
By employing Eqs. \eqref{eq:C2asvarianceofsigma2} and \eqref{eq:C1statasvar}, we obtain, for the HC
\begin{equation}\label{eq:C1OUprocess}
\mathcal C_{1}^H=4\,L_H^4\,
\mathcal{F}_1^H\!\left(\sqrt{\frac{r}{\mu}}\right),
\qquad
\mathcal{F}_1^H(\nu_H)=
\frac{5\nu_H^2+8}{(\nu_H^2+2)^{2}(\nu_H^2+4)},
\end{equation}
and
\begin{equation}\label{eq:C2OUprocess}
\mathcal C_{2}^H=4\,L_H^4\,
\mathcal{F}_2^H\!\left(\sqrt{\frac{r}{\mu}}\right),
\qquad
\mathcal{F}_2^H(\nu_H)=
\frac{\nu_H^2}{(\nu_H^2+2)^{2}(\nu_H^2+4)},
\end{equation}
where we recall $L_H=\sqrt{\frac{D}{\mu}}$.
Analogously, for the BC, we obtain
\begin{equation}\label{eq:F1Boxdefinition}
\mathcal{C}_1^B=L^4
\mathcal{F}_1^B\!\left(\frac{1}{2} \sqrt{\frac{rL^2}{D}}\right),
\qquad
\mathcal{F}_1^B(\nu_B)=
\frac{1}{4\nu_B^4}\left[5-
\frac{\nu_B(\nu_B^2+4)}{\sinh \nu_B}-
\frac{\nu_B^2}{\sinh^2 \nu_B}\right],
\end{equation}
and
\begin{equation}\label{eq:F2Boxdefinition}
\mathcal{C}_2^B
=
L^4
\mathcal{F}_2^B\!\left( \frac{1}{2} \sqrt{\frac{rL^2}{D}}\right),
\qquad
\mathcal{F}_2^B(\nu_B)
=
\frac{1}{4\nu_B^4}
\left[
1
-
\frac{\nu_B^2}{\sinh^2 \nu_B}
-
\frac{\nu_B^3}{3\sinh \nu_B}
-
\frac{4\nu_B^4}{\pi^3} S(\nu_B)
\right],
\end{equation}
where
\begin{equation}
S(\nu_B)=
\sum_{m=1}^{\infty}
\frac{(-1)^m}{
m^2 \sqrt{m^2 + (\nu_B/\pi)^2}
\sinh\!\left(
\pi \sqrt{m^2 + (\nu_B/\pi)^2}
\right)}.
\end{equation}
\begin{figure}[h!] 
    \centering \includegraphics[width=0.75\textwidth]{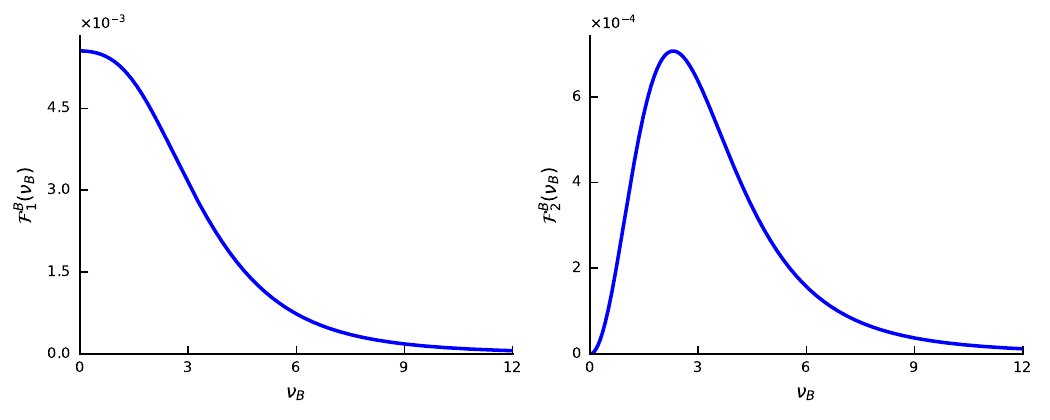} \caption{Plot of the two functions $\mathcal{F}_1^B(\nu_B)$ (left) and $\mathcal{F}_2^B(\nu_B)$ (right) given respectively in \eqref{eq:F1Boxdefinition} and \eqref{eq:F2Boxdefinition}. The function $\mathcal{F}_2^B(\nu_B)$ has a maximum at a value $\nu_B^{*}\approx  2.3007\dots$ which was determined numerically.} 
    \label{fig:F1_F2_BOX} 
\end{figure}
These functions behave as
\begin{equation}\label{eq:F1_asympt}
\begin{aligned}
\mathcal F_1^{H}(\nu_H)&\approx
\begin{cases}
\dfrac{1}{2}
-\dfrac{5}{16}\,\nu_H^2,
& \nu_H\to 0,\\[8pt]
\dfrac{5}{\nu_H^{4}},
& \nu_H\to \infty,
\end{cases}
\qquad \text{and} \qquad
\mathcal F_1^B(\nu_B)\approx
\begin{cases}
\dfrac{1}{180}
-\dfrac{\nu_B^{2}}{6048},
& \nu_B\to 0,\\[8pt]
\dfrac{5}{4\,\nu_B^{4}},
& \nu_B\to \infty,
\end{cases}
\end{aligned}
\end{equation}
while
\begin{equation}\label{eq:F2_asympt}
\begin{aligned}
\mathcal{F}_2^H(\nu_H)
\approx
\begin{cases}
\dfrac{\nu_H^2}{16}, & \nu_H\to 0,\\[6pt]
\dfrac{1}{\nu_H^4}, & \nu_H\to \infty,
\end{cases}
\qquad \text{and} \qquad
\mathcal{F}_2^B(\nu_B)
\approx
\begin{cases}
c\,\nu_B^2, & \nu_B\to 0,\\[6pt]
\dfrac{1}{4\nu_B^4}, & \nu_B\to \infty,
\end{cases}
\end{aligned}
\end{equation}
where
\begin{equation}\label{eq:defofc}
c=\frac{31}{30240}+\frac{S_2}{2\pi^4},
\qquad
S_2=\frac{1}{\pi}\sum_{m=1}^{\infty}
\frac{(-1)^m}{m^5\sinh(\pi m)}+\sum_{m=1}^{\infty}
\frac{(-1)^m\cosh(\pi m)}{m^4\sinh^2(\pi m)}.
\end{equation}
Numerically we find $c= 4.389...\times10^{-4}$.

Interestingly, the harmonic and box geometries display the same qualitative behavior at the level of the unnormalized $\mathcal{C}_1$ and $\mathcal{C}_2$. As follows from Eqs.~\eqref{eq:F1_asympt} and \eqref{eq:F2_asympt}, the scaling functions associated with $\mathcal{C}_1$ and $\mathcal{C}_2$ have the same leading asymptotic structure in the two confinements, once written in terms of the appropriate dimensionless parameters $\nu_H$ and $\nu_B$. In particular, as shown in Figs.~\ref{fig:F1_F2_Harmonic} and \ref{fig:F1_F2_BOX}, both $\mathcal{F}_2^H(\nu_H)$ and $\mathcal{F}_2^B(\nu_B)$ vanish for weak and strong resetting, and exhibit a maximum at intermediate resetting rates. This reflects the fact that $\mathcal{C}_2$ is the variance, with respect to the reset age $\tau$, of the reset-free squared displacement $\langle x^2(\tau)\rangle_0$.
The origin of this non-monotonicity is in fact the competition between correlations and fluctuations. For small $r$, the typical reset time $t_r = 1/r$ is much larger than the relaxation time of the confined dynamics. The system therefore equilibrates between two resets, $\langle x^2(\tau)\rangle_0$ becomes essentially independent of $\tau$, and the particles become effectively uncorrelated. Hence $\mathcal{C}_2 \to 0$ as $r \to 0$. Conversely, for large $r$, resets occur so frequently that particles remain localized close to the origin. Fluctuations are then strongly suppressed, and the variance of $\langle x^2(\tau)\rangle_0$ again vanishes, giving $\mathcal{C}_2 \to 0$ as $r \to \infty$. At intermediate values of $r$, resetting is frequent enough to prevent equilibration but not so frequent as to suppress diffusion. The squared displacement then fluctuates strongly with the reset age, producing a maximum of $\mathcal{C}_2$. A similar competition between correlations and fluctuations was recently observed in the context of a new resetting protocol called batch resetting, introduced in Ref. \cite{deMauro2026}.

It is now useful to discuss the asymptotics of $\mathcal C_2$. We have found
\begin{eqnarray}\label{eq:exactasymptunnormcorrel}
    \mathcal{C}_2^B \approx c L^4 \nu_B^2
    = 16c\, \frac{r L_B^6}{D}, 
    \qquad
    \mathcal{C}_2^H \approx \frac{D^2\nu_H^2}{4\mu^2}
    = \frac{1}{4} \frac{rL_H^6}{D},
\end{eqnarray}
where $c$ is defined in \eqref{eq:defofc}.
Thus, in both geometries, $\mathcal{C}_2$ vanishes linearly with $r$ and
scales as the sixth power of the confinement length.
This scaling has a simple origin. In the small-$\nu_B$ regime, reached for
instance by taking $r$ small at fixed confinement length, resets are rare.
Most trajectories have an age $\tau$ much larger than the confinement
relaxation time $t^\ast$ (see Sec. \ref{sec:statstateandcontrolparam} for a definition), and have therefore relaxed to the factorized
equilibrium state. Therefore, these trajectories cannot contribute to inter-particle correlations. The contribution to $\mathcal{C}_2$ comes only from the rare trajectories with $\tau\lesssim t^\ast$. Since the reset age is exponentially distributed as $r e^{-r\tau}$, the probability of such trajectories is
\begin{equation}
    \mathbb{P}(\tau<t^\ast)=1-e^{-r t^\ast}\approx r t^\ast
\end{equation}
for $r t^\ast\ll 1$. Their typical contribution is of order
$L_B^4$ or $L_H^4$, because $\mathcal{C}_2$ involves $x_i^2x_j^2$. Hence
\begin{equation}
    \mathcal{C}_2^B \sim (r t^\ast_B)L_B^4,
    \qquad
    \mathcal{C}_2^H \sim (r t^\ast_H)L_H^4. 
\end{equation}
Using $t_B^\ast= \frac{L_B^2}{\pi^2D}$ in the box (see Eq. \eqref{eq:BCrelaxtime}) and
$t_H^\ast= \frac{1}{2\mu}=\frac{L_H^2}{2D}$ in the harmonic trap (see Eq. \eqref{eq:HCrelaxtime}), gives
\begin{equation}
    \mathcal{C}_2^B \sim \frac{1}{\pi^2}\frac{rL_B^6}{D},
    \qquad
    \mathcal{C}_2^H \sim \frac{1}{2}\frac{rL_H^6}{D},
\end{equation}
in agreement, up to numerical constants, with the exact asymptotics in Eq. \eqref{eq:exactasymptunnormcorrel}.

In the opposite limit, $r\to\infty$, the time between resets is too short for the particles to feel the confinement. Both geometries therefore reduce to the unconfined resetting gas. Consistently,
\begin{eqnarray}
    \mathcal{C}_2^B \approx \frac{L^4}{4\nu_B^4}
    = \frac{4D^2}{r^2}, 
    \qquad
    \mathcal{C}_2^H \approx \frac{4D^2}{\mu^2\nu_H^4}
    = \frac{4D^2}{r^2}.
\end{eqnarray}
Thus, the small-$r$ regime is controlled by the relaxation time of the
confinement, while the large-$r$ regime is universal and independent of the
confining geometry.

\subsection{Normalized correlations}\label{subsec:normalizedcorrel_dim1}

We now specialize the normalized correlation coefficient defined in Eq.~\eqref{eq:def_of_a} to the two confinement geometries. Following the notation introduced above, we write it as $A_H(\nu_H)$ in the HC and as $A_B(\nu_B)$ in the BC. For the HC, this gives
\begin{equation}\label{eq:a(w)OUprocess}
    A_H(\nu_H)=\frac{\mathcal{F}_2^H(\nu_H)}{\mathcal{F}_1^H(\nu_H)}=\frac{\nu_H^2}{5\nu_H^2+8}, \qquad \nu_H=\frac{L_H}{\ell_r}=\sqrt{\frac{r}{\mu}}.
\end{equation}
where $\mathcal{F}_1^H(\nu_H)$ and $\mathcal{F}_2^H(\nu_H)$ are given, respectively, in \eqref{eq:C1OUprocess} and \eqref{eq:C2OUprocess}.
Its asymptotics are
\begin{equation}\label{eq:asymptaHARM}
A_H(\nu_H)\approx
\begin{cases}
\dfrac{\nu_H^2}{8},
& \nu_H\to 0,
\\[6pt]
\dfrac{1}{5}-\dfrac{8}{25}\dfrac{1}{\nu_H^2},
& \nu_H\to \infty.
\end{cases}
\end{equation}
For the BC, instead, it reads
\begin{equation}\label{eq:aydimension1}
    A_B(\nu_B)=\frac{\mathcal{F}_2^B(\nu_B)}{\mathcal{F}_1^B(\nu_B)}, \qquad \nu_B=\frac{L_B}{\ell_r}=\frac{1}{2} \sqrt{\frac{rL^2}{D}},
\end{equation}
where $\mathcal{F}_1^B(\nu_B)$ and $\mathcal{F}_2^B(\nu_B)$ are given, respectively, in \eqref{eq:F1Boxdefinition} and \eqref{eq:F2Boxdefinition}.
The asymptotics are
\begin{equation}\label{eq:asymptaBOX}
A_B(\nu_B)\approx
\begin{cases}
(180c)\,\nu_B^2,
& \nu_B\to 0,
\\[6pt]
\dfrac{1}{5}
+\left(\dfrac{2}{25}\nu_B^3-\dfrac{2}{5}\nu_B^2-\dfrac{2}{25}\nu_B\right)e^{-\nu_B},
& \nu_B\to \infty,
\end{cases}
\end{equation}
where $c$ is defined in \eqref{eq:defofc}.

\begin{figure}[h!]
    \centering

    \begin{minipage}{0.48\textwidth}
        \centering
        \includegraphics[width=\textwidth]{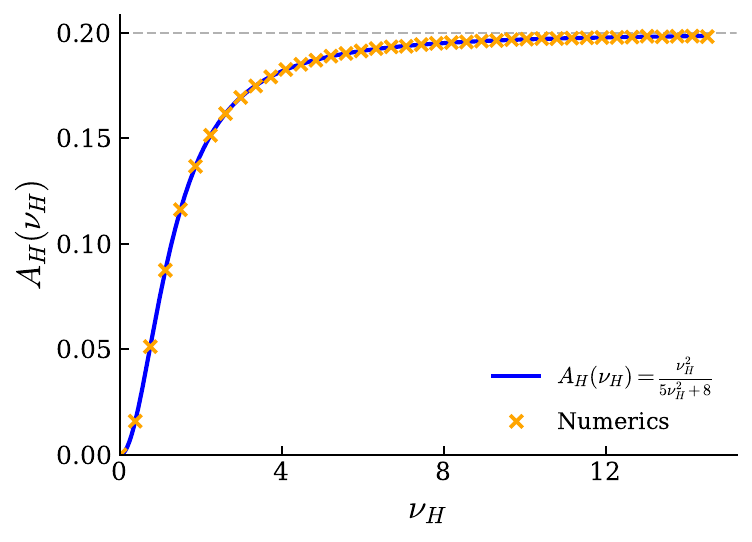}

        \vspace{0.1cm}
        {\small \textbf{(a)} Harmonic confinement: $A_H(\nu_H)$ given in Eq.~\eqref{eq:a(w)OUprocess}.}
    \end{minipage}
    \hfill
    \begin{minipage}{0.48\textwidth}
        \centering
        \includegraphics[width=\textwidth]{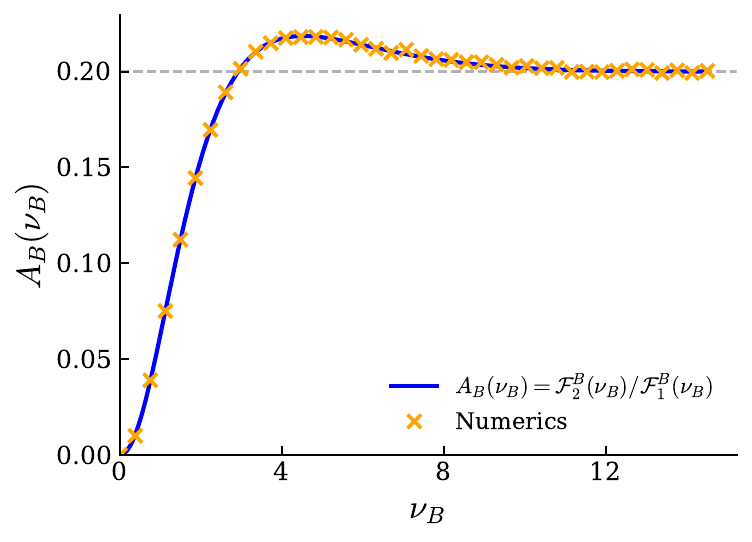}

        \vspace{0.1cm}
        {\small \textbf{(b)} Box confinement: $A_B(\nu_B)$ given in Eq.~\eqref{eq:aydimension1}.}
    \end{minipage}

    \caption{Plots of the normalized correlation coefficient in the HC (left) and in the BC (right), along with numerical simulation results (orange crosses). The function $A_B(\nu_B)$ on the right shows an overshoot with a maximum at the value $\nu_B^*\approx4.5180...$ determined numerically.}
    \label{fig:a_Harmonic_vs_Box}
\end{figure}

As shown in Fig.~\ref{fig:a_Harmonic_vs_Box}, and as follows from the asymptotic behaviors in Eqs. \eqref{eq:asymptaHARM} and \eqref{eq:asymptaBOX}, the normalized correlation coefficient behaves very differently in the two geometries. This contrasts with the unnormalized correlators $\mathcal{C}_2^H$ and $\mathcal{C}_2^B$, which display the same qualitative trends in both the HC and the BC.
For both geometries, the large-$\nu_H$ or large-$\nu_B$ limit corresponds to unconfined resetting. Accordingly, the normalized correlation coefficient approaches the same value, $A_H(\nu_H\to\infty)=A_B(\nu_B\to\infty)=1/5$ \cite{deMauro2026}.
This result is completely universal. In fact, for any confining potential $V(x)$, the normalized correlation coefficient must tend to $1/5$ when the resetting length becomes much smaller than the typical equilibrium confinement
length scale. Equivalently, this corresponds to the limit in which the ratio between the confinement length and the resetting length goes to infinity. This follows from the fact that, in this limit, the particles do not feel the confinement between two consecutive resets, and the dynamics reduces to the unconfined resetting problem. 

However, the approach to this limit is qualitatively different. In the HC, $A_H(\nu_H)$ increases monotonically toward $1/5$, whereas in the BC, $A_B(\nu_B)$ is non-monotonic and overshoots the unconfined value (see Fig. \ref{fig:a_Harmonic_vs_Box}). This is unexpected at first sight: if $\nu_B$ is increased by increasing $r$ at fixed $L$, one might expect correlations to grow monotonically, since resetting is the only mechanism that correlates different particles.

The origin of the overshoot in the BC can be understood by recalling that $A_B(\nu_B)$ is a normalized measure of correlations: it quantifies the fraction of the total fluctuations that is collective.  We can now reason as follows. Imagine fixing $r$ and varying $\nu_B$ by varying the confinement length $L_B$. In the unconfined resetting gas ($L_B\to\infty$), once the reset age $\tau$ is fixed, most particles typically lie at distances of the same order, $\sqrt{D\tau}$. Fluctuations of the common reset age from one realization to another therefore produce collective fluctuations of the squared displacements, and hence of the spatial extent of the whole cloud. At the same time, rare trajectories may occur in which some particles make very large excursions between two resets, far beyond the typical scale $\sqrt{D\tau}$. These atypical trajectories are poorly correlated with the bulk of the gas and therefore reduce the collective fraction of the total fluctuations. Now consider a box with large but finite $L_B$. The hard walls cut off precisely these rare large excursions: trajectories that would go very far in the unconfined system are reflected back into the box. In this way, the box removes part of the non-collective fluctuations while preserving most of the collective ones. As a result, the normalized coefficient $A_B(\nu_B)$ can become larger than its unconfined value. However, if $L_B$ is decreased further, confinement starts to suppress also the trajectories that carry useful information about the last reset. Being close to the origin is then no longer a clear signature of a recent reset, because a particle can also return near the origin after reflections at the walls. The reset-induced correlations are therefore weakened, and $A_B(\nu_B)$ decreases, eventually vanishing as $\nu_B\to0$.

In the HC, the same selective suppression of rare trajectories is not strong enough to produce an overshoot. The harmonic force does reduce large excursions, but not strongly enough relative to its effect on typical trajectories to eliminate the decorrelating part of the fluctuations. Consequently, the rare trajectories that reduce the collective fraction of the fluctuations remain relevant, and $A_H(\nu_H)$ approaches the unconfined value $1/5$ from below, without developing an overshoot.

We test this interpretation in two complementary ways. First, in the following subsection, we consider the family of potentials defined in Eq.~\eqref{eq:familypotent}, $V(x)=\kappa |x|^\alpha$, with $\alpha>0$. Increasing $\alpha$ makes the restoring force grow more rapidly at large distances, so that rare large excursions are suppressed more strongly relative to typical trajectories. This allows us to identify a critical value of $\alpha$ above which this selective suppression becomes strong enough to produce a non-monotonic behavior of $A_\alpha(\nu_\alpha)$. This is shown in Fig. \ref{fig:Anualpha}.

Second, we keep the BC geometry and increase the spatial dimension. We compute $A_B^{(d)}(\nu_B)$ for particles confined in a $d$ dimensional hypersphere, with $d=2$ and $d=3$. As shown in Fig.~\ref{fig:aind1d2d3}, the overshoot becomes stronger as $d$ increases. This is consistent with the rare trajectory picture, since large excursions have a larger statistical weight in higher dimensions \cite{Gardiner2009}, and cutting them off with hard walls has a stronger effect on the collective fraction of the fluctuations.

\subsubsection{Correlations for a general $\alpha$}\label{sec:correl_generalalpha}

In order to further test the physical interpretation of the non-monotonic
behavior observed in the BC, we now study the stationary correlations for the
broader family of confining potentials
\begin{equation}\label{eq:Vgeneralalpha_2}
V(x)=\kappa |x|^{\alpha},
\qquad
\alpha>0.
\end{equation}
In the absence of resetting, this potential leads to the equilibrium
single-particle distribution
\begin{equation}\label{eq:equilgeneralalpha}
    p_0^{\rm eq}(x)
    =
    \frac{1}{Z_0}
    \exp\left[-\frac{\kappa}{D}|x|^{\alpha}\right],
\end{equation}
which will be useful as a reference state and to identify the natural
confinement length scale. In the following, however, we always consider the
system in the presence of simultaneous resetting with rate $r>0$. The
stationary state is therefore not the equilibrium distribution above, but the
nonequilibrium stationary state
\begin{equation}
P_{r}^{\rm st}(\mathbf x)
=
r\int_0^\infty d\tau\, e^{-r\tau}
\prod_{i=1}^N p_{0}(x_i,\tau),
\end{equation}
where $p_{0}(x,\tau)$ is the reset-free propagator in the potential
$V(x)=\kappa |x|^\alpha$, starting from the origin.
This family of potentials smoothly interpolates between the HC case for
$\alpha=2$, with $\kappa=\mu/2$, and the BC in the limit
$\alpha\to\infty$, understood in rescaled units with half-box size
$L_B=1$. To define the analogue of $\nu_H$ and $\nu_B$ for a general
$\alpha$, we introduce the confining length scale
\begin{equation}\label{eq:conflengenericalpha}
L_{\alpha}=
\left(\frac{D}{\kappa}\right)^{1/\alpha},
\qquad
\alpha>0,
\end{equation}
which can be read off directly from Eq.~\eqref{eq:equilgeneralalpha}.
The resetting length scale is still $\ell_r=\sqrt{\frac{D}{r}},$
and we define the dimensionless control parameter
\begin{equation}
\nu_{\alpha}
=
\frac{L_{\alpha}}{\ell_r}.
\end{equation}
This parameter controls the physics of the system.
\begin{figure}[h!]
\centering
\includegraphics[width=0.6\textwidth]{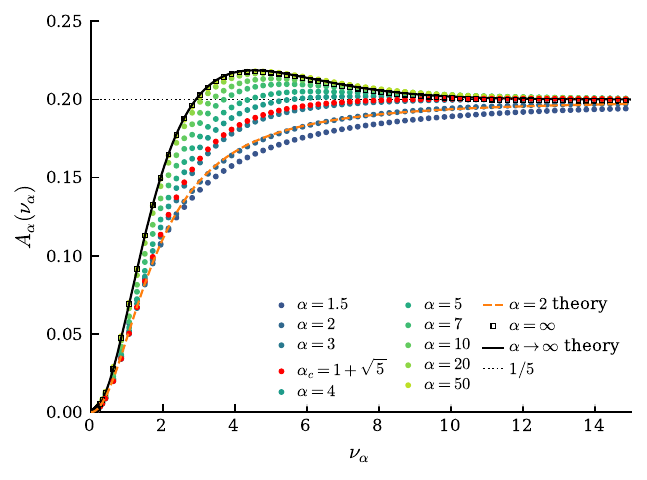}
\caption{Normalized correlation coefficient $A_{\alpha}(\nu_\alpha)$ for the confining potential $V(x)=\kappa |x|^\alpha$. Symbols correspond to numerical simulations for different values of $\alpha$. The solid curves show the theoretical predictions for the harmonic case, $\alpha=2$ (see Eq.~\eqref{eq:a(w)OUprocess}), and for the hard wall limit, $\alpha\to\infty$ (see Eq.~\eqref{eq:aydimension1}). The red dots correspond to the critical value $\alpha=\alpha_c=1+\sqrt{5}$.}
\label{fig:Anualpha}
\end{figure}
Let us first make explicit how this definition connects to the harmonic case.
For $\alpha=2$, the potential becomes $V(x)=\kappa x^2$. To recover the harmonic confinement used above, $V(x)=\frac{1}{2}\mu x^2$, one has to identify $\kappa=\mu/2$. Therefore,
\begin{equation}
L_{\alpha=2}=\left(\frac{2D}{\mu}\right)^{1/2}
=\sqrt{2}L_H,
\end{equation}
where $L_H=\sqrt{D/\mu}$ is the harmonic confinement length introduced previously. Consequently, $\nu_{\alpha=2}=\sqrt{2}\nu_H$.

The BC case is recovered instead by taking the limit $\alpha\to\infty$. In this limit, one has $L_{\alpha\to\infty}\to1$, and the potential approaches a hard wall at $|x|=1$. Thus, in these rescaled units, the limiting box has half-size $L_B=1$. In this limit, $\nu_{\alpha}$ coincides with the box parameter $\nu_B$, since for $L_B=1$ one has
\begin{equation}
\nu_B=\frac{L_B}{\ell_r}=\frac{1}{\ell_r}.
\end{equation}
Even though the half-size of the box is fixed in these units, $\nu_B$ can still be varied by changing the resetting rate $r$.

We can therefore define the normalized correlation coefficient for a general $\alpha$ as
\begin{equation}\label{eq:defAalpha}
A_{\alpha}(\nu_\alpha)=
\frac{\langle x_i^{2}x_j^{2}\rangle
-\langle x_i^{2}\rangle\langle x_j^{2}\rangle}
{\langle x_i^{4}\rangle-\langle x_i^{2}\rangle^2},
\qquad i\neq j,
\end{equation}
where the averages are taken over the stationary JPDF corresponding to the potential $V(x)=\kappa |x|^\alpha$. This definition is the direct generalization of the normalized correlators $A_H(\nu_H)$ and $A_B(\nu_B)$ introduced above.

In Fig.~\ref{fig:Anualpha}, we show the results of numerical simulations of the normalized correlation coefficient for different values of $\alpha$. The black solid line corresponds to the theoretical prediction for the BC, $A_B(\nu_B)$, given in Eq.~\eqref{eq:aydimension1}, while the orange dotted line corresponds to the harmonic prediction $A_H(\nu_2/\sqrt{2})$, where $A_H(\nu_H)$ is given in Eq.~\eqref{eq:a(w)OUprocess}.

The numerical results in Fig.~\ref{fig:Anualpha} show that, as expected, for all $\alpha$
\begin{equation}
A_\alpha(\nu_\alpha)\xrightarrow[\nu_\alpha\to\infty]{}\frac{1}{5}.
\end{equation}
The value of $\alpha$ only affects how this limit is approached. In fact, Fig.~\ref{fig:Anualpha} shows that the behavior of $A_{\alpha}(\nu_\alpha)$
changes qualitatively as $\alpha$ is increased. In Appendix~\ref{sec:criticalalpha_APP}, we also analyze analytically the large-$\nu_\alpha$ expansion of $A_{\alpha}(\nu_\alpha)$ and show that the first correction to the unconfined value $1/5$ changes sign at
\begin{equation}
\alpha_c=1+\sqrt{5}=3.236\dots.
\end{equation}
For $0<\alpha<\alpha_c$, we find that $A_{\alpha}(\nu_\alpha)$ approaches $1/5$ from below at large $\nu_\alpha$, while for $\alpha>\alpha_c$ it approaches $1/5$ from above. At $\alpha=\alpha_c$, the leading correction vanishes and the behavior is controlled by higher-order terms, which we do not analyze here. Since $A_{\alpha}(\nu_\alpha)$ vanishes in the limit $\nu_{\alpha}\to0$ (i.e. $r\to0$), the approach from above for $\alpha>\alpha_c$ necessarily implies an overshoot. Together with the numerical results shown in Fig.~\ref{fig:Anualpha}, this identifies $\alpha_c$ as the critical value separating the monotonic ($0<\alpha<\alpha_c$) and non-monotonic ($\alpha>\alpha_c$) regimes. 

This supports the interpretation proposed above for the BC. As $\alpha$ increases, the restoring force associated with the potential, $-V'(x)=-\alpha\kappa\,x|x|^{\alpha-2}$,
acts more strongly on particles that are far from the origin, while its effect near the origin becomes comparatively weaker. Therefore, typical trajectories close to the bulk of the gas are only weakly modified, whereas rare large excursions are increasingly suppressed. Since these rare trajectories are precisely those that contribute most strongly to the noncollective, decorrelating part of the fluctuations, suppressing them enhances the normalized collective correlations. For $\alpha>\alpha_c$, this suppression becomes strong enough to produce an overshoot of $A_{\alpha}(\nu_\alpha)$, as previously observed in the BC.

\subsubsection{Generalization to \texorpdfstring{$d$}{d} dimensions}\label{sec:ddimensionalcorrel}

We now fix the BC geometry and extend the analysis to arbitrary spatial dimension $d$ to test the robustness of the effect discussed in Sec.~\ref{subsec:normalizedcorrel_dim1}. In particular, we show that the overshoot of $A_B^{(d)}(\nu_B)$ becomes more pronounced as $d$ increases, consistently with the physical interpretation given above.
We consider $N$ particles with positions $\mathbf{x}_i \in \mathbb{R}^d$, undergoing free diffusion inside a $d$-dimensional sphere of radius $L/2$ with reflecting (Neumann) boundary conditions. At random times, with Poisson rate $r$, all particles are simultaneously reset to the origin $\mathbf{x}=(0,\dots,0)$.
\begin{figure}[h!] 
    \centering
    \includegraphics[width=0.6\textwidth]{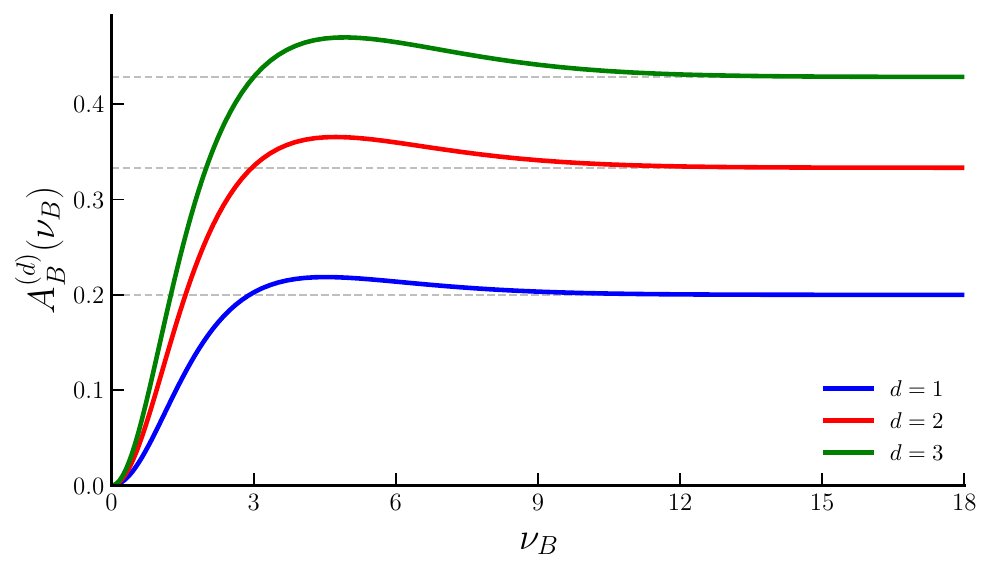}
    \caption{Plot of the correlation coefficient $A_B^{(d)}(\nu_B)$ in dimension $d=1$, $d=2$ and $d=3$. The corresponding expressions are given, respectively, in \eqref{eq:aydimension1}, \eqref{eq:a_nu_dimension2_APP} and \eqref{eq:a_nu_dimension3_APP}.}
    \label{fig:aind1d2d3}
\end{figure}
To quantify pair correlations in higher dimensions, we generalize Eqs.~\eqref{eq:C2C1_def_together} and \eqref{eq:def_of_a} and define
\begin{equation}\label{eq:defainmoredim}
A_B^{(d)}(\nu_B)=\frac{\mathcal{C}_2^{(d)}(\nu_B)}{\mathcal{C}_1^{(d)}(\nu_B)}=\frac{\langle |\mathbf{x}_i|^2 |\mathbf{x}_j|^2 \rangle - \langle |\mathbf{x}_i|^2 \rangle \langle |\mathbf{x}_j|^2 \rangle}{\langle |\mathbf{x}_i|^4 \rangle - \langle |\mathbf{x}_i|^2 \rangle^2},
\end{equation}
where $|\mathbf{x}|^2 = x_1^2 + \dots + x_d^2$.
In Appendix~\ref{sec:dDimensions_APP}, we perform calculations similar to those previously done for $d=1$ and derive the expressions for $A_B^{(2)}(\nu_B)$ and $A_B^{(3)}(\nu_B)$.
As in one dimension, for $d=2,3$ we can write
\begin{equation}\label{eq:aydimension2}
    A_B^{(2)}(\nu_B) = \frac{\mathcal{F}_2^{(2)}(\nu_B)}{\mathcal{F}_1^{(2)}(\nu_B)},
    \qquad
    A_B^{(3)}(\nu_B) = \frac{\mathcal{F}_2^{(3)}(\nu_B)}{\mathcal{F}_1^{(3)}(\nu_B)}.
\end{equation}
and their explicit expressions are given in Eqs.~\eqref{eq:a_nu_dimension2_APP} and \eqref{eq:a_nu_dimension3_APP}.
They are plotted, together with the one dimensional result \eqref{eq:aydimension1}, in Fig.~\ref{fig:aind1d2d3}. The figure clearly shows that the effect is robust upon increasing the spatial dimension: the non-monotonic behavior persists in $d = 2$ and $d = 3$. Moreover, the amplitude of the overshoot becomes more pronounced as $d$ increases. This is consistent with our physical interpretation, since rare large excursions have a larger statistical weight in higher dimensions \cite{Gardiner2009} and contribute more strongly to the noncollective part of the fluctuations. The hard boundary suppresses precisely these decorrelating excursions, and its effect on the normalized collective correlations therefore becomes stronger as $d$ increases.

\section{Edge observables: extreme value and gap statistics}\label{sec:observables}

We now turn to the EVS and gap statistics of the system. We denote by $M_1$ and $M_2$ the positions of the rightmost and second rightmost particles, respectively. The EVS describe the probability distribution of $M_1$, while the gap statistics describe the distribution of the spacing $d_1=M_1-M_2$. These observables are particularly informative because they probe the edge of the gas, where the effect of confinement is expected to be most pronounced.

Such quantities have been extensively studied for independent and identically distributed (IID) random variables \cite{Gumbel1958,Leadbetter1983,FortinClusel2015,Arnold1992,DavidNagaraja2003,MPSch2019,MajumdarSchehr2024}, whereas exact results for strongly correlated systems are much less common \cite{MPSch2019,Biroli2023,BLMS2024,Derrida1985,Derrida1986,Tracy1994,Tracy1996,Dean2001,MPLK2003,MC2004,Bertin2006}. Recently, Ref.~\cite{Biroli2023} introduced one of the few analytically tractable examples with strong long-range correlations, for which exact EVS and gap statistics could be derived.
Here, we investigate how confinement modifies the unconfined resetting picture of Ref.~\cite{Biroli2023}. We show that confinement has a strong impact on these edge observables.
Moreover, we show that the different DEC structure in the HC with respect to BC is also reflected in the statistics of $M_1$ and $d_1$. 
As a result, the EVS and gap statistics differ not only from the unconfined case, but also sharply between the two confinement geometries.
The starting point of our analysis is the CIID structure of the resetting gas \eqref{eq:NESS}. Indeed, once the time elapsed since the last reset is fixed, the particles are IID according to the corresponding reset-free propagator. It is therefore useful to first recall some standard results for the EVS and gap statistics of IID random variables, which will then be averaged over the random reset age.

Consider $N$ random variables ${x_1,\dots,x_N}$, drawn independently from a probability density $p(x)$, and let $Q(x)=\int_{-\infty}^{x} p(y)\,dy$ be the associated cumulative distribution function. We are interested in the maximum $M_1=\max\{x_1,\dots,x_N\}$. For finite $N$, its probability density is exactly
\begin{equation}\label{eq:finiteNEVS}
    \operatorname{Prob}\{M_1=m\}=N\,Q(m)^{N-1}p(m).
\end{equation}
The interpretation is straightforward: one variable must take the value $m$, while the remaining $N-1$ variables must all be smaller than $m$, and there are $N$ possible choices for the variable realizing the maximum.

Our main interest lies in the large $N$ limit, where universal behavior emerges. In this regime, one introduces a typical location $a_N$ and a fluctuation scale $b_N$ such that the rescaled random variable
\begin{equation}
    Z=\frac{M_1-a_N}{b_N}
\end{equation}
remains of order $O(1)$ as $N\to\infty$.
The centered and scaled variable $Z$ converges, as $N\to\infty$, to an
$N$-independent limiting distribution, whose possible forms are classified by
the Fisher--Tippett--Gnedenko theorem \cite{MPSch2019,MajumdarSchehr2024}. 
Remarkably, only three universality classes are possible for IID variables: Gumbel, Fréchet, and Weibull, depending only on the tail behavior of $p(x)$. We denote by $G_\rho(z)$ the limiting distribution of $Z$ for large $N$, where $\rho=I,II,III$ corresponds to the Gumbel, Fréchet, and Weibull classes, respectively.
The Gumbel class corresponds to parent distributions $p(x)$ with rapidly decaying tails, for instance $p(x)\sim e^{-c x^\delta}$ as $x\to\infty$, with $c>0$ and $\delta>0$. The Fréchet class corresponds to heavy-tailed parent distributions, such as $p(x)\sim x^{-1-\beta}$ as $x\to\infty$, with $\beta>0$. Finally, the Weibull class corresponds to parent distributions with bounded support. If $\bar x$ denotes the upper edge of the support, then $p(x)\sim (\bar x-x)^{\gamma-1}$ as $x\to \bar x^{-}$, with $\gamma>0$.
The three limiting distributions read
\begin{equation}\label{eq:IIDEVSResults}
    G_{\rho}(z)=
    \begin{cases}
        \exp\!\left[-(z+e^{-z})\right],
        & \text{(Gumbel, $\rho=I$)}, \\[8pt]
        \Theta(z)\,\beta\, z^{-1-\beta} e^{-z^{-\beta}},
        & \text{(Fréchet, $\rho=II$)}, \\[8pt]
        \Theta(-z)\,\gamma\, (-z)^{\gamma-1} e^{-(-z)^\gamma},
        & \text{(Weibull, $\rho=III$)}.
    \end{cases}
\end{equation}
The centering and scaling constants $a_N$ and $b_N$ are instead non-universal and they may be determined from (see \cite{MPSch2019,MajumdarSchehr2024})
\begin{equation}\label{eq:defaNbN}
    \int_{a_N}^{\infty} p(x)\,dx=\frac{1}{N},
    \qquad
    b_N=N\int_{a_N}^{\infty}(x-a_N)p(x)\,dx.
\end{equation}
The first relation in Eq. \eqref{eq:defaNbN} defines $a_N$ as the typical location of the maximum. Indeed, if $N$ independent variables are drawn from $p(x)$, the expected number of variables larger than $a_N$ is
\begin{equation}
N\int_{a_N}^{\infty}p(x)\,dx=1.
\end{equation}
Thus, $a_N$ is the point beyond which one typically finds only one particle. The second relation defines the typical scale of the fluctuations around this point. Since $\int_{a_N}^{\infty}p(x)\,dx=1/N$, one can rewrite it as
\begin{equation}
b_N=
\frac{\int_{a_N}^{\infty}(x-a_N)p(x)\,dx}
{\int_{a_N}^{\infty}p(x)\,dx},
\end{equation}
showing that $b_N$ can be interpreted as the average fluctuation of the maximum
around $a_N$, conditioned on the fact that there is a single variable in $[a_N,\infty)$.
With these definitions, for large $N$, the maximum can be written as $M_1=a_N+b_N Z$, where $Z$ is an $O(1)$ random variable distributed according to one of the three universal laws in \eqref{eq:IIDEVSResults}. In the following, we will use $Z$ generically to denote an $O(1)$ rescaled random variable, whose distribution will be specified in each case.

In the present work, two specific parent distributions are particularly relevant. For a Gaussian parent distribution $p(x)=\frac{1}{\sqrt{2\pi}\,\sigma}\exp\!\left(-\frac{x^2}{2\sigma^2}\right)$, which belongs to the Gumbel class, with zero mean and variance $\sigma^2$, one finds
\begin{equation}\label{eq:gumbelIIDaNbN}
    a_N=\sqrt{2}\sigma\,\operatorname{erfc}^{-1}\!\left(\frac{2}{N}\right)
    \approx
    \sigma\sqrt{2\ln N},
    \qquad
    b_N\approx
    \frac{\sigma}{\sqrt{2\ln N}};
\end{equation}
where $\operatorname{erfc}^{-1}(x)$ is the inverse of the complementary error function, defined as
$\operatorname{erfc}(x)=\frac{2}{\sqrt{\pi}}\int_x^{\infty} e^{-t^2}\,dt$.
For later convenience, we introduce the shorthand notation
\begin{equation}\label{eq:defdiuN}
    u_N=\operatorname{erfc}^{-1}\!\left(2/N\right).
\end{equation}
In the large-$N$ limit, it behaves at leading order as $u_N\approx \sqrt{\ln N}$.
The second important case is a bounded distribution with upper edge $\bar x$, such that $p(x)\to c_0$ as $x\to\bar x^{-}$, with $c_0>0$. This belongs to the Weibull class. In this case,
\begin{equation}
    a_N\approx \bar x,
    \qquad
    b_N\approx \frac{1}{c_0N}.
\end{equation}

The same reasoning extends naturally to gap statistics. Let $d_1=M_1-M_2$, where $M_2$ is the second largest value. For large $N$, one has $M_1=a_N+b_N Z_1$ and $M_2=a_N+b_N Z_2$, where $Z_1$ and $Z_2$ are the rescaled random variables associated with the largest and second largest values. They are both of order $O(1)$, but they are in general not identically distributed; nevertheless, the same centering $a_N$ and scaling $b_N$ apply to both of them \cite{MajumdarSchehr2024}. Therefore,
\begin{equation}\label{eq:gapniceform}
    d_1=b_N(Z_1-Z_2)=b_N Y,
\end{equation}
where $Y=Z_1-Z_2$ is a random variable of order $O(1)$. Thus, unlike the maximum, the gap is independent of the centering constant $a_N$ and depends only on the fluctuation scale $b_N$. It is therefore a more direct probe of edge fluctuations.
As in standard EVS, $b_N$ is nonuniversal and depends on the parent distribution $p(x)$, whereas the rescaled variable $Y$ converges to a universal limiting law. This law again depends only on the tail behavior of $p(x)$, with the same three classes \cite{MajumdarSchehr2024}. The corresponding gap scaling functions are
\begin{equation}
    f_{\rho}(y)=
    \begin{cases}
        \Theta(y)\,e^{-y},
        & \text{(Gumbel, $\rho=I$)}, \\[8pt]
        \Theta(y)\,\beta^2
        \displaystyle\int_0^\infty
        e^{-x^{-\beta}}
        x^{-\beta-1}(x+y)^{-\beta-1}\,dx,
        & \text{(Fréchet, $\rho=II$)}, \\[10pt]
        \Theta(y)\,\gamma^2
        \displaystyle\int_0^\infty
        (x+y)^{\gamma-1}e^{-(x+y)^\gamma}
        x^{\gamma-1}\,dx,
        & \text{(Weibull, $\rho=III$)}.
    \end{cases}
\end{equation}

We now apply these IID results to the resetting gas. At fixed reset age $\tau$,
Eq.~\eqref{eq:NESS_CIIDform} shows that the particles are IID with common
density $p_0(x,\tau)$. Therefore, the maximum $M_1(\tau)$ conditioned on a certain value of $\tau$ takes the form
\begin{equation}
    M_1(\tau)=a_N(\tau)+b_N(\tau)Z,
\end{equation}
where $Z$ is distributed as one of the three cases in Eq. \eqref{eq:IIDEVSResults}.
The stationary EVS is then obtained by averaging the fixed-$\tau$ IID result over the exponentially distributed reset age:
\begin{equation}\label{eq:general_formula_EVS}
    \operatorname{Prob}\{M_1=m\}=
    r\int_0^\infty d\tau\,  e^{-r\tau}\,
    \operatorname{Prob}\{M_1(\tau)=m\}.
\end{equation} 
The same reasoning extends to gap statistics \cite{MPSch2019,MajumdarSchehr2024}. If $d_1(\tau)$ denotes the gap between the two rightmost particles at fixed reset age $\tau$, then its stationary distribution is obtained through the same renewal average:
\begin{equation}\label{eq:gapstatsverygeneral}
    \operatorname{Prob}\{d_1=g\}=
    r\int_0^\infty d\tau\,  e^{-r\tau}\,
    \operatorname{Prob}\{d_1(\tau)=g\}.
\end{equation}
Thus, the computation of EVS and gap statistics in our resetting gas proceeds in two steps. First, one solves the IID problem at fixed reset age $\tau$, using the appropriate reset-free propagator $p_0(x,\tau)$. Second, one averages the resulting fixed-$\tau$ distributions over the random reset age. This final average is precisely where the correlations induced by simultaneous resetting enter.

In the following, we will continue to use the notation $M_1(\tau)$ and $d_1(\tau)$ for the random variables corresponding to the maximum and to the first gap conditioned on a fixed value of the reset age $\tau$. By contrast, $M_1$ and $d_1$ will denote the corresponding stationary random variables, obtained after averaging over the exponentially distributed reset age. To keep the notation simple, we will use the same symbols $M_1$ and $d_1$ in both the HC and BC cases, specifying explicitly which geometry is being considered whenever necessary.

As already mentioned in the Introduction, the HC results can also be obtained as a limiting case of the switching-trap model studied in Ref.~\cite{BKMS2024}. In the following, however, we rederive the EVS and gap statistics directly, since this particular limit was neither explicitly extracted nor analyzed in Ref.~\cite{BKMS2024}.

\subsection{Extreme Value Statistics (EVS)}\label{sec:EVS}

As anticipated above, the exact EVS can be obtained using Eq. \eqref{eq:general_formula_EVS},
where $\operatorname{Prob}\{M_1(\tau)=m\}$ is the distribution of the rightmost particle conditioned on a time $\tau$ since the last reset.
Since we are dealing with confined systems, it is convenient to introduce a
dimensionless time variable adapted to each geometry. We define the corresponding
diffusive time scales as
\begin{equation}
t_H=\frac{L_H^2}{D},
\qquad
t_B=\frac{L_B^2}{D}.
\end{equation}
These time scales are proportional to the equilibrium relaxation time scales
defined in Sec.~\ref{sec:statstateandcontrolparam}. We then define the rescaled time
\begin{equation}
T=\frac{\tau}{t_{H/B}}=\frac{D\tau}{L_{H/B}^2},
\end{equation}
where the subscript $H/B$ indicates either HC or BC.
With this change of variables, Eq.~\eqref{eq:general_formula_EVS} takes the unified form
\begin{equation}\label{eq:general_formula_EVS_T}
\operatorname{Prob}\{M_1=m\}=
\nu^2\int_0^\infty \, dT\, e^{-\nu^2 T}\,
\operatorname{Prob}\{M_1(T)=m\}.
\end{equation}
Here, with a slight abuse of notation, $\nu$ denotes the dimensionless control parameter appropriate to the chosen geometry,
\begin{equation}
\nu =
\begin{cases}
\nu_H=\sqrt{\dfrac{r}{\mu}}, & \mathrm{HC},\\[0.8em]
\nu_B=\dfrac{1}{2}\sqrt{\dfrac{rL^2}{D}}, & \mathrm{BC}.
\end{cases}
\end{equation}
Similarly, the rescaled time $T$ denotes the time elapsed since the last reset, measured in units of the diffusive time associated with the corresponding confinement length,
\begin{equation}
T =
\begin{cases}
\dfrac{D\tau}{L_H^2}=\mu\tau, & \mathrm{HC},\\[0.8em]
\dfrac{D\tau}{L_B^2}=\dfrac{4D\tau}{L^2}, & \mathrm{BC}.
\end{cases}
\end{equation}
With this convention, $T$ is exponentially distributed with rate $\nu_H^2$ in the HC and with rate $\nu_B^2$ in the BC. In this sense, the rate of the dimensionless reset age already incorporates the geometrical information of the corresponding confinement.

At fixed $T$, the conditional EVS is an ordinary IID problem with parent
density given by the reset-free propagator at that time. Therefore, the
relevant quantities are the corresponding centering and scaling parameters
$a_N(T)$ and $b_N(T)$, defined as in Eq.~\eqref{eq:defaNbN}. The nonstandard
stationary EVS then arises from the final average over the common random age
$T$ in Eq.~\eqref{eq:general_formula_EVS_T}. Our main focus will be the
large-$N$ behavior of $a_N(T)$ and its dependence on $T$.

We start with the HC case, where the analysis is relatively straightforward, and then turn to the BC case, which is more involved.

\subsubsection{Harmonic confinement (HC)}\label{sec:EVS_HC}

\begin{figure}[htbp]
\centering

\begin{minipage}{0.48\textwidth}
    \centering
    \includegraphics[width=\linewidth]{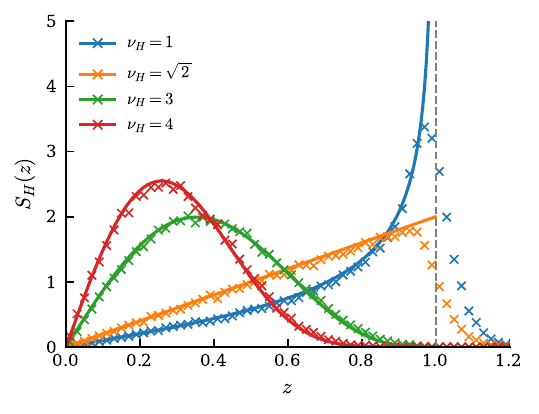}
\end{minipage}
\hfill
\begin{minipage}{0.48\textwidth}
    \centering
    \includegraphics[width=\linewidth]{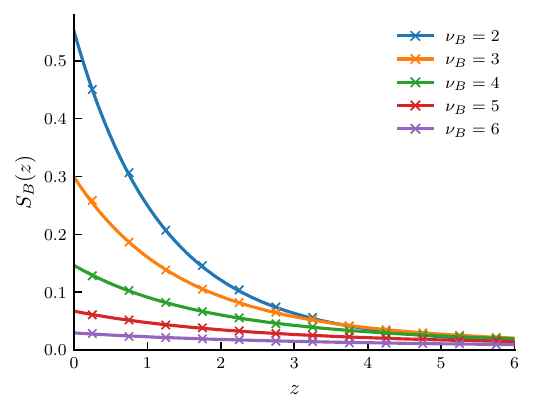}
\end{minipage}

\caption{EVS in the HC (left) and in the BC (right). Crosses show numerical simulations with $N=10^6$ particles, while solid lines correspond to the asymptotic scaling predictions. \textbf{Left:} HC scaling function $S_H(z)$, given in Eq.~\eqref{eq:SHforHC}, plotted as a function of $z=m/\tilde{\Lambda}_N$, with $\tilde{\Lambda}_N$ defined in Eq.~\eqref{eq:tildelambda}. The dashed vertical line marks the upper edge $z=1$ of the limiting support, where $S_H(z)$ undergoes a shape transition. For $\nu_H=1<\nu_H^c$, the asymptotic prediction diverges as $z\to1^{-}$, although this divergence is not resolved in the simulations because of finite $N$ effects. \textbf{Right:} BC scaling function $S_B(z)$, given in Eq.~\eqref{eq:defSBEVS}, for different values of $\nu_B$. Here $z=N\left(\frac{1}{2}-\frac{m}{L}\right)$ is the rescaled distance of the maximum from the right boundary.}
\label{fig:EVS}
\end{figure}

We now analyze the EVS in the HC and show that they are described by
\begin{equation}\label{eq:pdfHCEVS}
\operatorname{Prob}\{M_1=m\}=
\frac{1}{\Lambda_N}
S_H\!\left(\frac{m}{\Lambda_N}\right),
\end{equation}
with
\begin{equation}\label{eq:SHforHC}
S_H(z)
=
\nu_H^2 z (1-z^2)^{\frac{\nu_H^2}{2}-1},
\qquad
z\in[0,1],
\end{equation}
where $\Lambda_N=L_H\sqrt{2\ln N}=\sqrt{(2D\ln N)/\mu}$ and $\nu_H=\sqrt{r/\mu}$. This function is plotted in Fig.~\ref{fig:EVS} on the left, together with the results of numerical simulations. This result shows that the maximum $M_1$ is typically located at a distance $O(\sqrt{\ln N})$ from the origin, and that its fluctuations are described by the scaling function $S_H(z)$ in Eq.~\eqref{eq:SHforHC}. Remarkably, this scaling function has bounded support, even though the particles are allowed to diffuse over the whole real line. As discussed in more detail below, $S_H(z)$ exhibits a shape transition at the right boundary $z=1$ as the parameter $\nu_H$ is varied.

The starting point to derive Eq. \eqref{eq:pdfHCEVS} is Eq.~\eqref{eq:general_formula_EVS}, where the reset-free single-particle propagator is given by $p_0^{(H)}(x,t)$ in Eq.~\eqref{eq:OU_propagator}. As discussed above, this propagator is Gaussian at all times, with variance $\sigma_H^2(t)$ given in Eq.~\eqref{eq:OU_variance}. In terms of the rescaled time $T=\frac{D\tau}{L_H^2}=\mu\tau$, introduced above, it reads
\begin{equation}\label{eq:propagHCtermsofT}
p^{(H)}_0(x,T)
=\frac{1}{\sqrt{2\pi\,\sigma_H^{2}(T)}}
\exp\!\left[-\frac{x^{2}}{2\sigma_H^{2}(T)}\right],
\qquad
\sigma_H^{2}(T)=L_H^2\left(1-\mathrm{e}^{-2T}\right),
\end{equation}
where we recall that $L_H=\sqrt{D/\mu}$.
The next step is to analyze $M_1(T)$ in this geometry, namely the position of the rightmost particle at fixed rescaled time $T$ elapsed since the last reset event. Equivalently, we want to determine the maximum of $N$ independent particles whose common distribution is the propagator in Eq.~\eqref{eq:propagHCtermsofT}. Since this propagator is Gaussian, the associated IID extreme value problem belongs to the Gumbel universality class.
Using the standard IID results recalled in \eqref{eq:gumbelIIDaNbN}, one can write
\begin{equation}
M_1(T)=a_N(T)+b_N(T)\,Z,
\end{equation}
where the random variable $Z$ follows the Gumbel law, $Z\sim G_I(z)=e^{-z-e^{-z}}$, and where the centering and fluctuation scales are
\begin{equation}\label{eq:aNexplicitHC}
a_N(T)
\approx \sigma_H(T)\sqrt{2\ln N},
\qquad
b_N(T)\approx \frac{\sigma_H(T)}{\sqrt{2\ln N}}.
\end{equation}
The stationary distribution of the maximum is then obtained by averaging over the random age $T$ since the last reset:
\begin{equation}\label{eq:EVS_HC_intermediate_Gumbel}
\operatorname{Prob}\{M_1=m\}=\nu_H^2 
\int_0^\infty dT\,e^{-\nu_H^2 T}\,
\tfrac{1}{b_N(T)}G_I\!\left(\tfrac{m-a_N(T)}{b_N(T)}\right).
\end{equation}
A crucial simplification, first pointed out in Ref.~\cite{Biroli2023}, comes from separating the two distinct sources of fluctuations in $M_1$. The first source is the intrinsic extreme value fluctuation at fixed $T$, encoded in the term $b_N(T)Z$. The second source comes from the randomness of the age $T$ itself, which changes from one realization to another.
In the present problem, $T$ is precisely the time elapsed between the last resetting event and the observation time. It is therefore a random variable distributed as $\nu_H^2 e^{-\nu_H^2 T}$. A fluctuation $\Delta T$ of the variable $T$ produces a shift in the typical maximum of order
\begin{equation}
\Delta M_1 \sim a_N'(T)\,\Delta T,
\end{equation}
where we have defined $a_N'(T)= \frac{da_N(T)}{dT}$.
Since $a_N'(T)=O(\sqrt{\ln N})$, these fluctuations are of order $O(\sqrt{\ln N})$. By contrast, the fluctuations at fixed $T$ are controlled by $b_N(T)=O\!\left(1/\sqrt{\ln N}\right)$,
which is much smaller than $a_N'(T)$ for large $N$. Hence, in the limit $N\to\infty$, the dominant fluctuations come entirely from the fluctuations of $T$, while the Gumbel fluctuations at fixed $T$ become negligible.
One may therefore replace the conditional Gumbel law by a delta peak:
\begin{equation}\label{eq:approxwithdeltaHC}
\tfrac{1}{b_N(T)}G_I\!\left(\tfrac{m-a_N(T)}{b_N(T)}\right)
\to
\delta\!\left(m-a_N(T)\right),
\end{equation}
which gives
\begin{equation}
\operatorname{Prob}\{M_1=m\}\approx
\nu_H^2\int_0^\infty dT\,e^{-\nu_H^2T}\,
\delta\!\left(m-a_N(T)\right).
\end{equation}
Using the standard identity for a delta function under change of variables, one obtains
\begin{equation}\label{eq:HCEVSintermTstar}
\operatorname{Prob}\{M_1=m\}\approx
\nu_H^2 e^{-\nu_H^2 T^*(m)}
\left|\frac{dT^*(m)}{dm}\right|,
\end{equation}
where $T^*(m)$ is defined by the condition $m=a_N(T)$.
Substituting the large $N$ form of $a_N(T)$ from Eq.~\eqref{eq:aNexplicitHC}, one finds
\begin{equation}\label{eq:Tstarofm}
T^*(m)=
-\frac{1}{2}
\ln\!\left(
1-\frac{m^2}{2L_H^2\ln N}
\right).
\end{equation}
Substituting the above expression into Eq. \eqref{eq:HCEVSintermTstar} leads to Eq. \eqref{eq:pdfHCEVS}.

The convergence to the asymptotic scaling form in Eq.~\eqref{eq:pdfHCEVS} is, however, rather slow. This slow approach is inherited from the Gaussian extreme value problem at fixed $T$, where the leading scale $\sqrt{\ln N}$ receives significant corrections even at large but finite $N$ (see Refs. \cite{Gyorgyi2008,Gyorgyi2010,Bertin2010}). To reduce these finite $N$ effects in the comparison with simulations, we rescale the numerical data with the more accurate scale
\begin{equation}\label{eq:tildelambda}
\tilde{\Lambda}_N = \sqrt{2} L_H \operatorname{erfc}^{-1}(2/N).
\end{equation}
This choice does not modify the asymptotic result, since $\operatorname{erfc}^{-1}(2/N)\sim\sqrt{\ln N}$ as $N\to\infty$, and therefore $\tilde{\Lambda}_N\sim\Lambda_N$. For the system sizes accessible in our simulations, for instance $N=10^6$, this corrected scale accounts for the dominant finite $N$ correction and gives a much better collapse onto the limiting scaling function $S_H(z)$ (see the left panel in Fig. \ref{fig:EVS}).

The result in \eqref{eq:pdfHCEVS} shows that, for large $N$, the rightmost particle is typically located at a distance $O(\sqrt{\ln N})$ from the origin.
Using the same notation as in standard IID extreme value theory, this corresponds to a centering factor $a_N=0$ and a fluctuation scale $b_N=L_H\sqrt{2\ln N}$, so that
\begin{equation}
M_1=L_H\sqrt{2\ln N}\,Z,
\qquad
Z\sim S_H(z).
\end{equation}
In this case, the notation $Z\sim S_H(z)$ means that the random variable $Z$
has probability density function $S_H(z)$.
This scaling form is clearly different from the three standard IID universality classes and therefore provides a direct signature of the strong correlations induced by resetting.

Equation~\eqref{eq:pdfHCEVS} also differs qualitatively from the unconfined resetting gas, where the EVS take the form
\begin{equation}
M_1=\sqrt{\frac{4D\ln N}{r}}\,Z
\end{equation}
where $Z$ is an $O(1)$ random variable with probability density $2z e^{-z^2}\Theta(z)$.
Thus, although the typical scale of $M_1$ is of the same order in $N$ in the HC confinement, the scaling function is qualitatively different: in the unconfined case it has unbounded support, whereas in the HC the scaling function $S_H(z)$ has compact support, $0\leq z\leq1$.

Interestingly, Eq.~\eqref{eq:SHforHC} reveals a shape transition at the boundary $z=1$. There exists a critical value $\nu_H^c=\sqrt{2}$,
such that for $\nu_H>\nu_H^c$, the function $S_H(z)$ vanishes as $z\to1$, whereas for $\nu_H<\nu_H^c$ it diverges. Exactly at $\nu_H=\nu_H^c$, it remains finite at the edge (see Fig. \ref{fig:EVS}). This critical value $\nu_H^c=\sqrt{2}$ arises precisely when the equilibrium relaxation time of the harmonic trap (given in Eq.~\eqref{eq:HCrelaxtime}) matches the typical resetting timescale $t^\ast_r=1/r$ (see Eq.~\eqref{eq:relaxandnus}).
This shape transition is a new feature induced by HC, with no analogue in the unconfined resetting gas~\cite{Biroli2023}.
It is also remarkable that the scaling function $S_H(z)$ has compact support, even though each individual particle can in principle explore the whole real line.

Finally, we observe in Fig.~\ref{fig:EVS} that the agreement between simulations and the $N \to \infty$ prediction is excellent only up to a finite-$N$ cutoff below the upper boundary $z=1$. Consequently, the simulations do not fully capture the shape transition at the right edge. This occurs because convergence toward Eq.~\eqref{eq:SHforHC} is progressive: it sets in first at small values of $z$ and only later extends toward the upper edge as $N$ increases.
A simple way to estimate the finite-$N$ cutoff $z^*(N)<1$ is to determine when the approximation performed in Eq.~\eqref{eq:approxwithdeltaHC} ceases to be accurate. As discussed above, this approximation neglects the intrinsic Gumbel fluctuations at fixed reset age $T$, whose scale is
\begin{equation}\label{eq:bN_HC_boundary_layer}
b_N(T)\approx\frac{L_H\sqrt{1-e^{-2T}}}{\sqrt{2\ln N}}.
\end{equation}
These fluctuations should be small compared with the fluctuations of the typical maximum induced by the random reset age, which are of order
\begin{equation}
    a_N'(T)=\frac{d a_N(T)}{dT}.
\end{equation}
Thus, the condition for the delta approximation to be valid is
\begin{equation}\label{eq:ratio_boundary_layer_HC}
\frac{b_N(T^*)}{a_N'(T^*)}\ll1.
\end{equation}
Using
\begin{equation}\label{eq:aN_HC_boundary_layer}
a_N(T)=\Lambda_N\sqrt{1-e^{-2T}},
\qquad
\Lambda_N=L_H\sqrt{2\ln N},
\end{equation}
one obtains
\begin{equation}\label{eq:derivative_aN_HC_boundary_layer}
a_N'(T)=
\frac{d a_N(T)}{dT}
=
\Lambda_N\frac{e^{-2T}}{\sqrt{1-e^{-2T}}}.
\end{equation}
For a given value of $z=m/\Lambda_N$, the delta approximation in Eq.~\eqref{eq:approxwithdeltaHC} selects the reset age $T^*(z)$ such that $m=a_N(T^*)$. Therefore, using \eqref{eq:aN_HC_boundary_layer}, we obtain
\begin{equation}\label{eq:Tstar_HC_boundary_layer}
z=\sqrt{1-e^{-2T^*(z)}},
\qquad
T^*(z)=-\frac{1}{2}\ln(1-z^2).
\end{equation}
Evaluating the condition \eqref{eq:ratio_boundary_layer_HC} at this value of $T=T^*(z)$, one finds
\begin{equation}\label{eq:ratio_boundary_layer_HC_explicit}
\frac{b_N(T^*)}{a_N'(T^*)}
=
\frac{z^2}{2\ln N\,(1-z^2)}
\ll 1.
\end{equation}
The cutoff $z^*(N)$ is then estimated by setting this ratio to one. This gives
\begin{equation}\label{eq:zstar_boundary_layer_HC}
z^*(N)=
\sqrt{\frac{2\ln N}{1+2\ln N}}
\approx
1-\frac{1}{4\ln N}.
\end{equation}
This suggests that the EVS should be well described by the asymptotic result in Eq.~\eqref{eq:SHforHC} for $0<z<z^*(N)$, while finite-$N$ effects remain important in the boundary layer $z^*(N)<z\leq1$. The width of this layer is of order $1/\ln N$, which explains why convergence near the edge is very slow. For instance, for $N=10^6$, Eq.~\eqref{eq:zstar_boundary_layer_HC} gives $z^*(N)\approx0.982$.

\subsubsection{Box confinement (BC)}\label{sec:BoxConfinementEVS}

We now turn to the BC. In this case, we study both the typical and the large deviation regime of the maximum $M_1$ and we show that its probability density reads, for large $N$,
\begin{equation}\label{eq:FULLEVSBCwithLD_M1}
    \operatorname{Prob}\{M_1=m\}
    =
    \begin{cases}
        \dfrac{N}{L}\,S_B(N\delta),
        & \delta=O(1/N),\\[3mm]
        \dfrac{1}{L\ln N}\,S_2(\ln N\,\delta),
        & \delta=O(1/\ln N),\\[3mm]
        \dfrac{1}{L\ln N}\,S_3(\delta),
        & \delta=O(1),
    \end{cases}
    \qquad
    \delta=\frac{1}{2}-\frac{m}{L},
\end{equation}
where the first line describes the typical regime. The three scaling functions read
\begin{equation}\label{eq:defSBEVS_2}
S_B(z)=\nu_B^2\int_0^\infty dT\,
e^{-\nu_B^2T}\,\vartheta(T)\,e^{-\vartheta(T)z}
\approx
\begin{cases}
c_1(\nu_B)-c_2(\nu_B)z, & z\to 0,\\[1mm]
\frac{\nu_B^2}{4}\,\frac{1}{z(\ln z)^2}, & z\to \infty,
\end{cases}
\end{equation}
with
\begin{equation}
c_1(\nu_B)=\frac{\nu_B}{\sinh \nu_B},
\qquad
c_2(\nu_B)=\nu_B^2\int_0^\infty dT\,e^{-\nu_B^2T}\vartheta(T)^2,
\end{equation}
and 
\begin{equation}\label{eq:ThetaTfirstdef_2}
\vartheta(T)=1+2\sum_{n=1}^{\infty}(-1)^n e^{-\pi^2 n^2 T}=
\frac{1}{\sqrt{\pi T}}
\sum_{k=-\infty}^{+\infty}
\exp\!\left[-\frac{(2k+1)^2}{4T}\right],
\end{equation}
\begin{figure}[h!] 
    \centering
    \includegraphics[width=0.6\textwidth]{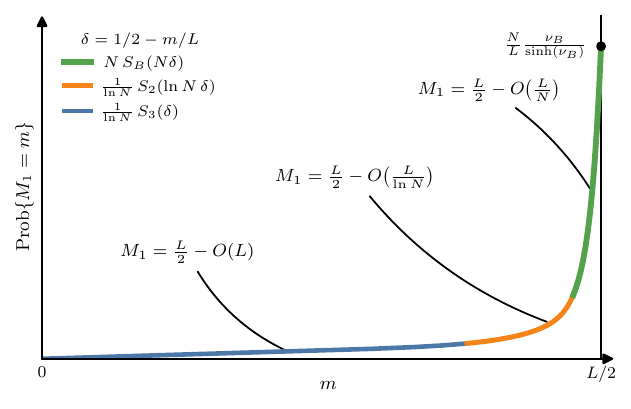}
    \caption{
    Schematic representation of the EVS in the BC case, summarized in Eq.~\eqref{eq:FULLEVSBCwithLD_M1}.
    The three colors indicate the three asymptotic regimes of the rightmost particle $M_1$.
    Since in the BC the rightmost particle is supported on $0\leq M_1\leq L/2$ for large $N$, the equivalent distance from the right boundary, $\Delta=1/2-M_1/L$, takes values in $[0,1/2]$.
    The green region corresponds to the typical regime, where $M_1$ lies at a distance $O(L/N)$ from the right boundary $L/2$, and is described by the scaling function $S_B(z)$ (see \eqref{eq:defSBEVS_2}).
    The orange region corresponds to the intermediate crossover regime, where this distance is of order $O(L/\ln N)$, and is described by $S_2(z)$ (see \eqref{eq:S2EVSBC_2}).
    The blue region corresponds to the large deviation regime, where $M_1$ lies at a macroscopic distance $O(L)$ from the boundary, and is described by $S_3(z)$ (see \eqref{eq:S3firstdefinition}).
    As $N$ increases, the probability mass concentrates in the green typical regime, while the intermediate and large deviation sectors are suppressed.}
    \label{fig:EVSBC_FULLL}
\end{figure}
while
\begin{equation}\label{eq:S2EVSBC_2}
S_2(z)=\nu_B^2\operatorname{coth}(4z)\approx
\begin{cases}
\frac{\nu_B^2}{4z}, & z\to 0,\\[1mm]
\nu_B^2, & z\to \infty,
\end{cases}
\end{equation}
and
\begin{equation}\label{eq:S3firstdefinition}
    S_3(z)=\nu_B^2(1-2z),
    \qquad
    0\leq z\leq \frac{1}{2}.
\end{equation}
A schematic representation of the three regimes of \eqref{eq:FULLEVSBCwithLD_M1} is shown in Fig. \ref{fig:EVSBC_FULLL}.

Equation~\eqref{eq:FULLEVSBCwithLD_M1} should be interpreted as follows. In the BC, the random variable $M_1$ describing the position of the rightmost particle in the stationary state is typically very close to the boundary $L_B=L/2$, at a distance $O(1/N)$. However, rare events may occur in which its distance from the boundary is much larger than its typical value. These events are only weakly suppressed, as a consequence of the fat tail of the typical scaling function $S_B(z)$. More precisely, for large values of its argument, $S_B(z)$ decays only as $1/[z(\ln z)^2]$. This tail is normalizable, but it is sufficiently broad that all its positive moments diverge, including the mean. When the distance from the boundary grows beyond the typical scale $O(1/N)$, the EVS crosses over first to the intermediate regime, in which $M_1$ lies at a distance $O(1/\ln N)$ from the boundary and is described by $S_2(z)$, and then to the large-deviation regime, in which $M_1$ lies at a distance $O(1)$ from the boundary and is described by $S_3(z)$. Thus, the three lines in Eq.~\eqref{eq:FULLEVSBCwithLD_M1} describe, respectively, the typical fluctuations close to the wall, the intermediate crossover regime, and the rare configurations in which the maximum lies at a macroscopic distance from the boundary.

We now derive all these results and analyze them in more detail.
As anticipated, the analysis is more involved than in the HC. In the HC case, the typical position of the maximum is controlled by a single $N$ dependent scale, i.e. $a_N(T)$ has the same $N$ dependence for all $T$. In the box geometry, by contrast, the dependence of $a_N(T)$ on both $N$ and $T$ is more intricate, and several asymptotic regimes emerge. Here the natural rescaled time is
\begin{equation}
T=\frac{D\tau}{L_B^2}=\frac{4D\tau}{L^2}.
\end{equation}
Our goal is again to study Eq.~\eqref{eq:general_formula_EVS_T}. To do so, we first need the conditional distribution $\operatorname{Prob}\{M_1(T)=m\}$, namely, the IID extreme value problem at fixed $T$. Indeed, conditioned on the elapsed time $T$ since the last reset, the particles are IID, because they all evolve with the same reset-free propagator given in Eqs.~\eqref{eq:propag_1} and \eqref{eq:propag_3}. In terms of the rescaled variable $T$, this propagator can be written as
\begin{equation}\label{eq:propag_1_TTT}
p_0^{(B)}(x,T)=\frac{1}{L}
+\frac{2}{L}\sum_{n=1}^{\infty}
\cos\left(\frac{2\pi n x}{L}\right)\,
\exp\!\left[-\pi^2 n^2 T\right],
\end{equation}
or, equivalently, using the Poisson summation formula, as
\begin{equation}\label{eq:propag_3_TTT}
    p_0^{(B)}(x,T)=\frac{1}{L\sqrt{\pi T}}\sum_{k=-\infty}^{\infty}
    \exp\!\left[-\frac{(x/L-k)^2}{T}\right].
\end{equation}
Both these representations will be useful. Eq.~\eqref{eq:propag_1_TTT} is convenient when $T$ is large, where only the first few Fourier modes contribute appreciably. Eq.~\eqref{eq:propag_3_TTT} is instead better suited to small $T$, where again only a few image terms matter. In particular, for sufficiently small $T$ one has, from Eq. \eqref{eq:propag_3_TTT},
\begin{equation}
p_0^{(B)}(x,T)\approx \frac{1}{L\sqrt{\pi T}}\exp\!\left[-\frac{x^2}{L^2T}\right],
\end{equation}
which is just the diffusive Gaussian propagator. For large $T$, on the other hand, the propagator converges exponentially fast to the uniform distribution,
\begin{equation}
p_0^{(B)}(x,T)\to \frac{1}{L}.
\end{equation}
This is precisely what makes the problem harder than in the HC case. In Eq.~\eqref{eq:general_formula_EVS_T}, one must integrate over all values of $T$, from very small to very large, but the form of $p_0^{(B)}(x,T)$ changes markedly across these regimes. For small $T$, the parent distribution is approximately Gaussian and therefore belongs to the Gumbel class. For large $T$, it is approximately uniform with bounded support at $x=L/2$, and thus belongs to the Weibull class. As a result, the behavior of the maximum is very different from that in HC. 

There is also a second important difference with respect to HC. In the HC, knowing the typical value $a_N(T)$ is enough to reconstruct the full distribution of $M_1$ in the large $N$ limit. Here this will no longer be true: also the fluctuation given by $b_N(T)$ will be relevant.
However, first of all, to make the physical picture transparent, we start by analyzing the typical position of the maximum, $a_N(T)$, defined implicitly by the equation
\begin{equation}\label{eq:aNdefBC}
\int_{a_N(T)}^{L/2} p_0^{(B)}(x,T)\,dx=\frac{1}{N}.
\end{equation}
Let us therefore fix $T$ and ask how the typical value of the maximum
$M_1(T)\sim a_N(T)$ behaves. When $T$ is small, $p_0^{(B)}(x,T)$ is approximately
Gaussian. Thus, Eq.~\eqref{eq:aNdefBC} gives
\begin{equation}
a_N(T)\approx L\sqrt{T}\,u_N,
\end{equation}
where $u_N$ is given in \eqref{eq:defdiuN}. At leading order, this gives
\begin{equation}
a_N(T)\approx L\sqrt{T\ln N}.
\end{equation}
Physically, this means that if only a short time $T$ has elapsed since the last reset, then even the rightmost particle has not had enough time to reach the boundary. The system is therefore effectively unconfined, and the standard Gaussian diffusive picture applies.
If $T$ is increased, however, the particles, and in particular the rightmost one, eventually reach the boundary at $x=L/2$. Thus, in this case, the maximum is expected to sit extremely close to the boundary, so that
\begin{equation}
a_N(T)\sim \frac{L}{2}.
\end{equation}
The Gaussian regime should therefore remain valid only as long as the typical position of the rightmost particle is still well inside the box. The crossover occurs when
\begin{equation}\label{eq:critical_condition}
L\sqrt{T}u_N\sim \frac{L}{2}.
\end{equation}
This defines the critical time
\begin{equation}
T_c(N)=\frac{1}{4u_N^2}\approx \frac{1}{4\ln N}.
\end{equation}
Thus, for $T<T_c(N)$, typically, not even the rightmost particle has reached the boundary, and the Gaussian approximation is expected to work well. For $T>T_c(N)$ instead, the rightmost particle has typically already reached the edge, and the behavior crosses over to a boundary dominated regime. The fact that $T_c(N)$ decreases as $N$ increases is natural: the larger the number of particles, the earlier the rightmost one reaches the boundary.
From the same condition in Eq.~\eqref{eq:critical_condition}, one may also define a critical number of particles,
\begin{equation}
N_c(T)\approx e^{\frac{1}{4T}},
\end{equation}
such that, at fixed $T$, the system is in the Gaussian regime when $N<N_c(T)$, whereas for $N>N_c(T)$ the rightmost particle has typically already reached the boundary.
Since our goal is to study the EVS in the large $N$ limit, there are then two distinct ways of taking this limit. The first is
\begin{equation}
1\ll N \ll N_c\!\left(T=1/\nu_B^2\right)=e^{\nu_B^2/4},
\end{equation}
where we have replaced $T$ by its typical value, of order $1/\nu_B^2$. This corresponds to the limit of a very large box, $L\to\infty$, in which confinement becomes irrelevant. In this regime one recovers the results of the unconfined resetting gas obtained in Ref.~\cite{Biroli2023}.
In the present work, however, we are interested in the opposite regime,
\begin{equation}
N\gg N_c\!\left(T=1/\nu_B^2\right),
\end{equation}
that is, we keep $\nu_B=O(1)$ so that the box size remains comparable to the typical resetting length, while taking $N$ large. This is precisely the regime in which the confinement affects non-trivially our system, or, at least, the rightmost particle.
In the following, we first analyze the typical regime of the EVS in the BC, and then turn to the corresponding large-deviation regime.\\

\paragraph{\textbf{Typical regime}}

Now that the relevant scaling regime has been identified, let us return to Eq.~\eqref{eq:general_formula_EVS_T}. A natural first idea is to split the integration domain into two parts, $[0,T_c(N)]$ and $[T_c(N),\infty)$. In the first region, $\operatorname{Prob}\{M_1(T)=m\}$ is approximately of Gumbel type, while in the second it belongs to the Weibull class. However, in the large $N$ limit of interest, $T_c(N)\approx\frac{1}{4\ln N}$ is very small. Therefore, at leading order, the contribution coming from $T<T_c(N)$ can be neglected.
The leading contribution thus comes from the boundary-dominated regime $T>T_c(N)$. In this regime, however, an important difference with respect to the HC case appears. There, the fluctuations at fixed $T$ were negligible compared with those induced by the randomness of $T$, which allowed us to replace $\operatorname{Prob}\{M_1(T)=m\}$ by $\delta(m-a_N(T))$. This is no longer true here.
Indeed, when $T=O(1)$, the IID Weibull results (see Eq.~\eqref{eq:IIDEVSResults}) give
\begin{equation}
M_1(T)=a_N(T)+b_N(T)\,Z,
\qquad
a_N(T)=\frac{L}{2},
\qquad
b_N(T)=\frac{1}{N\,p_0^{(B)}(L/2,T)}.
\end{equation}
where $Z$ is distributed as $Z\sim G_{III}(z)=e^{z}\Theta(-z).$
Let us again compare the two sources of fluctuations. For a typical age $T=O(1)$, one has $a_N(T)=L/2$, which is independent of $T$. Hence the fluctuations induced by the randomness of $T$ are now of order $\Delta M_1 \sim b_N'(T)\,\Delta T = O\!\left( 1/N\right),$
while the fluctuations at fixed $T$ are themselves of order $b_N(T)=O\!\left( 1/N\right)$.
Thus, unlike the HC case, the two contributions are of the same order and both must be retained. As a consequence, one cannot replace $\operatorname{Prob}\{M_1(T)=m\}$ by a delta peak. One must instead keep the full Weibull scaling form, given by
\begin{equation}\label{eq:conditionalWeibullBC}
\operatorname{Prob}\{M_1(T)=m\}=
\frac{1}{b_N(T)}
G_{III}\!\left(\frac{m-a_N(T)}{b_N(T)}\right),
\qquad
m<a_N(T),
\end{equation}
where $G_{III}(z)=\Theta(-z)e^{z}$, and substitute it into Eq.~\eqref{eq:general_formula_EVS_T}.
As in \eqref{eq:ThetaTfirstdef_2}, we now define the function
\begin{equation}\label{eq:ThetaTfirstdef}
\vartheta(T)=1+2\sum_{n=1}^{\infty}(-1)^n e^{-\pi^2 n^2 T}=
\frac{1}{\sqrt{\pi T}}
\sum_{k=-\infty}^{+\infty}
\exp\!\left[-\frac{(2k+1)^2}{4T}\right],
\end{equation}
so that, using Eq.~\eqref{eq:propag_1_TTT} we have $p_0^{(B)}(L/2,T)=\frac{1}{L}\vartheta(T)$. It follows that
\begin{equation}
b_N(T)=\frac{L}{N\vartheta(T)}.
\end{equation}
It is now convenient to introduce the dimensionless distance of the maximum
from the right boundary,
\begin{equation}
    \Delta=\frac{1}{2}-\frac{M_1}{L},
\end{equation}
which is a random variable defined in $[0,1]$.
We denote by $\delta$ the value taken by this random variable when $M_1=m$, namely
\begin{equation}
    \delta=\frac{1}{2}-\frac{m}{L}
\end{equation}
Substituting Eq. \eqref{eq:conditionalWeibullBC} into Eq.~\eqref{eq:general_formula_EVS_T}, one finally obtains
\begin{equation}\label{eq:typicalEVSBC}
\operatorname{Prob}\{M_1=m\}=
\frac{N}{L}\,S_B(N\delta),
\end{equation}
where
\begin{equation}\label{eq:defSBEVS}
S_B(z)=\nu_B^2\int_0^\infty dT\,
e^{-\nu_B^2T}\,\vartheta(T)\,e^{-\vartheta(T)z}
\approx
\begin{cases}
c_1(\nu_B)-c_2(\nu_B)z, & z\to 0,\\[1mm]
\frac{\nu_B^2}{4}\,\frac{1}{z(\ln z)^2}, & z\to \infty,
\end{cases}
\end{equation}
with
\begin{equation}
c_1(\nu_B)=\frac{\nu_B}{\sinh \nu_B},
\qquad
c_2(\nu_B)=\nu_B^2\int_0^\infty dT\,e^{-\nu_B^2T}\vartheta(T)^2.
\end{equation}
The function $S_B(z)$, together with the results of numerical simulations, is shown in Fig.~\ref{fig:EVS} on the right panel.
We notice here that the fluctuations of the scaled distance $N\Delta$ are extremely broad. Indeed, we see from Eq. \eqref{eq:defSBEVS} that $S_B(z)\sim \frac{\nu_B^2}{4z(\ln z)^2}$ for large $z$.
As a consequence, all moments of the distribution $S_B(z)$ diverge, including the mean.

In the same spirit as for the IID case, this result can be written as
\begin{equation}
M_1=\frac{L}{2}-\frac{L}{N}Z,
\end{equation}
where the rescaled random variable
\begin{equation}
Z=\frac{N}{L}\left(\frac{L}{2}-M_1\right)=N\Delta
\end{equation}
is of order $O(1)$ and has probability density $S_B(z)$.
This form is especially instructive when compared with equilibrium. In fact, in the absence of resetting, the system would simply relax to the uniform equilibrium stationary distribution, which factorizes. The corresponding EVS $M_1^{eq}$ is then the standard Weibull one, where
\begin{equation}
M_1^{eq}=\frac{L}{2}+\frac{L}{N}Z,
\qquad
Z\sim G_{\rm III}(z)=e^{z}\Theta(-z),
\end{equation}
or, equivalently
\begin{equation}
M_1^{eq}=\frac{L}{2}-\frac{L}{N}Z,
\qquad
Z\sim e^{-z}\Theta(z).
\end{equation}
Hence, the DEC generated by resetting do not modify the leading centering and scaling parameters $a_N=L/2$ and $b_N=L/N$, which remain the same as in equilibrium. What they completely change is the statistics of the rescaled variable
\begin{equation}
Z=\frac{a_N-M_1}{b_N}.
\end{equation}
Indeed, the exponential tail of the equilibrium Weibull case is replaced by an extremely broad heavy tail in the resetting system. Thus, while the typical distance of the maximum from the boundary remains of order $1/N$, the fluctuations around this typical value are drastically enhanced by correlations. This confirms the physical picture of a "fluffy" gas.\\

\paragraph{\textbf{Large deviation regime}}

We have seen that, in the BC geometry, the distribution of the distance of the maximum $M_1$ from the boundary at $L/2$ has an extremely broad tail, so broad that even its first moment diverges. This means that, although the maximum is typically very close to the boundary, configurations in which it lies anomalously far from it are much more likely than in the equilibrium case. 
\begin{figure}[h!] 
    \centering
    \includegraphics[width=0.75\textwidth]{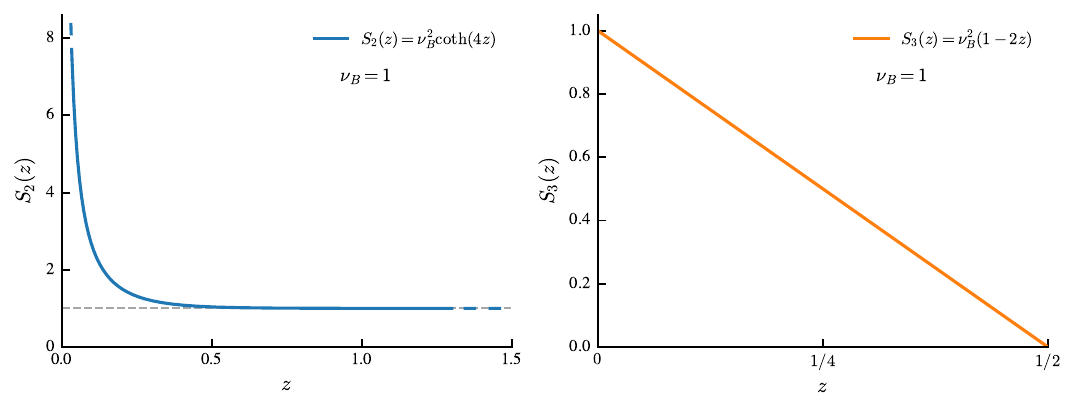}
    \caption{Plot of the scaling functions $S_2(z)$ and $S_3(z)$, given in Eqs.~\eqref{eq:S2EVSBC} and \eqref{eq:S3EVSBC}, respectively.
    They describe the intermediate crossover and large-deviation regimes of the EVS in the BC case.}
    \label{fig:S2S3_largedevEVS_BC}
\end{figure}
In terms of the dimensionless distance
\begin{equation}
\delta=\frac{1}{2}-\frac{m}{L},
\end{equation}
this corresponds to atypically large values of $\delta$. We now study the large deviations of this variable.
The origin of this regime can be traced back to the small-time regime of the integral in Eq. \eqref{eq:general_formula_EVS_T}. In the previous section, the interval $[0,T_c(N)]$ was neglected because
\begin{equation}\label{eq:TcNagain}
T_c(N)\approx \frac{1}{4\ln N}
\end{equation}
vanishes as $N\to\infty$. However, for very large but finite $N$, there is still a small probability that two consecutive resets occur very close to each other, so that the elapsed time satisfies $T<T_c(N)$.
When this condition is verified, as discussed above, the propagator is well approximated by a Gaussian. Therefore, at fixed $T$, the maximum behaves as
\begin{equation}\label{eq:maxgaussian_nf}
M_1(T)=L\sqrt{T\ln N}+L\sqrt{\frac{T}{4\ln N}}\,Z,
\end{equation}
where $Z$ is a Gumbel random variable. As in the HC case, the fluctuations at
fixed $T$, which are encoded in the second term on the right-hand side of
Eq.~\eqref{eq:maxgaussian_nf}, are negligible compared with those induced by the
randomness of $T$. We may therefore replace the conditional distribution by a delta peak,
\begin{equation}
\operatorname{Prob}\{M_1(T)=m\}\to \delta\!\left(m-a_N(T)\right),
\qquad
a_N(T)=L\sqrt{T\ln N}.
\end{equation}
Substituting this approximation into Eq.~\eqref{eq:general_formula_EVS_T}, restricted to the small-time interval, gives
\begin{equation}
\operatorname{Prob}\{M_1=m\}\approx
\nu_B^2\int_0^{T_c(N)} dT\, e^{-\nu_B^2T}\,
\delta\!\left(m-L\sqrt{T\ln N}\right),
\end{equation}
where $T_c(N)$ is given in \eqref{eq:TcNagain}.
Performing the integral and keeping only the leading contribution at large $N$, one obtains
\begin{equation}
\operatorname{Prob}\{M_1=m\}
=
\frac{2\nu_B^2}{L^2\ln N}\,m,
\qquad
0\leq m\leq \frac{L}{2}.
\end{equation}
Equivalently, rewriting the result in terms of the dimensionless distance $\delta=1/2-m/L$, we find
\begin{equation}\label{eq:S3EVSBC}
\operatorname{Prob}\{M_1=m\}
=
\frac{1}{L\ln N}\,S_3(\delta),
\qquad
S_3(z)=\nu_B^2(1-2z),
\qquad
0\leq z\leq \frac{1}{2}.
\end{equation}
We refer to this regime as the large deviation regime, while Eq.~\eqref{eq:typicalEVSBC} describes the typical regime. The function $S_3(z)$ is shown in Fig. \ref{fig:S2S3_largedevEVS_BC}.

We show in the Appendix \ref{sec:intermregmeEVS_BCcase_APP} that there is also an intermediate regime connecting these two limits. This matching regime appears when $\delta=O\!\left(1/\ln N\right)$, where we recall $\delta=1/2-m/L$.
In this regime, the distribution takes the form
\begin{equation}
\operatorname{Prob}\{M_1=m\}=
\frac{1}{L\ln N}\,
S_2(\ln N\,\delta),
\end{equation}
with
\begin{equation}\label{eq:S2EVSBC}
S_2(z)=\nu_B^2\operatorname{coth}(4z)\approx
\begin{cases}
\frac{\nu_B^2}{4z}, & z\to 0,\\[1mm]
\nu_B^2, & z\to \infty,
\end{cases}
\end{equation}
which is shown in Fig. \ref{fig:S2S3_largedevEVS_BC}.
It is useful to notice that this intermediate regime provides the matching between the typical and large-deviation sectors. Indeed, in the overlap region with $N\delta\gg 1$ and $\delta\ln N\ll 1$, the large-$z$ tail of $S_B(z)$ matches the small-$z$ behavior of $S_2(z)$. Similarly, in the overlap region with $\delta\ln N\gg 1$ and $\delta\ll 1$, the large-$z$ limit of $S_2(z)$ matches the small-$\delta$ behavior of $S_3(z)$.
The complete EVS in the BC geometry can therefore be summarized as
\begin{equation}\label{eq:FULLEVSBCwithLD}
\operatorname{Prob}\{\Delta=\delta\}=
\begin{cases}
N\,S_B(N\delta), & \delta=O(1/N),\\[2mm]
\dfrac{1}{\ln N}\,S_2(\ln N\,\delta), & \delta=O(1/\ln N),\\[2mm]
\dfrac{1}{\ln N}\,S_3(\delta), & \delta=O(1),
\end{cases}
\end{equation}
where $\Delta=1/2-M_1/L$. In principle, $M_1$ is supported on
$[-L/2,L/2]$, and therefore $\Delta$ is supported on $[0,1]$.
However, from Eq. \eqref{eq:FULLEVSBCwithLD} we see that, to leading order for large $N$, the asymptotic regimes described
by Eq.~\eqref{eq:FULLEVSBCwithLD} are supported on $0\leq M_1\leq L/2$,
or equivalently on $0\leq \Delta\leq 1/2$.
Here $S_B(z)$, $S_2(z)$, and $S_3(z)$ are the scaling functions given respectively in
\eqref{eq:defSBEVS}, \eqref{eq:S2EVSBC}, and \eqref{eq:S3EVSBC}. A schematic representation of this result is shown in Fig. \ref{fig:EVSBC_FULLL}. Equation~\eqref{eq:FULLEVSBCwithLD} is the counterpart of Eq.~\eqref{eq:FULLEVSBCwithLD_M1} written in terms of the dimensionless distance from the boundary,
$\Delta=1/2-M_1/L$, rather than in terms of the value of the maximum $M_1$ itself.

\subsection{Gap statistics}\label{sec:GAPstats}

We now turn to the gap statistics. The first gap is a particularly useful observable because, unlike the maximum itself, it is insensitive to the centering scale $a_N$, as it can be seen from Eq. \eqref{eq:gapniceform}. It therefore probes the fluctuations at the edge more directly. 
We have already seen in Eq. \eqref{eq:gapstatsverygeneral} that the probability distribution of the first gap $d_1$ can be written as
\begin{equation}
\operatorname{Prob}\{d_1=g\}=
r\int_0^\infty d\tau\,  e^{-r\tau}\,\operatorname{Prob}\{d_1(\tau)=g\}.
\end{equation}
As in the EVS analysis, it is convenient to introduce a rescaled time variable. 
In the HC case we use $T=\mu\tau$, while in the BC case we use $T=\frac{4D\tau}{L^2}$. With this change of variables, the gap distribution becomes
\begin{equation}\label{eq:GAP_general_base}
\operatorname{Prob}\{d_1=g\}=
\nu^2\int_0^\infty \, dT\, e^{-\nu^2T}\,
\operatorname{Prob}\{d_1(T)=g\}.
\end{equation}
Here, with a slight abuse of notation, $\nu$ and $T$ denote the control parameter and the rescaled time appropriate to the chosen geometry: $\nu=\nu_H$, $T=\mu\tau$ in the HC, and $\nu=\nu_B$, $T=4D\tau/L^2$ in the BC.
The strategy is the same as for the EVS: first solve the associated IID problem at fixed $T$, and then average over $T$. 
In Ref.~\cite{Biroli2023}, this large-$N$ form was derived for the unconfined
resetting gas. The derivation, however, only relies on the IID structure at fixed reset age and on the local Poisson description of the particles near the edge.
It can therefore be applied also in the present confined setting, provided that
$p_0$ is replaced by the appropriate reset-free confined propagator and that
$a_N(T)$ is determined from the corresponding geometry. This gives
\begin{equation}\label{eq:eqpartenzaGAP}
\operatorname{Prob}\{d_1=g\}=
\nu^2\int_0^\infty  dT\, e^{-\nu^2T}\,
N\,p_0\bigl(a_N(T),T\bigr)\,
e^{-N p_0(a_N(T),T)\,g}.
\end{equation}
In particular, we will use $p_0=p_0^{(H)}$ in the HC and $p_0=p_0^{(B)}$ in the BC. The function $p_0(a_N(T),T)$ is the corresponding propagator evaluated at the edge location $x=a_N(T)$, where $a_N(T)$ is defined by
\begin{equation}\label{eq:aNdefinGAP}
\int_{a_N(T)}^{\Lambda^*} p_0(x,T)\,dx=\frac{1}{N}.
\end{equation}
Here $\Lambda^*=\infty$ in the HC, while $\Lambda^*=L/2$ in the BC. Thus, $a_N(T)$ represents the typical location of the rightmost particle at fixed time $T$ elapsed since the last reset.
It is then natural to introduce the local edge density
\begin{equation}
    \varrho(T)=N p_0(a_N(T),T).
\end{equation}
This quantity is the particle density near the typical position of the maximum $a_N(T)$. From Eq. \eqref{eq:eqpartenzaGAP} we see that in terms of $\varrho(T)$, the conditional distribution of the gap at fixed $T$ is simply exponential, namely
\begin{equation}
    \operatorname{Prob}\{d_1(T)=g\}
    \approx
    \varrho(T)e^{-\varrho(T)g}.
\end{equation}
This has a simple interpretation. Near the edge, on the scale of the typical gap, the density profile varies slowly. Therefore, the particles can be locally approximated by a homogeneous Poisson point process with rate $\varrho(T)$. The spacing between the two rightmost particles is then exponentially distributed with this rate.
The full stationary gap distribution is obtained by averaging these conditional exponential laws over the random reset age $T$. Since $a_N(T)$ has already been determined in both HC and BC from the EVS analysis, the remaining task is to evaluate $p_0(a_N(T),T)$ in each geometry and insert it into Eq.~\eqref{eq:eqpartenzaGAP}. In the following we will use the notation $\varrho_H(T)$ in the HC and $\varrho_B(T)$ in the BC.

\subsubsection{Harmonic confinement (HC)}

\begin{figure}[htbp]
\centering

\begin{minipage}{0.48\textwidth}
    \centering
    \includegraphics[width=\linewidth]{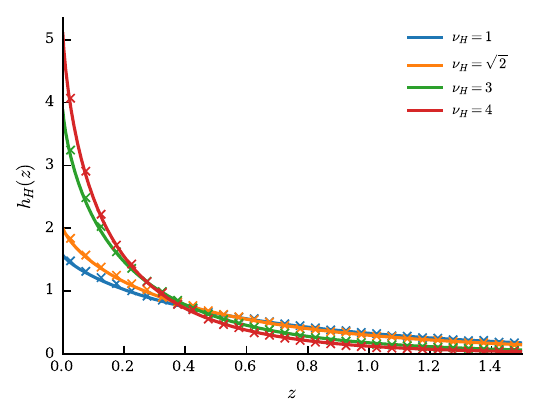}
\end{minipage}
\hfill
\begin{minipage}{0.48\textwidth}
    \centering
    \includegraphics[width=\linewidth]{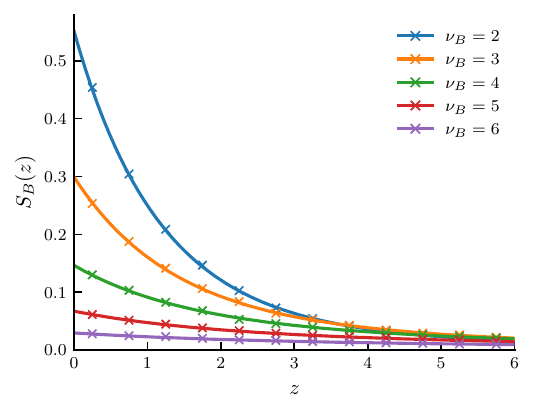}
\end{minipage}

\caption{Gap statistics in the HC (left) and in the BC (right).
Crosses show numerical simulations with $N=10^6$ particles, while solid lines correspond to the asymptotic scaling predictions.
\textbf{Left:} HC scaling function $h_H(z)$, given in Eq.~\eqref{eq:h_Hofz_GAPHC_begin}, for different values of $\nu_H$.
Here $z=g/\lambda_N$, with $\lambda_N=\sqrt{D/(2\mu\ln N)}$.
\textbf{Right:} BC scaling function $S_B(z)$, given in Eq.~\eqref{eq:typicalGAPBC}, for different values of $\nu_B$.
Here $z=Ng/L$ is the rescaled first gap.}
\label{fig:gap_statistics_plot}
\end{figure}

We now analyze the first gap statistics in the HC. We show that the probability density of $d_1$ takes the scaling form
\begin{equation}\label{eq:GAP_HC_scalform_begin}
\operatorname{Prob}\{d_1=g\}=
\frac{1}{\lambda_N}
h_H\left(\frac{g}{\lambda_N}\right),
\qquad
\lambda_N=
\frac{L_H}{\sqrt{2\ln N}},
\end{equation}
where
\begin{equation}\label{eq:h_Hofz_GAPHC_begin}
h_H(z)=
\nu_H^2
\int_0^1 dy\,
(1-y^2)^{\frac{\nu_H^2}{2}-1}
e^{-z/y}\approx
\begin{cases}
B_1+\nu_H^2 z\ln z, & z\to 0,\\
B_2 \, z^{-\nu_H^2/2}e^{-z}, & z\to\infty,
\end{cases}
\end{equation}
with $B_1=\sqrt{\pi}
\tfrac{\Gamma\!\left(1+\frac{\nu_H^2}{2}\right)}
{\Gamma\!\left(\frac{1}{2}+\frac{\nu_H^2}{2}\right)},$ and $B_2=2^{\nu_H^2/2}\Gamma\!\left(1+\frac{\nu_H^2}{2}\right)$ where $\Gamma$ denotes the Gamma function.
The scaling function $h_H(z)$ is shown in the left panel of Fig.~\ref{fig:gap_statistics_plot} for different values of $\nu_H$, together with numerical simulations. Eq.~\eqref{eq:GAP_HC_scalform_begin} shows that the first gap $d_1$ is typically of size $O(1/\sqrt{\ln N})$, while its fluctuations are described by the scaling function $h_H(z)$ in Eq.~\eqref{eq:h_Hofz_GAPHC_begin}.

We now derive these results in detail.
In this case, the reset-free single-particle propagator is Gaussian and reads
\begin{equation}
p_0^{(H)}(x,T)=
\frac{1}{\sqrt{2\pi \sigma_H^2(T)}}
\exp\!\left(-\frac{x^2}{2\sigma_H^2(T)}\right),
\qquad
\sigma_H^2(T)=L_H^2(1-e^{-2T}),
\end{equation}
where $L_H=\sqrt{\frac{D}{\mu}}$.
From the EVS analysis, we already know that the typical position of the rightmost particle at fixed age $T$ is
\begin{equation}
a_N(T)=\sqrt{2}\sigma_H(T)u_N,
\qquad
u_N=\operatorname{erfc}^{-1}\!\left(\frac{2}{N}\right),
\end{equation}
which, at leading order for large $N$, gives $a_N(T)\approx \sigma_H(T)\sqrt{2\ln N}$. We recall that $\operatorname{erfc}^{-1}(x)$ is the inverse function of $\operatorname{erfc}(x)=\frac{2}{\sqrt{\pi}}\int_x^{\infty} e^{-t^2}\,dt$.
To evaluate the gap distribution, we need the function
$p_0^{(H)}(a_N(T),T)$. Here one has to be careful: although the leading estimate $u_N\sim\sqrt{\ln N}$ is sufficient in most parts of the analysis, it is not enough when $u_N$ appears in an exponential factor. In that case, the first subleading correction contributes at leading order to the prefactor. We therefore have to use
\begin{equation}
u_N\approx\sqrt{\ln N-\frac{1}{2}\ln(4\pi\ln N)}.
\end{equation}
This gives
\begin{equation}
e^{-u_N^2}\approx
\frac{2\sqrt{\pi\ln N}}{N}.
\end{equation}
and therefore
\begin{equation}
p_0^{(H)}(a_N(T),T)=
\frac{2\ln N}{N a_N(T)}.
\end{equation}
Substituting this result into the general large $N$ expression for the first gap gives
\begin{equation}
\operatorname{Prob}\{d_1=g\}=\nu_H^2
\int_0^\infty
 dT\, e^{-\nu_H^2T}
\frac{2\ln N}{a_N(T)}
\exp\!\left(
-\frac{2\ln N}{a_N(T)}g
\right),
\end{equation}
where $\nu_H=\sqrt{\frac{r}{\mu}}$.
We now insert $a_N(T)\approx L_H\sqrt{2\ln N(1-e^{-2T})},$
and perform the change of variable $y=\sqrt{1-e^{-2T}}$.
This leads directly to the scaling form in Eq. \eqref{eq:GAP_HC_scalform_begin}.
Hence, all the dependence on $N$ is contained in the scale $\lambda_N$, while the fluctuations are described by the scaling function $h_H(z)$ given in \eqref{eq:h_Hofz_GAPHC_begin}.
This result tells us that, in the HC, the typical size of the first gap is $d_1=O\!\left(\frac{1}{\sqrt{\ln N}}\right)$, or, more precisely, that
\begin{equation}
d_1=\frac{L_H}{\sqrt{2\ln N}}\,Z,
\end{equation}
where the rescaled random variable $Z=d_1/\lambda_N$ is distributed according to $h_H(z)$.

Therefore, the HC does not modify the typical scale of the first gap with respect to the unconfined resetting gas of Ref.~\cite{Biroli2023}. Indeed, in both cases the typical first gap is of order $O(1/\sqrt{\ln N})$. However, the corresponding scaling functions are different. In the unconfined case, the scaling function is
\begin{equation}
h(z)=2\int_{0}^{\infty}\mathrm{d}u\,
\exp\!\left(-u^{2}-\frac{z}{u}\right),
\end{equation}
with asymptotic behaviors
\begin{equation}
h(z)\approx
\begin{cases}
\sqrt{\pi}+2z\ln z,
& z\to 0, \\[6pt]
2\sqrt{\dfrac{\pi}{3}}\,
\exp\!\left[-3\left(\dfrac{z}{2}\right)^{2/3}\right],
& z\to \infty.
\end{cases}
\end{equation}
Thus, the large-gap tail in the unconfined case is a stretched exponential with exponent $2/3$. By contrast, in the HC case, Eq.~\eqref{eq:h_Hofz_GAPHC_begin} gives
\begin{equation}
h_H(z)\approx B_2 z^{-\nu_H^2/2}e^{-z},
\qquad z\to\infty,
\end{equation}
so that the tail is essentially exponential, up to an algebraic prefactor. Harmonic confinement therefore leaves the leading gap scale unchanged, but makes large gaps less likely than in the unconfined resetting gas.

\subsubsection{Box confinement (BC)}

\begin{figure}[h!] 
    \centering
    \includegraphics[width=0.6\textwidth]{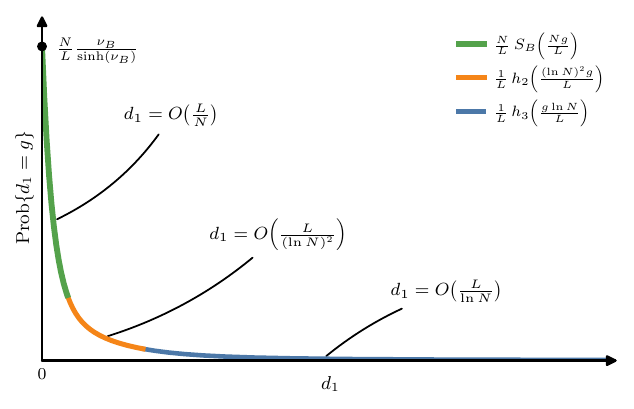}
    \caption{
    Schematic representation of the gap statistics in the BC case, summarized in Eq.~\eqref{eq:recapGAPBC_begin}.
    The three colors indicate the three asymptotic regimes of the first gap $d_1=M_1-M_2$.
    The green region corresponds to the typical regime, where the gap is of order $O(L/N)$, and is described by the scaling function $S_B(z)$ (see \eqref{eq:typicalGAPBC}).
    The orange region corresponds to the intermediate crossover regime, where the gap is of order $O(L/(\ln N)^2)$, and is described by $h_2(z)$ (see \eqref{eq:h2maintextGAP_begin}).
    The blue region corresponds to the large deviation regime, where the gap is of order $O(L/\ln N)$, and is described by $h_3(z)$ (see \eqref{eq:h3definit_begin}).
    As $N$ increases, the probability mass concentrates in the green typical regime, while the intermediate and large deviation sectors are suppressed.}
    \label{fig:GAPBC_FULLL}
\end{figure}

\begin{figure}[h!] 
    \centering
    \includegraphics[width=0.75\textwidth]{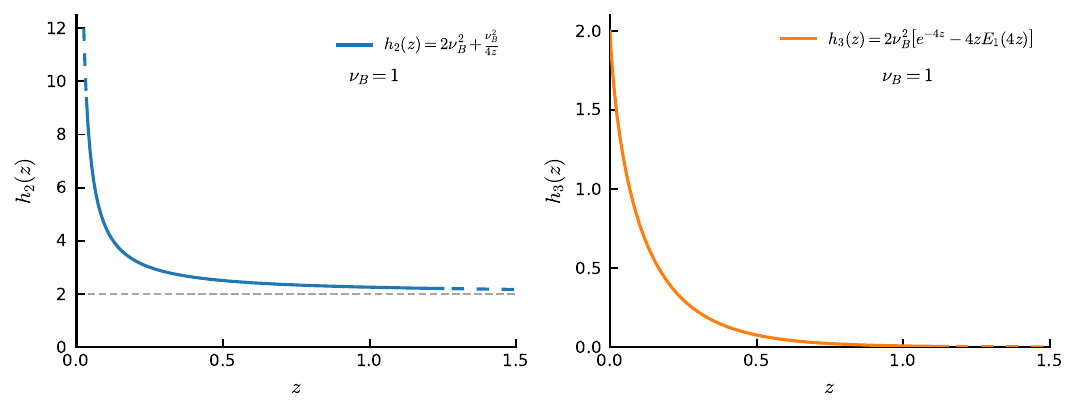}
    \caption{Plot of the scaling functions $h_2(z)$ and $h_3(z)$, given in Eqs.~\eqref{eq:h2maintextGAP_begin} and \eqref{eq:h3definit_begin}, respectively.
    They describe the intermediate crossover and large-deviation regimes of the gap statistics in the BC case. See also Eq. \eqref{eq:recapGAPBC_begin}.}
    \label{fig:h2h3_largedevGAP_BC}
\end{figure}

We now turn to the distribution of the first gap $d_1$ in the BC geometry. As for the EVS, in this confinement geometry we study not only the typical regime, but also the intermediate and large deviation regimes, since the typical scaling function has a very broad tail. The distribution of $d_1$ reads
\begin{equation}\label{eq:recapGAPBC_begin}
\mathrm{Prob}\{d_1=g\}=
\begin{cases}
\dfrac{N}{L}\,S_B\!\left(\dfrac{gN}{L}\right),
& g=O\!\left(\dfrac{L}{N}\right),\\[3mm]
\dfrac{1}{L}\,h_2\!\left(\dfrac{(\ln N)^2 g}{L}\right),
& g=O\!\left(\dfrac{L}{(\ln N)^2}\right),\\[3mm]
\dfrac{1}{L}\,h_3\!\left(\dfrac{g\ln N}{L}\right),
& g=O\!\left(\dfrac{L}{\ln N}\right).
\end{cases}
\end{equation}
The first line corresponds to the typical regime and the scaling function $S_B(z)$ is given in \eqref{eq:defSBEVS_2}, since it coincides with the scaling function of the typical regime of the EVS. The reason for this coincidence is explained below. The other two scaling functions, describing respectively the intermediate regime and the large deviation regime, read
\begin{equation}\label{eq:h2maintextGAP_begin}
h_2(z)=2\nu_B^2+\frac{\nu_B^2}{4z},
\end{equation}
and
\begin{equation}\label{eq:h3definit_begin}
h_3(z)=2\nu_B^2\left[e^{-4z}-4zE_1(4z)\right]\approx
\begin{cases}
2\nu_B^2, & z\to 0,\\[1mm]
\dfrac{\nu_B^2}{2}\dfrac{e^{-4z}}{z}, & z\to\infty,
\end{cases}
\end{equation}
where $E_1(z)$ is the exponential integral function, defined as $E_1(z)=\int_z^\infty dt\,\frac{e^{-t}}{t}$ for $z>0$. These three regimes are summarized schematically in Fig.~\ref{fig:GAPBC_FULLL}. The typical scaling function $S_B(z)$ is shown in the right panel of Fig.~\ref{fig:gap_statistics_plot}, while the intermediate and large-deviation scaling functions, $h_2(z)$ and $h_3(z)$, are plotted in Fig.~\ref{fig:h2h3_largedevGAP_BC}.

Equation~\eqref{eq:recapGAPBC_begin} should be interpreted as follows. In the BC, the first gap $d_1=M_1-M_2$ is typically very small, of order $O(1/N)$. However, rare events may occur in which the gap is much larger than its typical value, due to the fat tail of the scaling function $S_B(z)$. More precisely, for large values of its argument, $S_B(z)$ decays only as $1/[z(\ln z)^2]$. This tail is normalizable, but it is sufficiently broad that all its moments diverge, including the mean. When the gap grows beyond the typical scale $O(1/N)$, the distribution crosses over first to the intermediate regime, in which $d_1=O(1/(\ln N)^2)$ and is described by $h_2(z)$, and then to the large-deviation regime, in which $d_1=O(1/\ln N)$ and is described by $h_3(z)$. Thus, the three lines in Eq.~\eqref{eq:recapGAPBC_begin} describe, respectively, the typical small-gap fluctuations close to the edge, the intermediate crossover regime, and the rare configurations in which the first gap is anomalously large. 

We now derive the above results in detail.
As in the HC case, the central quantity is the local edge density
\begin{equation}\label{eq:densityrhoBC}
\varrho_B(T)=N\,p_0^{(B)}(a_N(T),T).
\end{equation}
We have already discussed in Sec.~\ref{sec:EVS} the behavior of $a_N(T)$ in the small- and large-$T$ regimes. A more refined analysis, reported in Appendix~\ref{sec:aNderivation_APP}, gives the complete asymptotic structure
\begin{equation}\label{eq:aNcompleteforGAP}
\frac{a_N(T)}{L}=
\begin{cases}
\sqrt{T\ln N}, & T\lesssim T_c(N),\\
\frac{1}{2}-\frac{1}{4\ln N}\operatorname{arsinh}(\frac{e^s}{2}),
& T=T_c(N)-\dfrac{s}{(2\ln N)^2},\\
\frac{1}{2}-\frac{1}{N\vartheta(T)},
& T\gtrsim T_c(N),
\end{cases}
\end{equation}
where $\vartheta(T)$ is given in \eqref{eq:ThetaTfirstdef}, $u_N=\operatorname{erfc}^{-1}(2/N)$ and
\begin{equation}
T_c(N)=\frac{1}{4u_N^2}\approx \frac{1}{4\ln N}.
\end{equation}
The three lines of Eq.~\eqref{eq:aNcompleteforGAP} correspond to the three relevant regimes of the reset age. For $T\lesssim T_c(N)$, the particles have not yet felt the boundary, and the edge statistics are governed by the Gaussian bulk form of the propagator. For $T\gtrsim T_c(N)$, the rightmost particles have already reached the boundary, and the statistics cross over to the Weibull edge regime. The intermediate line describes the narrow crossover window around $T_c(N)$, of width $O((\ln N)^{-2})$, where the scaled variable $s=(2\ln N)^2(T_c(N)-T)$ is of order $O(1)$ and $a_N(T)$ smoothly interpolates between the two regimes.

Let us briefly outline how this intermediate expression is obtained, leaving the complete derivation to Appendix~\ref{sec:aNderivation_APP}. The crossover regime corresponds to the time window in which the Gaussian estimate for the position of the rightmost particle first reaches the boundary. Equivalently, it is the regime where the bulk Gaussian description is just about to break down because the typical maximum becomes close to $L/2$. Starting from the definition of $a_N(T)$ in Eq.~\eqref{eq:aNdefinGAP}, we therefore focus on this region and write
\begin{equation}
\frac{a_N(T)}{L}=\frac{1}{2}-\epsilon,
\qquad \epsilon\ll 1.
\end{equation}
Since this regime occurs for $T\sim T_c(N)$ and $T_c(N)\approx 1/(4\ln N)$ is small for large $N$, it is convenient to use the representation of the box propagator in Eq.~\eqref{eq:propag_3_TTT}. Close to the right boundary, the leading contributions are the two terms $k=0$ and $k=1$. Keeping only these two terms in Eq.~\eqref{eq:aNdefinGAP} and solving the resulting equation in the scaling window
\begin{equation}
T=T_c(N)-\frac{s}{(2\ln N)^2},
\qquad s=O(1),
\end{equation}
gives the second line of Eq.~\eqref{eq:aNcompleteforGAP}. This crossover of $a_N(T)$ from the Gaussian bulk regime to the boundary-dominated regime is precisely what generates the intermediate gap sector.

We can now insert these expressions for $a_N(T)$ into the $p_0^{(B)}(a_N(T),T)$ in \eqref{eq:densityrhoBC}, where $p_0^{(B)}(a_N(T),T)$ is given in \eqref{eq:propag_1_TTT} and \eqref{eq:propag_3_TTT}. For the small-time regimes, namely for $T\lesssim T_c(N)$ and in the crossover window $T=T_c(N)-\frac{s}{(2\ln N)^2},$
it is convenient to use the representation in Eq.~\eqref{eq:propag_3_TTT}. For $T\gtrsim T_c(N)$, instead, the boundary-dominated regime is more conveniently described by the representation in Eq.~\eqref{eq:propag_1_TTT}. This gives the following behavior for the local edge density $\varrho_B(T)=N p_0^{(B)}(a_N(T),T)$:
\begin{equation}\label{eq:rhodiTvariregimi}
\varrho_B(T)\approx
\begin{cases}
\displaystyle
\frac{2\sqrt{\ln N}}{L \sqrt{T}},
& T \lesssim T_c(N),
\\[10pt]
\displaystyle
\frac{4\ln N}{L}\,\sqrt{1+4e^{-2s}},
& T=T_c(N)-\frac{s}{(2\ln N)^2},
\\[10pt]
\displaystyle
N\frac{\vartheta(T)}{L},
& T \gtrsim T_c(N).
\end{cases}
\end{equation}
where the function $\vartheta(T)$ is defined in \eqref{eq:ThetaTfirstdef_2}.
Thus, different scaling regimes emerge depending on the value of the reset age $T$.

\paragraph{\textbf{Typical regime}}

As for the EVS in the BC, the typical regime here is controlled by reset ages $T>T_c(N)$, since $T_c(N)\to0$ for large $N$. In this region, the local edge density reads $\varrho_B(T)=N\,p_0^{(B)}(a_N(T),T)=\frac{N}{L}\vartheta(T)$. Substituting this expression into Eq.~\eqref{eq:eqpartenzaGAP} and retaining the leading order for large $N$ yields directly
\begin{equation}\label{eq:typicalGAPBC}
\operatorname{Prob}\{d_1=g\}=
\frac{N}{L}\,
S_B\!\left(\frac{Ng}{L}\right),
\qquad
S_B(z)=\nu_B^2\int_0^\infty dT\,
e^{-\nu_B^2 T}\,\vartheta(T)\,e^{-\vartheta(T)\,z}.
\end{equation}
where $\vartheta(T)$ is given in \eqref{eq:ThetaTfirstdef_2}.
Here $S_B(z)$ is exactly the same scaling function that appeared in the EVS problem for the BC case (see Eq.~\eqref{eq:defSBEVS}).
The reason for this is the following. In the typical regime, and at fixed $T$, the particles near the right boundary can be locally approximated by a homogeneous Poisson point process with rate $\varrho_B(T)=N\vartheta(T)/L$. Therefore, both the distance between the boundary and the rightmost particle, $L/2-M_1$, and the first gap, $d_1=M_1-M_2$, are exponentially distributed with the same rate $\varrho_B(T)$. Since the same average over the common reset age $T$ is then performed, the two stationary distributions are governed by the same scaling function $S_B(z)$.

The result in Eq. \eqref{eq:typicalGAPBC} shows that, in the BC geometry, the typical first gap scales as $d_1=O(1/N)$, which is much smaller than in the HC case, where the corresponding typical scale is $d_1=O(1/\sqrt{\ln N})$. Equivalently, Eq.~\eqref{eq:typicalGAPBC} can be written as
\begin{equation}
d_1=\frac{L}{N}\,Z,
\end{equation}
where the rescaled variable $Z$ has density $S_B(z)$. Importantly, as already pointed out in Eq.~\eqref{eq:defSBEVS}, this scaling function has a very broad tail
\begin{equation}\label{eq:solocodaSB}
S_B(z)\approx \frac{\nu_B^2}{4}\frac{1}{z(\ln z)^2},
\qquad z\to\infty.
\end{equation}
Hence, although the typical gap is of order $O(1/N)$, realizations with $d_1$ much larger than its typical value remain possible. By contrast, in the HC case we have found
\begin{equation}
d_1=\frac{L_H}{\sqrt{2\ln N}}\,Z,
\end{equation}
where $Z$ has density $h_H(z)$ given in \eqref{eq:h_Hofz_GAPHC_begin}, whose tail decays exponentially fast. Therefore, large deviations of the gap are much rarer in HC. In summary, the BC geometry produces a much smaller typical gap, but also much stronger fluctuations. The HC geometry instead leads to a larger typical gap, together with a much more concentrated distribution around that scale.

It is also worth comparing our result with the corresponding equilibrium problem. Without resetting, the gas would simply equilibrate inside the box. In that case, one would still have the same leading scale,
\begin{equation}
d_1=\frac{L}{N}Z,
\end{equation}
but the rescaled random variable $Z$ would be exponentially distributed, corresponding to a much thinner tail. By contrast, in the resetting case, the same leading scale $L/N$ is accompanied by the broad tail in Eq.~\eqref{eq:solocodaSB}, which is a direct consequence of the correlations induced by the common reset age.\\

\paragraph{\textbf{Large deviation regime}}

Even though, as we have seen, the typical value of the first gap $d_1$ is of order $O(1/N)$, anomalously large gaps can still be observed for very large but finite $N$. In the previous paragraph, we split the integration domain over $T$ into the two intervals $[0,T_c(N)]$ and $[T_c(N),\infty)$, where
\begin{equation}
T_c(N)=\frac{1}{4u_N^2}\approx \frac{1}{4\ln N}.
\end{equation}
The leading contribution to the typical gap statistics comes from the interval $[T_c(N),\infty)$. We now focus instead on the complementary interval $[0,T_c(N)]$, which is responsible for anomalously large values of the first gap.
In this regime, from Eq.~\eqref{eq:rhodiTvariregimi} we know that the local edge density $\varrho_B(T)=N\,p_0^{(B)}(a_N(T),T)$ behaves as
\begin{equation}
\varrho_B(T)=\frac{2\sqrt{\ln N}}{L\sqrt{T}}.
\end{equation}
Substituting this expression into Eq.~\eqref{eq:eqpartenzaGAP}, we obtain
\begin{equation}
\operatorname{Prob}\{d_1=g\}\approx
\frac{2\nu_B^2\sqrt{\ln N}}{L}
\int_{0}^{T_c(N)}
\frac{dT}{\sqrt{T}}\,
e^{-\nu_B^2 T}
\exp\!\left[
-\frac{2g\sqrt{\ln N}}{L\sqrt{T}}
\right].
\end{equation}
We now perform the change of variable $T=y/\ln N$. Using $T_c(N)\approx 1/(4\ln N)$, this gives
\begin{equation}
\operatorname{Prob}\{d_1=g\}\approx
\frac{2\nu_B^2}{L}
\int_{0}^{1/4}
\frac{dy}{\sqrt{y}}\,
\exp\!\left[
-\frac{\nu_B^2}{\ln N}\,y
-\frac{2g\ln N}{L\sqrt{y}}
\right].
\end{equation}
Keeping only the leading contribution for large $N$, performing the integral
exactly, and introducing the scaling variable $g\ln N/L$, we obtain
\begin{equation}\label{eq:verylargedevBCGAP}
\operatorname{Prob}\{d_1=g\}=
\frac{1}{L}h_3\left(\frac{g\ln N}{L}\right),
\end{equation}
where $h_3(z)$ is given in \eqref{eq:h3definit_begin}.
Therefore, the small-$T$ interval produces a large deviation sector in which the gap is of order $g=O\!\left(L/\ln N\right)$, which is much larger than the typical scale $g=O(L/N)$.
In the Appendix \ref{sec:intermregimeGAPBC_APP}, we show that inserting the intermediate regime of Eq. \eqref{eq:rhodiTvariregimi}, namely
\begin{equation}
\varrho_B(T)=\frac{4\ln N}{L}\sqrt{1+4e^{-2s}},
\end{equation}
into \eqref{eq:eqpartenzaGAP} leads to a third scaling regime, connecting the typical regime in Eq.~\eqref{eq:typicalGAPBC} to the large deviation tail in Eq.~\eqref{eq:verylargedevBCGAP}. Altogether, the complete asymptotic structure of the gap distribution reads as in Eq. \eqref{eq:recapGAPBC_begin} and it is shown in Fig. \ref{fig:GAPBC_FULLL}.

This result has a simple probabilistic interpretation. For large $N$, most realizations belong to the typical sector, in which the first gap has size
\begin{equation}
d_1=\frac{L}{N}Z,
\end{equation}
with $Z=O(1)$ and probability density $S_B(z)$. This sector carries the dominant probability mass. However, rare realizations can produce gaps much larger than the typical value of order $O(1/N)$. The first rare sector is the intermediate regime, where the gap is of order $d_1=O\!\left(1/(\ln N)^2\right)\gg O\!\left(1/N\right)$. In this regime, introducing the scaling variable
\begin{equation}
z=\frac{(\ln N)^2}{L}g,
\end{equation}
the contribution to the density of $d_1$ is described by the scaling function $h_2(z)$. Finally, the far large deviation regime corresponds to even larger gaps, of order $d_1=O\!\left(L/\ln N\right)\gg O\!\left(L/(\ln N)^2\right)\gg O\!\left(L/N\right)$. In this sector, using the scaling variable
\begin{equation}
z=\frac{g\ln N}{L},
\end{equation}
the density is described by the scaling function $h_3(z)$ given in \eqref{eq:h3definit_begin}.

We also note here that, even though the typical regimes of the EVS and gap
statistics in the BC are described by the same scaling function $S_B(z)$, the
large deviations are rather different for the two observables (see
Eqs.~\eqref{eq:FULLEVSBCwithLD} and \eqref{eq:recapGAPBC_begin}). For example,
the large-deviation scaling function $S_3(z)$ of the EVS, given in
Eq.~\eqref{eq:S3EVSBC}, has bounded support, $z\in[0,1/2]$, and it is linear in $z$. By contrast, the large-deviation scaling function of the gap
statistics, $h_3(z)$, given in Eq.~\eqref{eq:h3definit_begin}, is supported on
$z\in[0,\infty)$ and decays exponentially fast as $z$ increases.

\section{EVS for a general confinement $V(x)=\kappa |x|^{\alpha}$}\label{sec:GeneralPotential}

In this final section, we present additional results, numerical simulations, and heuristic arguments aimed at extending the previous analysis of EVS in Sec.~\ref{sec:EVS} to a generic confining potential of the form $V(x)=\kappa |x|^{\alpha}$, with $\alpha\geq0$. This family of potentials smoothly interpolates between the cases studied above. Indeed, setting $\alpha=2$ recovers the HC case, with $\kappa=\mu/2$, while in the limit $\alpha\to\infty$ one recovers the BC case, with half-box size $L_B=1$. Moreover, setting $\alpha=0$ gives the unconfined resetting gas studied previously in Ref.~\cite{Biroli2023}.

Between two consecutive reset events, each particle evolves independently according to the overdamped Langevin dynamics given in Eq.~\eqref{eq:general_lang_eq}. As in the rest of the paper, all particles are simultaneously reset to the origin with rate $r$. For a general value of $\alpha$, we present arguments suggesting that the maximum $M_1$ scales as
\begin{equation}\label{eq:recapgeneralalphascaling}
M_1=
\begin{cases}
O(\sqrt{\ln N}), & 0\leq \alpha \leq 1,\\[4pt]
O((\ln N)^{1/\alpha}), & 1<\alpha<\infty,\\[4pt]
1-O(1/N), & \alpha=\infty.
\end{cases}
\end{equation}
This means that there are mainly two EVS universality classes: one for $0\leq \alpha\leq 1$, and one for $1<\alpha<\infty$. The BC limit, corresponding to $\alpha\to\infty$, represents a separate case.

The picture in Eq.~\eqref{eq:recapgeneralalphascaling} can be motivated by the large-distance behavior of the force associated with the potential. The potential $V(x)=\kappa |x|^{\alpha}$ gives rise to the force $-V'(x)=-\alpha \kappa \, x |x|^{\alpha-2}$ acting on each particle. Since we are interested in extreme values, the relevant regime is the large-$x$ behavior. 
It is clear that at $\alpha=1$ the behavior of the force $-V'(x)$ changes: for $\alpha<1$, the force decreases and vanishes as $x\to\infty$, whereas for $\alpha>1$ it keeps increasing indefinitely as the distance from the origin increases. We argue that this change of behavior is reflected in the EVS and results in \eqref{eq:recapgeneralalphascaling}.

We first analyze the class $0\leq \alpha \leq 1$. For $\alpha<1$, the
confining force becomes negligible far from the origin, precisely in the region explored by the extreme particles. Therefore, we expect the EVS to behave as in the unconfined case $\alpha=0$ studied in Ref.~\cite{Biroli2023},
namely
\begin{equation}\label{eq:resultEVSunconfined}
\operatorname{Prob}\{M_1=m\}=
\frac{1}{L_N}
S\left(\frac{m}{L_N}\right),
\qquad
L_N=\sqrt{\frac{4D}{r}\ln N},
\qquad
S(z)=2ze^{-z^2}.
\end{equation}
The marginal case $\alpha=1$, corresponding to a force of constant magnitude, must be treated separately, since the force remains finite at large distances. In this case, however, an explicit expression for the reset-free single-particle propagator is available (see Eq.~\eqref{eq:propag_alpha1_APP}). This allows us to perform an exact analytical calculation, reported in Appendix~\ref{sec:alphaequal1_APP}. We find that, also for $\alpha=1$, the maximum $M_1$ is distributed according to Eq.~\eqref{eq:resultEVSunconfined}. Therefore, the result is exact at the two endpoints $\alpha=0$ and $\alpha=1$: the former was established in Ref.~\cite{Biroli2023}, while the latter is derived here. Together with the physical argument given above, this supports the expectation that the same EVS universality class extends throughout the whole interval $0\leq \alpha \leq 1$.
\begin{figure}[htbp]
\centering

\begin{minipage}{0.42\textwidth}
    \centering
    \includegraphics[width=\linewidth]{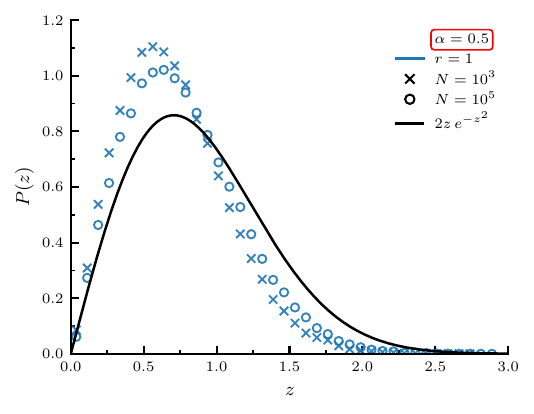}
\end{minipage}
\hspace{0.02\textwidth}
\begin{minipage}{0.42\textwidth}
    \centering
    \includegraphics[width=\linewidth]{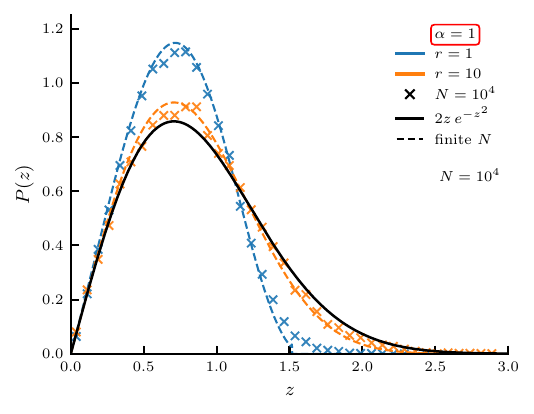}
\end{minipage}

\vspace{0.15cm}

\begin{minipage}{0.42\textwidth}
    \centering
    \includegraphics[width=\linewidth]{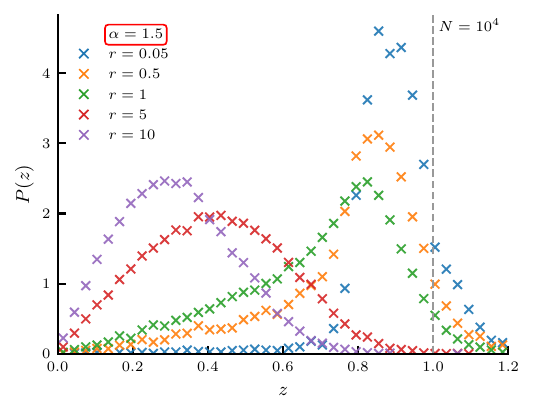}
\end{minipage}
\hspace{0.02\textwidth}
\begin{minipage}{0.42\textwidth}
    \centering
    \includegraphics[width=\linewidth]{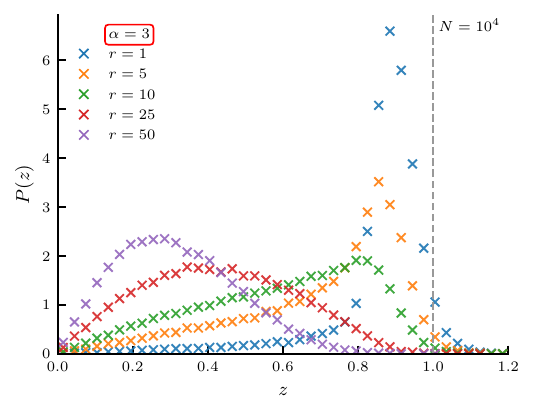}
\end{minipage}

\caption{Numerical simulations for the confining potential
$V(x)=\kappa |x|^{\alpha}$, with
$\alpha=0.5,1,1.5,3$. In each panel, $z$ denotes the rescaled maximum, with the rescaling chosen according to the corresponding regime summarized in Table~\ref{tab:EVS_general_alpha}. More explicitly, we use
$z=M_1/\sqrt{(4D/r)\ln N}$ for $\alpha=0.5$ and $\alpha=1$, while
$z=M_1/\left[(D/\kappa)\ln N\right]^{1/\alpha}$ for $\alpha=1.5$ and $\alpha=3$.
The vertical axis, denoted generically by $P(z)$, represents the probability density of the rescaled variable, estimated from the normalized histogram. The crosses correspond to numerical simulations performed with $\kappa=1$ and $D=1$.}
\label{fig:generalpotential}
\end{figure}

While the result in Eq.~\eqref{eq:resultEVSunconfined} was numerically verified for $\alpha=0$ in Ref.~\cite{Biroli2023}, here we present simulations for $\alpha=0.5$ and $\alpha=1$ in the top panels of Fig.~\ref{fig:generalpotential}. It is important to note that the convergence of the EVS to the asymptotic large-$N$ limit can be remarkably slow. As shown in Refs.~\cite{Gyorgyi2008,Gyorgyi2010,Bertin2010}, even for IID variables, this convergence is sometimes only logarithmic in $N$, and we expect a similarly slow approach in our system. The top-left panel of Fig.~\ref{fig:generalpotential} displays the simulation results for $\alpha=0.5$. Although there is a discrepancy with the asymptotic prediction depicted by the solid black curve, we attribute this to finite-$N$ effects, as the agreement visibly improves with increasing $N$. Further arguments supporting this interpretation are provided in Appendix~\ref{sec:potenzsqrt_APP}.

Similarly, for $\alpha=1$, the initial agreement appears imperfect (see the top-right panel of Fig.~\ref{fig:generalpotential}). However, because we can perform exact calculations in this case, we can demonstrate that this mismatch is entirely due to finite-$N$ effects. In Appendix~\ref{sec:alphaequal1_APP}, we formally show that for large $N$, the distribution of the maximum $M_1$ for $\alpha=1$ converges to Eq.~\eqref{eq:resultEVSunconfined}. Furthermore, we calculate a finite-$N$ corrected distribution, given by Eq.~\eqref{eq:finiteNalpha1dist_APP}, and plotted with dotted lines in the top-right panel of Fig.~\ref{fig:generalpotential}. The numerical simulations match this corrected expression remarkably well. This confirms that, at least for $\alpha=1$, the convergence is exceptionally slow, yet the asymptotic limit remains valid. We are confident that this slow convergence holds true for $\alpha=0.5$ as well, and more generally across the entire range $0\leq \alpha \leq 1$.

Let us now consider the second class, $1<\alpha<\infty$. As noticed in Sec. \ref{sec:correl_generalalpha} one can define a natural confinement length scale as
\begin{equation}\label{eq:conflengenericalpha_2}
L_{\alpha}=\left(\frac{D}{\kappa}\right)^{1/\alpha}, \qquad \alpha>1.
\end{equation}
For $\alpha=2$, the generic potential reads $V(x)=\kappa x^2$. To recover the HC used above,
$V(x)=\frac{1}{2}\mu x^2$, one has to identify $\kappa=\mu/2$. Therefore,
\[
L_{\alpha=2}=\left(\frac{2D}{\mu}\right)^{1/2}
=\sqrt{2}\,L_H,
\]
where $L_H=\sqrt{D/\mu}$ is the HC length introduced above.
In this regime, we do not expect the full distribution of $M_1$ to be identical for all values of $\alpha$. 
However, following an argument similar to the ``blocking argument'' of Ref.~\cite{MMS2022}, we expect the maximum $M_1$ to be of order
$(\ln N)^{1/\alpha}$. The idea is the following. Between two consecutive
resetting events, the particles evolve independently and tend to relax towards
the equilibrium stationary state given in Eq.~\eqref{eq:equilgeneralalpha}.
As discussed in Sec.~\ref{sec:observables}, the typical maximum of $N$
independent particles drawn from this equilibrium distribution is set by the
scale $a_N$, defined in Eq.~\eqref{eq:defaNbN}. Using the large-$x$ tail of
Eq.~\eqref{eq:equilgeneralalpha}, this scale is obtained from
\begin{equation}
    N \exp\left[-\frac{\kappa a_N^\alpha}{D}\right]\sim 1,
\end{equation}
which gives
\begin{equation}
    a_N \sim \left(\frac{D}{\kappa}\ln N\right)^{1/\alpha}.
\end{equation}
This suggests that the same logarithmic scale controls the maximum also in the resetting problem.
Therefore, we expect the rescaled maximum
\begin{equation}
Z=\frac{M_1}{L_{\alpha}(\ln N)^{1/\alpha}}=
\frac{M_1}{\left(\frac{D}{\kappa}\ln N\right)^{1/\alpha}}
\end{equation}
to display a behavior qualitatively similar to that of the HC case. 
In particular, we expect the probability density $P(z)$ of the rescaled random variable $Z$ to have bounded support, $z\in[0,1]$. Moreover, by analogy with the HC case, we expect a shape transition at the upper edge $z=1$ when the resetting rate $r$ is varied.
For $\alpha=2$, we have shown analytically in Sec. \ref{sec:EVS_HC} that this transition is controlled by the parameter $\nu_H=L_H/\ell_r$, which measures the ratio between the harmonic confinement length and the resetting length (see Fig. \ref{fig:EVS}). For a general potential with $\alpha>1$, the resetting length remains $\ell_r=\sqrt{D/r}$, while the confinement length is $L_{\alpha}$. The natural analogue of $\nu_H$ is therefore
\begin{equation}
\nu_{\alpha}=\frac{L_{\alpha}}{\ell_r}.
\end{equation}
Thus, varying $r$ at fixed potential strength $\kappa$ changes the ratio between the resetting length and the confinement length, and is expected to drive a shape transition analogous to the one found in HC.
This expectation is supported by the numerical simulations shown in the bottom panels of Fig.~\ref{fig:generalpotential} for $\alpha=1.5$ and $\alpha=3$. In both cases, the rescaled distribution is approximately supported on the interval $z\in[0,1]$ and depends strongly on the resetting rate $r$. As $r$ is decreased, the peak of the distribution moves towards the upper edge $z=1$. Once the peak approximately reaches the edge, it remains pinned there, while its height continues to increase as $r$ keeps decreasing. This is the same qualitative behavior observed for $\alpha=2$, where the shape transition is known analytically.

\begin{table*}[h]
\centering
\renewcommand{\arraystretch}{1.6}
\begin{tabular}{|lll|} 
\hline
\parbox[t]{0.16\textwidth}{\vspace{1mm}Regime\vspace{1mm}}
&
\parbox[t]{0.20\textwidth}{\vspace{1mm}Typical scale of $M_1$\vspace{1mm}}
&
\parbox[t]{0.56\textwidth}{\vspace{1mm}Distribution of the rescaled maximum\vspace{1mm}}
\\[6pt]
\hline

\parbox[t]{0.16\textwidth}{\vspace{2mm}$0\leq \alpha \leq 1$\vspace{2mm}}
&
\parbox[t]{0.20\textwidth}{\vspace{2mm}$M_1=O(\sqrt{\ln N})$\vspace{2mm}}
&
\parbox[t]{0.56\textwidth}{\vspace{2mm}
Unconfined class. With
$L_N=\sqrt{(4D/r)\ln N}$ and $Z=M_1/L_N$, one has
$\mathrm{Prob}\{M_1=m\}=L_N^{-1}S(m/L_N)$, with
$S(z)=2ze^{-z^2}$.
\vspace{2mm}}
\\[22pt]
\hline

\parbox[t]{0.16\textwidth}{\vspace{2mm}$1<\alpha<\infty$\vspace{2mm}}
&
\parbox[t]{0.20\textwidth}{\vspace{2mm}$M_1=O((\ln N)^{1/\alpha})$\vspace{2mm}}
&
\parbox[t]{0.56\textwidth}{\vspace{2mm}
Soft-confinement class. With
$L_{\alpha}=(D/\kappa)^{1/\alpha}$ and
$Z=M_1/[L_{\alpha}(\ln N)^{1/\alpha}]$, the distribution is expected to have support
$z\in[0,1]$ and a shape transition at $z=1$.
\vspace{2mm}}
\\[22pt]
\hline

\parbox[t]{0.16\textwidth}{\vspace{2mm}$\alpha\to\infty$\vspace{2mm}}
&
\parbox[t]{0.20\textwidth}{\vspace{2mm}$M_1=L/2-O(1/N)$\vspace{2mm}}
&
\parbox[t]{0.56\textwidth}{\vspace{2mm}
Hard-confinement class. With
$\delta=1/2-m/L$, one has
$\mathrm{Prob}\{M_1=m\}=(N/L)S_B(N\delta)$. The scaling function $S_B(z)$ has a very broad tail
$S_B(z)\sim 1/[z(\ln z)^2]$ for large $z$.
\vspace{2mm}}
\\[22pt]
\hline
\end{tabular}
\caption{Summary of the EVS universality classes for the confining potential $V(x)=\kappa |x|^{\alpha}$.}
\label{tab:EVS_general_alpha}
\end{table*}

At the values of $N$ accessible in our simulations, however, we do not observe either a true divergence of the peak as $z\to1^-$ or a strictly finite support. This does not contradict the presence of a shape transition. Indeed, even in the HC case $\alpha=2$, where the transition is analytically established, the divergence is not clearly visible at comparable system sizes and would require much larger values of $N$. We therefore interpret both the absence of a visible divergence and the residual weight outside the interval $z\in[0,1]$ for $\alpha=1.5$ and $\alpha=3$ as finite-$N$ effects. Since the histograms for $\alpha>1$ display the same qualitative behavior as the analytically understood $\alpha=2$ case, we expect the shape transition to persist throughout the whole regime $1<\alpha<\infty$.

Finally, the case $\alpha\to\infty$ is different from all finite values of $\alpha$ because it is the only hard-confinement case, namely the only case with a strictly bounded accessible region. In contrast, every finite $\alpha$ corresponds to soft confinement: particles remain confined, but can still, in principle, explore arbitrarily large distances.
In Table \ref{tab:EVS_general_alpha} we recap the results for all the three classes.

\section{Conclusions}\label{sec:conclusions}

In this work, we have studied a gas of non-interacting Brownian particles confined by an external potential and subject to simultaneous stochastic resetting. Although the particles do not interact directly, the common resetting events generate DEC through the shared random time elapsed since the last reset. By exploiting the resulting CIID structure, we derived exact results for the stationary joint probability distribution, the density profile, several correlation measures, and the statistics of the rightmost particle and of the first gap. We focused mainly on two analytically tractable confinements, namely the harmonic confinement (HC) and the box confinement (BC), and showed that the stationary state is controlled by the competition between the confinement length scale and the resetting length scale. In both cases, the system interpolates between the equilibrium stationary state imposed by confinement and the NESS of the unconfined resetting gas. At the level of the density and of the unnormalized two-particle correlator, the two geometries display a similar qualitative behavior.
However, the normalized correlation coefficient reveals a sharper distinction. In the BC, $A_B(\nu_B)$ develops a non-monotonic overshoot above the unconfined value. More generally, for the family of potentials $V(x)=\kappa |x|^\alpha$, a similar behavior is observed for $\alpha>\alpha_c=1+\sqrt{5}=3.236\dots$. In this regime, the confining force suppresses rare trajectories that wander far from the origin more efficiently than typical trajectories in the bulk. Since these rare excursions mainly contribute to the non-collective part of the fluctuations, their suppression increases the normalized collective correlations and produces an overshoot. By contrast, for $0<\alpha<\alpha_c$, including the HC case $\alpha=2$, this selective suppression is not strong enough, and the normalized correlation coefficient increases monotonically towards the unconfined value.

We then showed that these differences are strongly reflected in edge observables. In the HC, the maximum scales as $M_1 = O(\sqrt{\ln N})$ and its rescaled distribution has a bounded support with a shape transition at a critical value of the ratio between the confinement and the resetting length scales. In the BC, instead, the maximum is typically located at a distance $O(1/N)$ from the boundary, as it would be in equilibrium, but the resetting induced correlations completely change the shape of the fluctuations, producing a broad tail with a divergent first moment. A similar distinction appears in the first gap: in the HC the typical scale is $O(1/\sqrt{\ln N})$ and the distribution has a tail which decays exponentially, whereas in the box the typical scale is $O(1/N)$ but anomalously large gaps have a much larger probability: the tail decays as a power law with logarithmic corrections. 

Finally, by considering the family of potentials $V(x) = \kappa |x|^\alpha$, we provided evidence for the existence of different universality classes in the EVS. For $0\leq\alpha \leq 1$, they appear to coincide with the EVS of the unconfined resetting gas ($\alpha=0$), while for $1<\alpha <\infty$ they share the same qualitative structure as in the HC ($\alpha=2$), such as a bounded support and a shape transition at the right boundary. The BC case, which is recovered in the limit $\alpha\to\infty$, constitutes a third distinct universality class.

Several perspectives remain open. A natural first direction would be to study other observables beyond the maximum and the first gap. For instance, one could consider full counting statistics, namely the distribution of the number of particles contained in a given spatial interval. Another interesting direction would be to investigate order statistics and gap statistics in the bulk of the gas. Order statistics generalize EVS by considering not only the rightmost particle, but the full sequence of ranked particle positions, and in particular the $k$-th rightmost particle. Similarly, bulk gap statistics describe the spacings between consecutive particles that are not necessarily close to the edge. These observables would make it possible to understand whether confinement mainly affects edge fluctuations, as emphasized in this work, or whether it also modifies the internal structure of the gas in a similar way.

In this work, we have focused on the gas in its NESS. It would be interesting to study how the observables considered here, and in particular the two-point correlations, evolve as a function of time.
Another promising perspective would be to explore possible experimental realizations of the confined resetting gas studied here. Stochastic resetting has already been implemented experimentally for single Brownian particles in colloidal systems suspended in water \cite{Besga2020,Tal-Friedman2020,GinotBechinger2025}, and simultaneous resetting of many Brownian particles has also recently been realized in colloidal experiments \cite{Faisant2021,Vatash2025,biroli2025exp}. Since harmonic traps generated by lasers are commonly used in these setups, the HC geometry studied here could provide a natural starting point for testing our predictions. An experimental realization of the BC geometry would also be particularly interesting, as it would allow one to directly probe the predicted differences between hard and soft confinement.

\begin{acknowledgments}
We thank A. K. Hartmann, L. Tonetti and T. Welker for useful discussions. 
We acknowledge support from ANR Grant No. ANR-23-CE30-0020-01 EDIPS.
Part of this work was completed while S.~N.~M. and G.~S. were visiting the Kavli Institute for Theoretical Physics (KITP). This research was supported in part by grant NSF PHY-2309135 to the KITP.
\end{acknowledgments}

\begin{appendix}

\section{Critical value $\alpha_c$ for non-monotonic correlations}\label{sec:criticalalpha_APP}

In this appendix, we derive the critical value $\alpha_c$ separating the monotonic and non-monotonic regimes of the normalized correlation coefficient $A_\alpha(\nu_\alpha)$ defined in Eq.~\eqref{eq:defAalpha}. The idea is to analyze the large-$\nu_\alpha$ limit, or equivalently the large-$r$ limit, and determine how $A_\alpha(\nu_\alpha)$ approaches the universal unconfined value $A_\alpha(\nu_\alpha=\infty)=1/5$. If the approach is from above, the leading correction to $1/5$ is positive and an overshoot is present. If the approach is from below, instead, this correction is negative. Therefore, the sign of the leading correction determines whether the overshoot occurs and allows us to identify the critical value $\alpha_c$.

Between two reset events, a single particle evolves according to Eq. \eqref{eq:general_lang_eq}, i.e.
\begin{equation}\label{eq:langev_alpha}
dx(t)=-\kappa\alpha\,x(t)|x(t)|^{\alpha-2}\,dt+\sqrt{2D}\,dW_t,
\end{equation}
where $W_t$ is a standard Brownian motion. We denote by
\begin{equation}
m_q(\tau)=\int_{-\infty}^{+\infty} dx\,x^q\,p_0(x,\tau)
\end{equation}
the $q$-th moment of the reset-free single-particle propagator at time $\tau$, starting from the origin. Using the CIID structure of the stationary state in Eq. \eqref{eq:NESS_CIIDform}, one has, for $i\neq j$,
\begin{align}
\langle x_i^2\rangle&=r\int_0^\infty d\tau\,\,e^{-r\tau}\,m_2(\tau),\\
\langle x_i^4\rangle&=r\int_0^\infty d\tau\,\,e^{-r\tau}\,m_4(\tau),\\
\langle x_i^2x_j^2\rangle&=r\int_0^\infty d\tau\,\,e^{-r\tau}\,m_2(\tau)^2.
\end{align}
Therefore, the normalized correlator defined in Eq.~\eqref{eq:defAalpha} can be rewritten as
\begin{equation}\label{eq:Anurewritten}
A_\alpha(\nu_{\alpha})
=
\frac{
r\int_0^\infty d\tau\,\,e^{-r\tau}\,m_2(\tau)^2
-
\left[
r\int_0^\infty d\tau\,e^{-r\tau}\,m_2(\tau)
\right]^2
}{
r\int_0^\infty d\tau\,e^{-r\tau}\,m_4(\tau)
-
\left[
r\int_0^\infty d\tau\,e^{-r\tau}\,m_2(\tau)
\right]^2
}.
\end{equation}
Since we are considering the large-$r$ limit, the integrals in Eq.~\eqref{eq:Anurewritten} are dominated by values of $\tau$ of order $\tau=O(1/r)$. We therefore need the small time expansions of $m_2(\tau)$ and $m_4(\tau)$.

We first derive the small time behavior of $m_2(t)$. Starting from Eq.~\eqref{eq:langev_alpha} and applying It\^o's formula to $x^2(t)$, we obtain
\begin{equation}
d\!\left[x^2(t)\right]
=
\left[2D-2\kappa\alpha\,|x(t)|^\alpha\right]dt
+
2\sqrt{2D}\,x(t)\,dW_t.
\end{equation}
Taking the average, the stochastic term drops out and we get
\begin{equation}\label{eq:diffeqform2}
\frac{d}{dt}m_2(t)=2D-2\kappa\alpha\,\left\langle |x(t)|^\alpha\right\rangle.
\end{equation}
The same procedure applied to $x^4(t)$ gives
\begin{equation}
d\!\left[x^4(t)\right]
=
\left[12D\,x^2(t)-4\kappa\alpha\,|x(t)|^{\alpha+2}\right]dt
+
4\sqrt{2D}\,x^3(t)\,dW_t.
\end{equation}
Averaging again, we obtain
\begin{equation}\label{eq:diffeqform4}
\frac{d}{dt}m_4(t)=12D\,m_2(t)-4\kappa\alpha\,\left\langle |x(t)|^{\alpha+2}\right\rangle.
\end{equation}

At very short times, the leading dynamics is controlled by diffusion. This can be seen directly from the integral form of Eq.~\eqref{eq:langev_alpha}, namely
\begin{equation}
x(t)=\sqrt{2D}\,W_t-\kappa\alpha\int_0^t x(s)|x(s)|^{\alpha-2}\,ds.
\end{equation}
The Brownian contribution is of order $t^{1/2}$. Estimating the drift term on this leading Brownian scale, $x(s)=O(s^{1/2})$, gives
\begin{equation}
\int_0^t x(s)|x(s)|^{\alpha-2}\,ds=O\!\left(t^{(\alpha+1)/2}\right).
\end{equation}
Hence, the ratio between the drift contribution and the diffusive contribution is of order $t^{\alpha/2}$, which vanishes as $t\to0$ for every $\alpha>0$. We can therefore introduce a Gaussian random variable $Z$ with zero mean and unit variance, and we can write, for small times
\begin{equation}
x(t)\approx \sqrt{2Dt}\,Z,
\end{equation}
so that
\begin{equation}\label{eq:smalltime_abs_moment}
\left\langle |x(t)|^q\right\rangle\approx (2Dt)^{q/2}g_q,
\qquad
g_q=\langle |Z|^q\rangle.
\end{equation}

Substituting Eq.~\eqref{eq:smalltime_abs_moment} into Eq.~\eqref{eq:diffeqform2} and integrating in time, we find
\begin{equation}\label{eq:m2smalltime_alpha}
m_2(t)=2Dt-b_2\,t^{1+\alpha/2}+\cdots,
\end{equation}
where
\begin{equation}\label{eq:b2def_alpha}
b_2=\frac{2\kappa\alpha\,(2D)^{\alpha/2}g_\alpha}{1+\frac{\alpha}{2}}.
\end{equation}
Similarly, substituting Eq.~\eqref{eq:m2smalltime_alpha} and the estimate
\begin{equation}
\left\langle |x(t)|^{\alpha+2}\right\rangle\approx (2Dt)^{1+\alpha/2}g_{\alpha+2}
\end{equation}
into Eq.~\eqref{eq:diffeqform4}, and integrating in time, gives
\begin{equation}\label{eq:m4smalltime_alpha}
m_4(t)=12D^2t^2-b_4\,t^{2+\alpha/2}+\cdots,
\end{equation}
with
\begin{equation}\label{eq:b4def_alpha_first}
b_4=
\frac{12D\,b_2+4\kappa\alpha\,(2D)^{1+\alpha/2}g_{\alpha+2}}{2+\frac{\alpha}{2}}.
\end{equation}
Using the Gaussian identity $g_{\alpha+2}=(\alpha+1)g_\alpha,$
the coefficient $b_4$ can be rewritten in terms of $b_2$ as
\begin{equation}\label{eq:b4def_alpha}
b_4=
\frac{
4D\,b_2
\left[
3+(\alpha+1)\left(1+\frac{\alpha}{2}\right)
\right]
}{
2+\frac{\alpha}{2}
}.
\end{equation}

We can now insert the small time expansions \eqref{eq:m2smalltime_alpha} and \eqref{eq:m4smalltime_alpha} into the resetting averages. Using
\begin{equation}
\int_0^\infty d\tau\,r\,e^{-r\tau}\,\tau^p=\frac{\Gamma(p+1)}{r^p},
\end{equation}
we obtain
\begin{align}
\int_0^\infty d\tau\,r\,e^{-r\tau}\,m_2(\tau)
&=
\frac{2D}{r}
-
\frac{b_2\Gamma\!\left(2+\frac{\alpha}{2}\right)}
{r^{1+\alpha/2}}
+
\cdots,
\\
\int_0^\infty d\tau\,r\,e^{-r\tau}\,m_4(\tau)
&=
\frac{24D^2}{r^2}
-
\frac{b_4\Gamma\!\left(3+\frac{\alpha}{2}\right)}
{r^{2+\alpha/2}}
+
\cdots,
\\
\int_0^\infty d\tau\,r\,e^{-r\tau}\,m_2(\tau)^2
&=
\frac{8D^2}{r^2}
-
\frac{4D\,b_2\Gamma\!\left(3+\frac{\alpha}{2}\right)}
{r^{2+\alpha/2}}
+
\cdots.
\end{align}
Moreover,
\begin{equation}
\left[
\int_0^\infty d\tau\,r\,e^{-r\tau}\,m_2(\tau)
\right]^2
=
\frac{4D^2}{r^2}
-
\frac{4D\,b_2\Gamma\!\left(2+\frac{\alpha}{2}\right)}
{r^{2+\alpha/2}}
+
\cdots.
\end{equation}
It follows that the numerator and denominator of $A_\alpha$ behave as
\begin{align}
\int_0^\infty d\tau\,r\,e^{-r\tau}\,m_2(\tau)^2
-
\left[
\int_0^\infty d\tau\,r\,e^{-r\tau}\,m_2(\tau)
\right]^2
&=
\frac{4D^2}{r^2}
+
\frac{B_\alpha}{r^{2+\alpha/2}}
+
\cdots,
\\
\int_0^\infty d\tau\,r\,e^{-r\tau}\,m_4(\tau)
-
\left[
\int_0^\infty d\tau\,r\,e^{-r\tau}\,m_2(\tau)
\right]^2
&=
\frac{20D^2}{r^2}
+
\frac{C_\alpha}{r^{2+\alpha/2}}
+
\cdots,
\end{align}
where
\begin{align}
B_\alpha
&=
4D\,b_2\Gamma\!\left(2+\frac{\alpha}{2}\right)
-
4D\,b_2\Gamma\!\left(3+\frac{\alpha}{2}\right),
\\
C_\alpha
&=
4D\,b_2\Gamma\!\left(2+\frac{\alpha}{2}\right)
-
b_4\Gamma\!\left(3+\frac{\alpha}{2}\right).
\end{align}
Using
\begin{equation}
\Gamma\!\left(3+\frac{\alpha}{2}\right)
=
\left(2+\frac{\alpha}{2}\right)
\Gamma\!\left(2+\frac{\alpha}{2}\right),
\end{equation}
together with Eq.~\eqref{eq:b4def_alpha}, we obtain
\begin{align}
B_\alpha
&=
-4D\,b_2\left(1+\frac{\alpha}{2}\right)
\Gamma\!\left(2+\frac{\alpha}{2}\right),
\\
C_\alpha
&=
-4D\,b_2
\Gamma\!\left(2+\frac{\alpha}{2}\right)
\left[
2+(\alpha+1)\left(1+\frac{\alpha}{2}\right)
\right].
\end{align}

We can now expand the ratio in Eq.~\eqref{eq:Anurewritten}. This gives
\begin{equation}
A_\alpha
=
\frac{
\frac{4D^2}{r^2}
+
\frac{B_\alpha}{r^{2+\alpha/2}}
+
\cdots
}{
\frac{20D^2}{r^2}
+
\frac{C_\alpha}{r^{2+\alpha/2}}
+
\cdots
},
\end{equation}
and therefore
\begin{equation}\label{eq:Aalpha_large_r_first}
A_\alpha
=
\frac{1}{5}
+
\frac{5B_\alpha-C_\alpha}{100D^2}\,r^{-\alpha/2}
+
\cdots.
\end{equation}
The prefactor of the first correction is
\begin{equation}
5B_\alpha-C_\alpha
=
4D\,b_2\Gamma\!\left(2+\frac{\alpha}{2}\right)
\left[
2+\left(1+\frac{\alpha}{2}\right)(\alpha-4)
\right].
\end{equation}
Since $D>0$, $b_2>0$, and $\Gamma\!\left(2+\frac{\alpha}{2}\right)>0$, the sign of the first correction is controlled only by
\begin{equation}
2+\left(1+\frac{\alpha}{2}\right)(\alpha-4).
\end{equation}
The critical value is therefore determined by
\begin{equation}
2+\left(1+\frac{\alpha}{2}\right)(\alpha-4)=0.
\end{equation}
Equivalently,
\begin{equation}
\alpha^2-2\alpha-4=0.
\end{equation}
The positive solution gives
\begin{equation}
\alpha_c=1+\sqrt{5}=3.236\dots.
\end{equation}

Finally, using $\nu_\alpha=L_\alpha/\ell_r$, with $L_\alpha=(D/\kappa)^{1/\alpha}$ and $\ell_r=\sqrt{D/r}$, the correction $r^{-\alpha/2}$ can be rewritten as a positive constant times $\nu_\alpha^{-\alpha}$. Thus, in the large-$\nu_\alpha$ limit,
\begin{equation}
A_\alpha(\nu_\alpha)
=
\frac{1}{5}
+
K_\alpha
\left[
2+\left(1+\frac{\alpha}{2}\right)(\alpha-4)
\right]
\nu_\alpha^{-\alpha}
+
\cdots,
\end{equation}
where
\begin{equation}
K_\alpha
=
\frac{
2^{1+\frac{\alpha}{2}}\alpha\,g_\alpha\,
\Gamma\!\left(1+\frac{\alpha}{2}\right)
}{
25
}
>0.
\end{equation}
The positivity follows from $\alpha>0$, $g_\alpha>0$, and $\Gamma\!\left(1+\frac{\alpha}{2}\right)>0$. Therefore, $A_\alpha(\nu_\alpha)$ approaches the unconfined value $1/5$ from below for $0<\alpha<\alpha_c$, while it approaches it from above for $\alpha>\alpha_c$. Since $A_\alpha(\nu_\alpha)$ vanishes in the opposite limit $\nu_\alpha\to0$, the approach from above for $\alpha>\alpha_c$ implies the presence of an overshoot. This identifies
\begin{equation}
\alpha_c=1+\sqrt{5}
\end{equation}
as the critical value separating the monotonic and non-monotonic regimes.

\section{Diffusion and Correlations inside a $d$-dimensional hypersphere}
\label{sec:dDimensions_APP}

In this appendix, we derive the results of Sec.~\ref{sec:ddimensionalcorrel} for $d=2$ and $d=3$. We consider a single Brownian particle with diffusion constant $D$ confined inside the $d$-dimensional hypersphere
\begin{equation}
    \Omega=\{\mathbf{x}\in\mathbb{R}^d:\ |\mathbf{x}|\le L_B\},
    \qquad
    L_B=\frac{L}{2},
\end{equation}
with reflecting boundary conditions at $|\mathbf{x}|=L_B$. Since this appendix deals only with the hard wall geometry, we omit the superscript $(B)$ and denote the propagator by $p_0^{(d)}(\mathbf{x},t\mid \mathbf{x}_0)$. It satisfies
\begin{equation}
\partial_t p_0^{(d)}(\mathbf{x},t\mid \mathbf{x}_0)
=
D\nabla^2 p_0^{(d)}(\mathbf{x},t\mid \mathbf{x}_0),
\qquad
\mathbf{x}\in\Omega,
\label{eq:FP_d_dim_Ball}
\end{equation}
with initial condition
\begin{equation}
    p_0^{(d)}(\mathbf{x},0\mid \mathbf{x}_0)
    =
    \delta^{(d)}(\mathbf{x}-\mathbf{x}_0).
\end{equation}
Here $\delta^{(d)}$ denotes the $d$-dimensional Dirac delta, i.e.
\begin{equation}
\delta^{(d)}(\mathbf{x}-\mathbf{x}_0)
=
\prod_{i=1}^d \delta(x_i-x_{0,i}),
\end{equation}
where $\delta$ is the one-dimensional Dirac delta.
The reflecting boundary condition corresponds to the vanishing of the normal probability current at the boundary,
\begin{equation}
    \mathbf{J}(\mathbf{x},t)
    =
    -D\nabla p_0^{(d)}(\mathbf{x},t\mid \mathbf{x}_0),
    \qquad
    \mathbf{n}\cdot \mathbf{J}\big|_{|\mathbf{x}|=L_B}=0,
\end{equation}
where $\mathbf{n}$ is the outward unit normal vector. We will only need the propagator for a particle starting from the origin. Since this initial condition is spherically symmetric, the propagator depends only on the radial coordinate $x=|\mathbf{x}|$. We therefore write
\begin{equation}
    p_0^{(d)}(\mathbf{x},t\mid \mathbf{0})
    \equiv
    p_0^{(d)}(x,t).
\end{equation}
In radial coordinates, the reflecting boundary condition reduces to
\begin{equation}
\partial_x p_0^{(d)}(x,t)\Big|_{x=L_B}=0.
\label{eq:Neumann_radial}
\end{equation}
For a radial function, the Laplacian reads
\begin{equation}
\nabla^2 p_0^{(d)}(x,t)
=
\partial_x^2 p_0^{(d)}(x,t)
+
\frac{d-1}{x}\partial_x p_0^{(d)}(x,t).
\end{equation}
Therefore Eq.~\eqref{eq:FP_d_dim_Ball} becomes
\begin{equation}
\partial_t p_0^{(d)}(x,t)
=
D\left[
\partial_x^2 p_0^{(d)}(x,t)
+
\frac{d-1}{x}\partial_x p_0^{(d)}(x,t)
\right],
\qquad
0<x<L_B,
\label{eq:FP_radial}
\end{equation}
supplemented with Eq.~\eqref{eq:Neumann_radial}.
We now solve the radial diffusion problem by separation of variables, using the spectral representation of the Neumann Laplacian in a ball \cite{Bickel2007,Grebenkov2021}. Setting
\begin{equation}
    p_0^{(d)}(x,t)=\Phi(x)T(t),
\end{equation}
one obtains
\begin{equation}
    \frac{T'(t)}{D T(t)}
=
\frac{\Phi''(x)}{\Phi(x)}
+
\frac{d-1}{x}\frac{\Phi'(x)}{\Phi(x)}
\equiv -k^2.
\end{equation}
Thus $T(t)=e^{-Dk^2t}$, while $\Phi$ solves
\begin{equation}
\Phi''(x)
+
\frac{d-1}{x}\Phi'(x)
+
k^2\Phi(x)=0.
\label{eq:radial_ODE}
\end{equation}
Introducing $\eta=\frac{d}{2}-1$ and writing $\Phi(x)=x^{-\eta}u(x)$, Eq.~\eqref{eq:radial_ODE} becomes Bessel's equation,
\begin{equation}
    u''(x)+\frac{1}{x}u'(x)+\left(k^2-\frac{\eta^2}{x^2}\right)u(x)=0.
\end{equation}
The solution regular at the origin is proportional to $J_\eta(kx)$, where $J_\eta$ is the Bessel function of the first kind. Hence the radial eigenfunctions can be written as
\begin{equation}
\Phi_k(x)
=
x^{-\eta}J_\eta(kx).
\label{eq:eigenfunctions}
\end{equation}
The reflecting boundary condition requires $\Phi_k'(L_B)=0$. Using the identity \cite{NIST2010}
\begin{equation}
\frac{d}{dx}
\left[
x^{-\eta}J_\eta(kx)
\right]
=
-kx^{-\eta}J_{\eta+1}(kx),
\label{eq:derivative_identity}
\end{equation}
one obtains
\begin{equation}
    J_{\eta+1}(kL_B)=0.
\end{equation}
Let $j_{\eta+1,n}$ denote the $n$-th positive zero of $J_{\eta+1}(x)$. The nonzero eigenvalues are therefore
\begin{equation}
    k_n=\frac{j_{\eta+1,n}}{L_B}
=
\frac{2j_{\eta+1,n}}{L},
\qquad
n\ge 1.
\end{equation}
In addition, there is a zero mode $k_0=0$, corresponding to the stationary uniform distribution.
The propagator can now be expanded over these radial eigenfunctions as
\begin{equation}\label{eq:propagator_spectral_general_APP}
p_0^{(d)}(x,t)
=
\frac{1}{V_d(L_B)}
+
\sum_{n\ge1}
\omega_n^{(d)}
x^{-\eta}
J_\eta(k_nx)
e^{-Dk_n^2t},
\qquad
0\le x\le L_B,
\end{equation}
where
\begin{equation}
V_d(L_B)
=
\frac{\pi^{d/2}}{\Gamma\!\left(\frac{d}{2}+1\right)}L_B^d
\label{eq:volume_Ball}
\end{equation}
is the volume of the $d$-dimensional hypersphere. The spectral weights
$\omega_n^{(d)}$ are fixed by the initial condition. In the following, we only need their explicit forms for $d=2$ and $d=3$.

This spectral representation allows us to compute the reset-free radial moments at fixed time $\tau$. These moments are the only ingredients needed to evaluate the correlation observables introduced in Sec.~\ref{sec:ddimensionalcorrel}. Indeed, using the CIID structure of the stationary state and denoting by $\langle\cdot\rangle_0^{(d)}$ averages over the reset-free propagator at fixed $\tau$ (given in Eq. \eqref{eq:propagator_spectral_general_APP}), one can rewrite them as
\begin{equation}
\mathcal{C}_1^{(d)}=
r\int_0^\infty d\tau\,e^{-r\tau}
\langle x^4(\tau)\rangle^{(d)}_0-
\left[
r\int_0^\infty d\tau\,e^{-r\tau}
\langle x^2(\tau)\rangle^{(d)}_0
\right]^2,
\label{eq:C1stat_dim_d}
\end{equation}
and
\begin{equation}
\mathcal{C}_2^{(d)}
=
r\int_0^\infty d\tau\,e^{-r\tau}
\left[
\langle x^2(\tau)\rangle^{(d)}_0
\right]^2
-
\left[
r\int_0^\infty d\tau\,e^{-r\tau}
\langle x^2(\tau)\rangle^{(d)}_0
\right]^2.
\label{eq:C2stat_dim_d}
\end{equation}

\subsection{Dimension $d=2$}

In two dimensions, one has
\begin{equation}
\eta=0,\qquad L_B=\frac{L}{2},\qquad V_2(L_B)=\pi L_B^2=\frac{\pi L^2}{4}.
\end{equation}
The radial eigenfunctions are therefore
\begin{equation}
\Phi_n(x)=J_0(k_n x),
\end{equation}
and the reflecting boundary condition selects the positive zeros of $J_1$. More explicitly,
\begin{equation}
J_1(j_{1,n})=0,\qquad k_n=\frac{j_{1,n}}{L_B}=\frac{2j_{1,n}}{L},\qquad n\ge 1.
\end{equation}
We now determine the coefficients of the spectral expansion. We write
\begin{equation}\label{eq:propagdeq2provv_APP}
p_0^{(2)}(x,t)=\frac{1}{\pi L_B^2}+\sum_{n\ge1}\omega_n^{(2)}J_0(k_nx)e^{-Dk_n^2t}.
\end{equation}
The coefficients $\omega_n^{(2)}$ are determined by the initial condition at the origin. To this end, we project $p_0^{(2)}(\mathbf{x},0\mid \mathbf{0})=\delta^{(2)}(\mathbf{x})$
onto the radial mode $J_0(k_mx)$. This gives
\begin{equation}\label{eq:coefficientprovspdim2_APP}
\int_{\Omega} d^2\mathbf{x}\,p_0^{(2)}(\mathbf{x},0\mid \mathbf{0})\,J_0(k_m x)=\int_{\Omega} d^2\mathbf{x}\,\delta^{(2)}(\mathbf{x})\,J_0(k_m x)=J_0(0)=1.
\end{equation}
Inserting \eqref{eq:propagdeq2provv_APP} into \eqref{eq:coefficientprovspdim2_APP} we obtain
\begin{equation}
1
=
\int_{\Omega} d^2\mathbf{x}\,
\left[
\frac{1}{\pi L_B^2}
+
\sum_{n\ge1}
\omega_n^{(2)}
J_0(k_nx)
\right]
J_0(k_mx).
\end{equation}
The contribution of the constant mode vanishes. Indeed,
\begin{equation}
\int_{\Omega} d^2\mathbf{x}\,J_0(k_mx)=2\pi\int_0^{L_B} x\,dx\,J_0(k_mx)=2\pi\frac{L_B}{k_m}J_1(k_mL_B)=0,
\end{equation}
because $k_mL_B=j_{1,m}$ and $J_1(j_{1,m})=0$. Therefore,
\begin{equation}
1=\sum_{n\ge1}
\omega_n^{(2)}
\int_{\Omega} d^2\mathbf{x}\,
J_0(k_nx)J_0(k_mx).
\end{equation}
We now use the orthogonality of the modes, i.e.
\begin{equation}
\int_{\Omega} d^2\mathbf{x}\,
J_0(k_nx)J_0(k_mx)
=
2\pi\int_0^{L_B} x\,dx\,
J_0(k_nx)J_0(k_mx)
=
\pi L_B^2J_0(j_{1,n})^2\delta_{nm}.
\label{eq:J0orthogonalityd2_APP}
\end{equation}
Thus only the term $n=m$ survives, and we get
\begin{equation}
1
=
\omega_m^{(2)}
\pi L_B^2J_0(j_{1,m})^2.
\end{equation}
Renaming $m$ into $n$, and using $L_B=L/2$, this gives
\begin{equation}
\omega_n^{(2)}=
\frac{1}{\pi L_B^2J_0(j_{1,n})^2}=
\frac{4}{\pi L^2J_0(j_{1,n})^2}.
\label{eq:omega_d2_APP}
\end{equation}
Substituting this result into Eq.~\eqref{eq:propagdeq2provv_APP}, we finally obtain
\begin{equation}
p_0^{(2)}(x,t)
=
\frac{1}{\pi L_B^2}
+
\sum_{n\ge1}
\frac{
J_0(k_nx)
}{
\pi L_B^2J_0(j_{1,n})^2
}
e^{-Dk_n^2t}.
\end{equation}
Equivalently, since $L_B=L/2$ and $k_n=2j_{1,n}/L$,
\begin{equation}
p_0^{(2)}(x,t)
=
\frac{4}{\pi L^2}
+
\frac{4}{\pi L^2}
\sum_{n\ge1}
\frac{
J_0\!\left(\frac{2j_{1,n}}{L}x\right)
}{
J_0(j_{1,n})^2
}
\exp\!\left[
-4D\frac{j_{1,n}^2}{L^2}t
\right],
\qquad
0\le x\le \frac{L}{2}.
\label{eq:propagator_d2}
\end{equation}
The radial moments are computed from
\begin{equation}
\langle x^q(\tau)\rangle^{(2)}_0=
\int_0^{L/2}2\pi x\,dx\,x^q\,p_0^{(2)}(x,\tau).
\end{equation}
Using the identities \cite{NIST2010}
\begin{align}
\int_0^{L/2}x^3J_0(k_nx)\,dx
&=
\frac{L^4}{8j_{1,n}^2}J_0(j_{1,n}),
\\[1ex]
\int_0^{L/2}x^5J_0(k_nx)\,dx
&=
\frac{L^6}{16j_{1,n}^4}
\left(j_{1,n}^2-8\right)J_0(j_{1,n}),
\end{align}
one obtains
\begin{align}
\langle x^2(\tau)\rangle^{(2)}_0
&=
\frac{L^2}{8}
+
L^2
\sum_{n\ge1}
\frac{1}{
j_{1,n}^2J_0(j_{1,n})
}
\exp\!\left(
-4D\frac{j_{1,n}^2}{L^2}\tau
\right),
\label{eq:moment2_d2}
\\[1ex]
\langle x^4(\tau)\rangle^{(2)}_0
&=
\frac{L^4}{48}
+
\frac{L^4}{2}
\sum_{n\ge1}
\frac{
j_{1,n}^2-8
}{
j_{1,n}^4J_0(j_{1,n})
}
\exp\!\left(
-4D\frac{j_{1,n}^2}{L^2}\tau
\right).
\label{eq:moment4_d2}
\end{align}
We recall the control parameter $\nu_B=\frac{L}{2}\sqrt{\frac{r}{D}}$, so that
\begin{equation}
    \int_0^\infty r\,d\tau\,e^{-r\tau}
\exp\!\left(
-4D\frac{j_{1,n}^2}{L^2}\tau
\right)
=
\frac{\nu_B^2}{\nu_B^2+j_{1,n}^2}.
\end{equation}
It is now useful to define
\begin{equation}
\beta_n^{(2)}(\nu_B)=
\frac{\nu_B^2}{\nu_B^2+j_{1,n}^2},
\qquad
b_n^{(2)}
=
\frac{4}{j_{1,n}^2J_0(j_{1,n})},
\qquad
c_n^{(2)}
=
\frac{8\left(j_{1,n}^2-8\right)}
{j_{1,n}^4J_0(j_{1,n})}.
\end{equation}
With these definitions we have
\begin{equation}
\langle x^2(\tau)\rangle^{(2)}_0=
\left(\frac{L}{2}\right)^2
\left[
\frac{1}{2}+
\sum_{n\ge1}
b_n^{(2)}
\exp\!\left(
-4D\frac{j_{1,n}^2}{L^2}\tau
\right)
\right],
\end{equation}
and
\begin{equation}
\langle x^4(\tau)\rangle^{(2)}_0
=
\left(\frac{L}{2}\right)^4
\left[
\frac{1}{3}
+
\sum_{n\ge1}
c_n^{(2)}
\exp\!\left(
-4D\frac{j_{1,n}^2}{L^2}\tau
\right)
\right].
\end{equation}
We now introduce
\begin{equation}
S_1^{(2)}(\nu_B)
=
\sum_{n\ge1}
b_n^{(2)}
\beta_n^{(2)}(\nu_B),
\qquad
S_2^{(2)}(\nu_B)
=
\sum_{n\ge1}
c_n^{(2)}
\beta_n^{(2)}(\nu_B).
\end{equation}
Substituting the moments into Eqs.~\eqref{eq:C1stat_dim_d} and \eqref{eq:C2stat_dim_d}, one obtains
\begin{equation}
\mathcal C_1^{(2)}(\nu_B)
=
\left(\frac{L}{2}\right)^4
\mathcal F_1^{(2)}(\nu_B),
\label{eq:C1_d2_scaled}
\end{equation}
with
\begin{equation}
\mathcal F_1^{(2)}(\nu_B)
=
\frac{1}{12}
+
S_2^{(2)}(\nu_B)
-
S_1^{(2)}(\nu_B)
-
\left[S_1^{(2)}(\nu_B)\right]^2.
\label{eq:F1_d2}
\end{equation}
Similarly,
\begin{equation}
\mathcal C_2^{(2)}(\nu_B)
=
\left(\frac{L}{2}\right)^4
\mathcal F_2^{(2)}(\nu_B),
\label{eq:C2_d2_scaled}
\end{equation}
where
\begin{equation}
\mathcal F_2^{(2)}(\nu_B)
=
\sum_{n,m\ge1}
b_n^{(2)}b_m^{(2)}
\frac{
\nu_B^2j_{1,n}^2j_{1,m}^2
}{
\left(\nu_B^2+j_{1,n}^2+j_{1,m}^2\right)
\left(\nu_B^2+j_{1,n}^2\right)
\left(\nu_B^2+j_{1,m}^2\right)
}.
\label{eq:F2_d2}
\end{equation}
Finally, the normalized correlation coefficient in two dimensions is
\begin{equation}
A_B^{(2)}(\nu_B)
=
\frac{\mathcal F_2^{(2)}(\nu_B)}
{\mathcal F_1^{(2)}(\nu_B)}.
\label{eq:a_nu_dimension2_APP}
\end{equation}
This function is shown in red in Fig.~\ref{fig:aind1d2d3}.

\subsection{Dimension $d=3$}

In three dimensions, one has
\begin{equation}
\eta=\frac{1}{2},\qquad L_B=\frac{L}{2},\qquad V_3(L_B)=\frac{4\pi L_B^3}{3}=\frac{\pi L^3}{6}.
\end{equation}
The regular radial eigenfunctions are proportional to $x^{-1/2}J_{1/2}(k_nx)$. Using the identity $J_{1/2}(x)=\sqrt{\frac{2}{\pi x}}\sin x$ \cite{NIST2010},
we can equivalently write eigenfunctions as
\begin{equation}
\Phi_n(x)=\frac{\sin(k_nx)}{x}.
\end{equation}
The reflecting boundary condition selects the positive zeros of $J_{3/2}$. Equivalently, defining
\begin{equation}
\alpha_n:=j_{3/2,n},\qquad k_n=\frac{\alpha_n}{L_B}=\frac{2\alpha_n}{L},
\end{equation}
the numbers $\alpha_n$ satisfy $J_{3/2}(\alpha_n)=0$, which is equivalent to the condition $\tan\alpha_n=\alpha_n$.
We now determine the coefficients of the spectral expansion. We write
\begin{equation}\label{eq:propagdeq3provv_APP}
p_0^{(3)}(x,t)=\frac{1}{V_3(L_B)}+\sum_{n\ge1}\omega_n^{(3)}\frac{\sin(k_nx)}{x}e^{-Dk_n^2t}.
\end{equation}
The coefficients $\omega_n^{(3)}$ are fixed by the initial condition at the origin. To determine them, we project the initial condition $p_0^{(3)}(\mathbf{x},0\mid \mathbf{0})=\delta^{(3)}(\mathbf{x})$ onto the mode $\sin(k_mx)/x$. This gives
\begin{equation}\label{eq:coefficientprovspdim3_APP}
\int_{\Omega} d^3\mathbf{x}\,p_0^{(3)}(\mathbf{x},0\mid \mathbf{0})\,\frac{\sin(k_mx)}{x}=\int_{\Omega} d^3\mathbf{x}\,\delta^{(3)}(\mathbf{x})\,\frac{\sin(k_mx)}{x}=k_m.
\end{equation}
Inserting Eq.~\eqref{eq:propagdeq3provv_APP} into Eq.~\eqref{eq:coefficientprovspdim3_APP}, we obtain
\begin{equation}
k_m=\int_{\Omega}d^3\mathbf{x}\left[\frac{1}{V_3(L_B)}+\sum_{n\ge1}\omega_n^{(3)}\frac{\sin(k_nx)}{x}\right]\frac{\sin(k_mx)}{x}.
\end{equation}
The contribution of the constant mode vanishes. Indeed,
\begin{equation}
\int_{\Omega}d^3\mathbf{x}\,\frac{\sin(k_mx)}{x}=4\pi\int_0^{L_B} x\sin(k_mx)\,dx=4\pi\left[-\frac{L_B\cos(k_mL_B)}{k_m}+\frac{\sin(k_mL_B)}{k_m^2}\right]=0,
\end{equation}
where in the last equality we used $k_mL_B=\alpha_m$ and $\tan\alpha_m=\alpha_m$. Therefore,
\begin{equation}
k_m=\sum_{n\ge1}\omega_n^{(3)}\int_{\Omega}d^3\mathbf{x}\,\frac{\sin(k_nx)}{x}\frac{\sin(k_mx)}{x}.
\end{equation}
We now use the orthogonality of the modes, which gives
\begin{equation}
\int_{\Omega}d^3\mathbf{x}\,\frac{\sin(k_nx)}{x}\frac{\sin(k_mx)}{x}=4\pi\int_0^{L_B} dx\,\sin(k_nx)\sin(k_mx)=2\pi L_B\sin^2\alpha_n\,\delta_{nm}.
\label{eq:sin_orthogonality_d3_APP}
\end{equation}
Thus only the term $n=m$ survives, and we get
\begin{equation}
k_m=\omega_m^{(3)}\,2\pi L_B\sin^2\alpha_m.
\end{equation}
Hence, using $k_m=\alpha_m/L_B$,
\begin{equation}
\omega_m^{(3)}=\frac{k_m}{2\pi L_B\sin^2\alpha_m}=\frac{\alpha_m}{2\pi L_B^2\sin^2\alpha_m}.
\end{equation}
Substituting this result into Eq.~\eqref{eq:propagdeq3provv_APP}, we finally obtain
\begin{equation}
p_0^{(3)}(x,t)=\frac{1}{V_3(L_B)}+\sum_{n\ge1}\frac{\alpha_n}{2\pi L_B^2}\frac{\sin(k_nx)}{x\sin^2\alpha_n}e^{-Dk_n^2t},\qquad 0\le x\le L_B.
\label{eq:propagator_d3}
\end{equation}
Equivalently, since $L_B=L/2$ and $Dk_n^2=4D\alpha_n^2/L^2$, we can write
\begin{equation}
p_0^{(3)}(x,t)=\frac{6}{\pi L^3}+\sum_{n\ge1}\frac{2\alpha_n}{\pi L^2}\frac{\sin\!\left(\frac{2\alpha_n}{L}x\right)}{x\sin^2\alpha_n}\exp\!\left(-\frac{4D\alpha_n^2}{L^2}t\right).
\label{eq:propagator_d3_L}
\end{equation}
The radial moments are computed as
\begin{equation}
\langle x^q(\tau)\rangle^{(3)}_0=\int_0^{L/2}4\pi x^2\,dx\,x^q\,p_0^{(3)}(x,\tau).
\label{eq:moments_d3_measure}
\end{equation}
The required integrals are
\begin{equation}
I_3(\alpha):=\int_0^{L_B} x^3\sin\!\left(\frac{\alpha}{L_B}x\right)\,dx=\frac{2L_B^4}{\alpha^2}\sin\alpha=\frac{L^4}{8\alpha^2}\sin\alpha.
\end{equation}
and
\begin{equation}
I_5(\alpha):=\int_0^{L_B} x^5\sin\!\left(\frac{\alpha}{L_B}x\right)\,dx=\frac{4L_B^6}{\alpha^4}\left(\alpha^2-10\right)\sin\alpha=\frac{L^6}{16\alpha^4}\left(\alpha^2-10\right)\sin\alpha.
\end{equation}
Using these expressions, one obtains
\begin{equation}
\langle x^2(\tau)\rangle^{(3)}_0=\frac{3}{5}\left(\frac{L}{2}\right)^2+\left(\frac{L}{2}\right)^2\sum_{n\ge1}b_n^{(3)}\exp\!\left(-\frac{4D\alpha_n^2}{L^2}\tau\right),
\label{eq:moment2_d3}
\end{equation}
and
\begin{equation}
\langle x^4(\tau)\rangle^{(3)}_0=\frac{3}{7}\left(\frac{L}{2}\right)^4+\left(\frac{L}{2}\right)^4\sum_{n\ge1}c_n^{(3)}\exp\!\left(-\frac{4D\alpha_n^2}{L^2}\tau\right),
\label{eq:moment4_d3}
\end{equation}
where we have defined
\begin{equation}
b_n^{(3)}=\frac{4}{\alpha_n\sin\alpha_n},\qquad c_n^{(3)}=\frac{8(\alpha_n^2-10)}{\alpha_n^3\sin\alpha_n}.
\label{eq:bn_cn_d3}
\end{equation}
Substituting Eqs.~\eqref{eq:moment2_d3} and \eqref{eq:moment4_d3} into Eqs.~\eqref{eq:C1stat_dim_d} and \eqref{eq:C2stat_dim_d}, we obtain
\begin{equation}
\mathcal C_1^{(3)}(\nu_B)=\left(\frac{L}{2}\right)^4\mathcal F_1^{(3)}(\nu_B),
\label{eq:C1_d3_scaled}
\end{equation}
with
\begin{equation}
\mathcal F_1^{(3)}(\nu_B)=\left[\frac{3}{7}+\sum_{n\ge1}c_n^{(3)}\frac{\nu_B^2}{\nu_B^2+\alpha_n^2}\right]-\left[\frac{3}{5}+\sum_{n\ge1}b_n^{(3)}\frac{\nu_B^2}{\nu_B^2+\alpha_n^2}\right]^2.
\label{eq:F1_d3}
\end{equation}
and
\begin{equation}
\mathcal C_2^{(3)}(\nu_B)=\left(\frac{L}{2}\right)^4\mathcal F_2^{(3)}(\nu_B),
\label{eq:C2_d3_scaled}
\end{equation}
where
\begin{equation}
\mathcal F_2^{(3)}(\nu_B)=\sum_{n,m\ge1}b_n^{(3)}b_m^{(3)}\frac{\nu_B^2\alpha_n^2\alpha_m^2}{\left(\nu_B^2+\alpha_n^2+\alpha_m^2\right)\left(\nu_B^2+\alpha_n^2\right)\left(\nu_B^2+\alpha_m^2\right)}.
\label{eq:F2_d3}
\end{equation}
The normalized correlation coefficient in three dimensions is therefore
\begin{equation}
A_B^{(3)}(\nu_B)=\frac{\mathcal F_2^{(3)}(\nu_B)}{\mathcal F_1^{(3)}(\nu_B)}.
\label{eq:a_nu_dimension3_APP}
\end{equation}
This function is plotted in green in Fig.~\ref{fig:aind1d2d3}.

\section{Analysis of $a_N(T)$ in the BC case}\label{sec:aNderivation_APP}

In this section, we derive Eq.~\eqref{eq:aNcompleteforGAP} of the main text. We begin by recalling the definition of $a_N(T)$:
\begin{equation}
\label{eq:defaNBC_APP}
\int_{a_N(T)}^{L/2} dx\, p_0^{(B)}(x,T)=\frac{1}{N},
\end{equation}
where $p_0^{(B)}(x,T)$ is the reset-free single-particle propagator in the BC geometry.
This propagator can be written in the two equivalent forms:
\begin{equation}\label{eq:propag_1_T_APP}
p_0^{(B)}(x,T)=\frac{1}{L}
+\frac{2}{L}\sum_{n=1}^{\infty}
\cos\left(\frac{2\pi n x}{L}\right)\,
\exp\!\left[-\pi^2 n^2 T\right],
\end{equation}
\begin{equation}\label{eq:propag_3_T_APP}
    p_0^{(B)}(x,T)=\frac{1}{L\sqrt{\pi T}}
    \sum_{k=-\infty}^{\infty}
    \exp\!\left[-\frac{(x/L-k)^2}{T}\right].
\end{equation}
Equation~\eqref{eq:propag_1_T_APP} is more convenient in the large $T$ regime, whereas Eq.~\eqref{eq:propag_3_T_APP} is better suited for small $T$.
Defining $y_N(T)=\frac{a_N(T)}{L}$ and substituting Eq.~\eqref{eq:propag_1_T_APP} into Eq.~\eqref{eq:defaNBC_APP}, we obtain
\begin{equation}\label{eq:aNwithexp_APP}
y_N(T)+\sum_{n=1}^{\infty}
\frac{1}{\pi n}
\sin(2\pi n \, y_N(T))\,
e^{-(\pi n)^2 T}
=
\frac{1}{2}-\frac{1}{N}.
\end{equation}
Alternatively, substituting Eq.~\eqref{eq:propag_3_T_APP} into Eq.~\eqref{eq:defaNBC_APP}, we find
\begin{equation}\label{eq:aNwitherfc_APP}
\frac{1}{2}\sum_{k\in\mathbb{Z}}
\left[\operatorname{erf}\!\left(\frac{1/2-k}{\sqrt{T}}\right)-
\operatorname{erf}\!\left(\frac{y_N(T)-k}{\sqrt{T}}\right)
\right]=\frac{1}{N}.
\end{equation}
These two equivalent implicit equations for $y_N(T)=a_N(T)/L$ provide the starting point for the asymptotic analysis in the different time regimes.\\

\paragraph{Small $T$ regime.}

When $T$ is small, namely smaller than $T_c(N)\approx \frac{1}{4\ln N}$, we can consider \eqref{eq:aNwitherfc_APP} and we notice that only the term $k=0$ survives. In fact, for $T$ small enough we expect $y_N(T)=\frac{a_N(T)}{L}$ to be much smaller than $\frac{1}{2}$, so that we can write
\begin{equation}
\frac{1}{2}
\left[\operatorname{erf}\!\left(\frac{1}{2\sqrt{T}}\right)-
\operatorname{erf}\!\left(\frac{y_N(T)}{\sqrt{T}}\right)
\right]\approx\frac{1}{N}.
\end{equation}
In this limit, $\frac{1}{2\sqrt{T}}\to\infty$, and therefore, since $\operatorname{erf}(x)\to 1$ as $x\to\infty$, we obtain
\begin{equation}
    \frac{1}{2}\left[1-\operatorname{erf}\!\left(\frac{y_N(T)}{\sqrt{T}}\right)\right]=\frac{1}{2}\operatorname{erfc}\!\left(\frac{y_N(T)}{\sqrt{T}}\right)\approx\frac{1}{N}.
\end{equation}
Inverting the above equation we obtain the result
\begin{equation}\label{eq:aNBCgausswithuN}
    y_N(T)=\frac{a_N(T)}{L}=\sqrt{T}u_N,
\end{equation}
where $u_N=\operatorname{erfc}^{-1}(2/N)$. At leading order for large $N$, using the expansion
\[
\operatorname{erfc}(x)\sim \frac{e^{-x^2}}{\sqrt{\pi}\,x}
\qquad (x\to\infty),
\]
we obtain $u_N\approx\sqrt{\ln N}$, and so, finally
\begin{equation}\label{eq:defofTc_APP}
    a_N(T)=L\sqrt{T\ln N}.
\end{equation}
The crossover time $T_c(N)$ is obtained by requiring the Gaussian estimate in Eq.~\eqref{eq:aNBCgausswithuN} to reach the boundary, namely 
\begin{equation}\label{eq:defofT_c_APP}
    L\sqrt{T_c}u_N=L/2
\end{equation}
This gives $T_c(N)=\frac{1}{4u_N^2}\approx \frac{1}{4\ln N}$.

\paragraph{Large $T$ regime.}

In the opposite limit, when $T$ is large, which means $T>T_c(N)$, we can consider Eq. \eqref{eq:aNwithexp_APP}. 
In this regime, we expect the typical value of the rightmost particle $a_N(T)$ to be very close to $\frac{L}{2}$. Thus, in terms of $y_N(T)$ we set $y_N(T)=\frac{1}{2}-\varepsilon$, where $\varepsilon\ll 1$, in \eqref{eq:aNwithexp_APP}.
We can now write
\begin{equation}
    \sin\!\left(2\pi n\left(\frac{1}{2}-\varepsilon\right)\right)=\sin(\pi n-2\pi n\varepsilon)=(-1)^{n+1}\sin(2\pi n\varepsilon),
\end{equation}
and, by expanding for small $\varepsilon$, we obtain
\begin{equation}\label{eq:intermedialargeTaN_APP}
\varepsilon-\sum_{n=1}^{\infty}\frac{(-1)^{n+1}}{\pi n}(2\pi n\varepsilon)\,e^{-(\pi n)^2 T}
\approx
\frac{1}{N}.
\end{equation}
We then recall the definition in Eq. \eqref{eq:ThetaTfirstdef} of the function $\vartheta(T)$, namely
\begin{equation}
\vartheta(T)=1-2\sum_{n=1}^{\infty}(-1)^{n+1}e^{-(\pi n)^2 T}=
1+2\sum_{n=1}^{\infty}(-1)^n e^{-(\pi n)^2 T}.
\end{equation}
Equation \eqref{eq:intermedialargeTaN_APP} therefore becomes
\begin{equation}
\varepsilon=\frac{1}{N\vartheta(T)}.
\end{equation}
Recalling that $y_N(T)=\frac{1}{2}-\varepsilon=\frac{a_N(T)}{L}$ we finally obtain
\begin{equation}\label{eq:boundarydominatedreg_APP}
a_N(T)=\frac{L}{2}-\frac{L}{N\vartheta(T)}.
\end{equation}

\paragraph{Crossover regime $(T\sim T_c(N))$.}\label{eq:aNcrossoverregion_APP}

We now analyze the narrow crossover region around $T_c(N)$, where the small-$T$ Gaussian regime and the large-$T$ boundary-dominated regime are matched. Since $T_c(N)$ is defined by the condition that the Gaussian estimate reaches the wall, given in Eq. \eqref{eq:defofT_c_APP}, the typical maximum is close to the right boundary in this regime. We therefore set
\begin{equation}
    y_N(T)=\frac{a_N(T)}{L}=\frac{1}{2}-\varepsilon,
    \qquad \varepsilon\ll 1.
\end{equation}
At the same time, since $T_c(N)\approx 1/(4\ln N)$, the crossover occurs at small $T$ for large $N$. It is therefore convenient to use the image representation in Eq.~\eqref{eq:aNwitherfc_APP}. Keeping the two dominant image terms, $k=0$ and $k=1$, we obtain
\begin{equation}
\frac{1}{N}\approx \frac{1}{2}\left[
\operatorname{erf}\!\left(\frac{1/2+\varepsilon}{\sqrt{T}}\right)-
\operatorname{erf}\!\left(\frac{1/2-\varepsilon}{\sqrt{T}}\right)
\right].
\end{equation}
Using the large-$x$ expansion
\begin{equation}
    \operatorname{erf}(x)\approx 1-\frac{e^{-x^2}}{\sqrt{\pi}\,x},
\end{equation}
we obtain
\begin{equation}
\frac{1}{N}\approx \frac{\sqrt{T}}{2\sqrt{\pi}}
\left[
\frac{e^{-(1/2-\varepsilon)^2/T}}{1/2-\varepsilon}-
\frac{e^{-(1/2+\varepsilon)^2/T}}{1/2+\varepsilon}
\right].
\end{equation}
Expanding at small $\varepsilon$, this becomes
\begin{equation}\label{eq:IDK_APP}
    \frac{1}{N}\approx 2\sqrt{\frac{T}{\pi}}\,e^{-1/(4T)}\sinh\!\left(\frac{\varepsilon}{T}\right).
\end{equation}
Recalling that $u_N=\operatorname{erfc}^{-1}(2/N)$, in the large-$N$ limit $u_N$ is large and
\begin{equation}
\operatorname{erfc}(u_N)\approx \frac{e^{-u_N^2}}{\sqrt{\pi}u_N}.
\end{equation}
Using the definition $\operatorname{erfc}(u_N)=2/N$, one obtains
\begin{equation}
\frac{2}{N}\approx \frac{e^{-u_N^2}}{\sqrt{\pi}u_N}.
\end{equation}
Moreover, from $T_c(N)=\frac{1}{4u_N^2}$, this can be rewritten as
\begin{equation}\label{eq:IDKpt2_APP}
\frac{1}{N}\approx \sqrt{\frac{T_c(N)}{\pi}}\,e^{-1/(4T_c)}.
\end{equation}
We can now divide Eq.~\eqref{eq:IDK_APP} by Eq.~\eqref{eq:IDKpt2_APP}, obtaining
\begin{equation}
    2\sqrt{\frac{T}{T_c(N)}}
    \exp\!\left(-\frac{1}{4T}+\frac{1}{4T_c}\right)
    \sinh\!\left(\frac{\varepsilon}{T}\right)\approx 1.
\end{equation}
We then zoom into the crossover window by writing
\begin{equation}
T = T_c(N) - \frac{s}{4u_N^4}
\approx T_c(N) - \frac{s}{(2\ln N)^2}.
\end{equation}
At leading order for large $N$, one has
\[
\frac{1}{4T}-\frac{1}{4T_c}\approx s,
\qquad
\sqrt{\frac{T}{T_c(N)}}\approx 1.
\]
Therefore, the previous equation reduces to
\begin{equation}
\sinh\!\left(\frac{\varepsilon}{T}\right)\approx \frac{e^{s}}{2}.
\end{equation}
Solving for $\varepsilon$ and recalling that $\frac{a_N(T)}{L}=\frac{1}{2}-\varepsilon$, we finally obtain
\begin{equation}\label{eq:aNintermregimefinal_APP}
\frac{a_N(T)}{L}
=
\frac{1}{2}-\varepsilon
\approx
\frac{1}{2}
-
\frac{1}{4\ln N}
\operatorname{arsinh}\!\left(\frac{e^{s}}{2}\right).
\end{equation}
This form is valid in a crossover window of width $O((\ln N)^{-2})$ around $T_c(N)$.

\section{Intermediate regime for the EVS in the BC}\label{sec:intermregmeEVS_BCcase_APP}

In this appendix, we derive the intermediate scaling regime for the EVS in the BC. This regime connects the typical regime, where the maximum lies at a distance $O(1/N)$ from the boundary, to the large-deviation regime, where it lies at a macroscopic distance $O(1)$ from the boundary. We start from Eq.~\eqref{eq:general_formula_EVS_T}, which we also report here:
\begin{equation}\label{eq:generalEVSmilliontimes_APP}
    \operatorname{Prob}\{M_1=m\}=
    \nu_B^2\int_0^\infty dT  e^{-\nu_B^2 T}
    \operatorname{Prob}\{M_1(T)=m\}.
\end{equation}
As discussed in the main text and in Appendix~\ref{sec:aNderivation_APP}, the crossover between the Gaussian regime and the boundary-dominated regime occurs at
\begin{equation}
    T_c(N)=\frac{1}{4u_N^2}
    \approx \frac{1}{4\ln N},
    \qquad
    u_N=\operatorname{erfc}^{-1}\left(\frac{2}{N}\right).
\end{equation}
In Appendix~\ref{sec:aNderivation_APP}, we showed that the appropriate crossover window is
\begin{equation}
    T=T_c(N)-\frac{s}{4u_N^4}
    \approx T_c(N)-\frac{s}{(2\ln N)^2},
\end{equation}
where $s=O(1)$. In this regime, the typical position of the maximum satisfies
\begin{equation}\label{eq:aNintermwiths_APP}
    \frac{a_N(T)}{L}=
    \frac{1}{2}-
    \frac{1}{4\ln N}\operatorname{arsinh}\left(\frac{e^s}{2}\right).
\end{equation}
Thus the distance of the extreme quantile from the boundary is
\begin{equation}
    \frac{L}{2}-a_N(T)=
    \frac{L}{4\ln N}
    \operatorname{arsinh}\left(\frac{e^s}{2}\right)=
    O\left(\frac{1}{\ln N}\right).
\end{equation}
It is important to note that in this case the fixed $T$ fluctuations of the maximum are of the same order, as can be checked using Eq.~\eqref{eq:defaNbN}. Hence, in this intermediate regime, the residual fluctuations of $M_1(T)$ at fixed $T$ are not subleading. In contrast with the HC case, or with the large-deviation regime of the BC, one cannot replace the conditional distribution $\operatorname{Prob}\{M_1(T)=m\}$ by a delta peak. We must instead keep the full fixed $T$ IID extreme-value form.
We now introduce the scaling variable
\begin{equation}\label{eq:defofz_APP}
    z=\delta\ln N,
    \qquad
    \delta=\frac{1}{2}-\frac{m}{L},
\end{equation}
or equivalently
\begin{equation}
    m=L\left(\frac{1}{2}-\frac{z}{\ln N}\right).
\end{equation}
At fixed $T$, the particles are IID with common density $p_0^{(B)}(x,T)$. Therefore, using Eq. \eqref{eq:finiteNEVS}, we can write
\begin{equation}
    \operatorname{Prob}\{M_1(T)=m\}=
    N p_0^{(B)}(m,T)
    \left[
        1-\int_m^{L/2}du\,p_0^{(B)}(u,T)
    \right]^{N-1}.
\end{equation}
In the large-$N$ limit, this becomes
\begin{equation}\label{eq:fixedT_EVS_intermediate_APP}
    \operatorname{Prob}\{M_1(T)=m\}
    \approx
    Np_0^{(B)}(m,T)\exp[-NQ(m,T)],
\end{equation}
where
\begin{equation}
    Q(m,T)=\int_m^{L/2}du\,p_0^{(B)}(u,T).
\end{equation}
We now evaluate Eq.~\eqref{eq:fixedT_EVS_intermediate_APP} in the crossover window. We write
\begin{equation}
    m=L\left(\frac{1}{2}-\frac{z}{\ln N}\right),
    \qquad
    T=T_c(N)-\frac{s}{4u_N^4}.
\end{equation}
Since $T=O(1/\ln N)$ and $m$ is close to $L/2$, the two relevant image terms in the propagator are the ones with $k=0$ and $k=1$. Thus,
\begin{equation}
    p_0^{(B)}(u,T)\approx
    \frac{1}{L\sqrt{\pi T}}
    \left[
    \exp\!\left(-\frac{(u/L)^2}{T}\right)
    +
    \exp\!\left(-\frac{(1-u/L)^2}{T}\right)
    \right].
\end{equation}
Setting $u=L(1/2-\xi/\ln N)$, one obtains
\begin{equation}
    p_0^{(B)}(u,T)
    \approx
    \frac{2}{L\sqrt{\pi T}}
    e^{-1/(4T)}
    \cosh\!\left(\frac{\xi}{T\ln N}\right),
\end{equation}
where corrections of relative order $1/\ln N$ have been neglected. In the crossover window that we are considering,
\begin{equation}
    \frac{1}{4T}-\frac{1}{4T_c}\approx s,
    \qquad
    \frac{\xi}{T\ln N}\approx 4\xi.
\end{equation}
Moreover, from $T_c(N)=1/(4u_N^2)$ and $\operatorname{erfc}(u_N)=2/N$, the large-$u_N$ expansion of the complementary error function gives
\begin{equation}
    \frac{1}{N}\approx
    \sqrt{\frac{T_c(N)}{\pi}}\,
    e^{-1/(4T_c)}.
\end{equation}
Using these relations, we find
\begin{equation}\label{eq:Np0_intermediate_APP}
    Np_0^{(B)}(m,T)
    \approx
    \frac{8\ln N}{L}e^{-s}\cosh(4z).
\end{equation}
We also need $NQ(m,T)$. Using again $u=L(1/2-\xi/\ln N)$, we have
\begin{equation}
    Q(m,T)
    =
    \frac{L}{\ln N}
    \int_0^z d\xi\,
    p_0^{(B)}\!\left(x=L\left(\frac{1}{2}-\frac{\xi}{\ln N}\right),T\right).
\end{equation}
Combining this expression with Eq.~\eqref{eq:Np0_intermediate_APP}, gives
\begin{equation}\label{eq:NQ_intermediate_APP}
    NQ(m,T)\approx2e^{-s}\sinh(4z).
\end{equation}
Substituting Eqs.~\eqref{eq:Np0_intermediate_APP} and \eqref{eq:NQ_intermediate_APP} into Eq.~\eqref{eq:fixedT_EVS_intermediate_APP}, we obtain
\begin{equation}
    \operatorname{Prob}\{M_1(T)=m\}
    \approx
    \frac{8\ln N}{L}
    e^{-s}\cosh(4z)
    \exp[-2e^{-s}\sinh(4z)].
\end{equation}
We now insert this expression into Eq.~\eqref{eq:generalEVSmilliontimes_APP}. Since $T=O(1/\ln N)$, we can write
\begin{equation}
    e^{-\nu_B^2T}=1+O\left(\frac{1}{\ln N}\right).
\end{equation}
Moreover,
\begin{equation}
    dT=-\frac{ds}{4u_N^4}
    \approx -\frac{ds}{4(\ln N)^2}.
\end{equation}
At finite $N$, the change of variable
$T=T_c(N)-s/(4u_N^4)$ maps the domain $T\in[0,\infty)$ onto
$s\in(-\infty,u_N^2)$, after reversing the orientation of the integral.
Since $u_N^2\sim \ln N$, the upper limit tends to $+\infty$ as
$N\to\infty$. Thus, at leading order, and for fixed $z>0$, we can replace
the upper limit by $+\infty$, obtaining
\begin{equation}
    \operatorname{Prob}\{M_1=m\}
    \approx
    \frac{2\nu_B^2}{L\ln N}\cosh(4z)
    \int_{-\infty}^{\infty}ds\,
    e^{-s}\exp[-2e^{-s}\sinh(4z)].
\end{equation}
The integral can now be performed with the change of variable
$w=2e^{-s}\sinh(4z)$, which gives
\begin{equation}
    \int_{-\infty}^{+\infty}ds\,
    e^{-s}\exp[-2e^{-s}\sinh(4z)]
    =
    \frac{1}{2\sinh(4z)}.
\end{equation}
Thus,
\begin{equation}
    \operatorname{Prob}\{M_1=m\}
    \approx
    \frac{\nu_B^2}{L\ln N}
    \coth(4z).
\end{equation}
We finally obtain the intermediate scaling form
\begin{equation}
    \operatorname{Prob}\{M_1=m\}
    \approx
    \frac{1}{L\ln N}S_2(\delta\ln N),
\end{equation}
with
\begin{equation}
    S_2(z)=\nu_B^2\coth(4z).
\end{equation}
This result also provides the matching between the typical and large-deviation regimes. For $z\to 0$, we have $S_2(z)\approx \frac{\nu_B^2}{4z}$, which matches the large-$z$ tail of the typical scaling function $S_B(z)$. For $z\to\infty$, we have $S_2(z)\to \nu_B^2$, which matches the small-$z$ limit of the large-deviation scaling function $S_3(z)=\nu_B^2(1-2z)$.

\section{Analysis of $p_0^{(B)}(a_N(T),T)$ in the BC case}

In this section, we show how to derive Eq. \eqref{eq:rhodiTvariregimi} of the main text. In Appendix \ref{sec:aNderivation_APP}, we derived the expression given in Eq. \eqref{eq:aNcompleteforGAP} for $a_N(T)$, which will be used throughout this analysis. The starting point is the reset-free single-particle propagator $p_0^{(B)}(x,T)$, given in Eqs. \eqref{eq:propag_1_T_APP} and \eqref{eq:propag_3_T_APP}.\\

\paragraph{Small $T$ regime.}

We consider here the regime $T<T_c(N) \approx \frac{1}{4\ln N}$. In Appendix \ref{sec:aNderivation_APP}, we have found that
\begin{equation}
a_N(T)=L\sqrt{T}\,u_N \approx L\sqrt{T\ln N}.
\end{equation}
At the same time, from Eq. \eqref{eq:propag_3_T_APP}, we see that for small $T$ the propagator $p_0^{(B)}(x,T)$ is well approximated by a Gaussian
\begin{equation}
p_0^{(B)}(x,T)\approx \frac{1}{L\sqrt{\pi T}}\exp\!\left[-\frac{x^2}{L^2T}\right].
\end{equation}
Evaluating this expression at $x=a_N(T)=L\sqrt{T}\,u_N$, one must be careful: keeping only the leading estimate $u_N\sim\sqrt{\ln N}$ is not sufficient, since the exponential factor is sensitive to subleading corrections. Using the standard expansion
\begin{equation}
u_N\approx
\sqrt{\ln N-\frac{1}{2}\ln(4\pi\ln N)},
\end{equation}
we obtain
\begin{equation}
    e^{-u_N^2}\approx\frac{2\sqrt{\pi\ln N}}{N}.
\end{equation}
Substituting into the Gaussian expression then gives
\begin{equation}
    p_0^{(B)}(a_N(T),T)=\frac{2\ln N}{N\,a_N(T)}=\frac{2}{LN}\sqrt{\frac{\ln N}{T}}.
\end{equation}\\

\paragraph{Large $T$ regime.}

We now consider the regime $T>T_c(N) \approx \frac{1}{4\ln N}$. From Appendix \ref{sec:aNderivation_APP}, we know that in this case
$a_N(T)=L\left(\frac{1}{2}-\varepsilon\right)$, where $\varepsilon=O\!\left(\frac{1}{N}\right)$ is very small. Substituting $a_N(T)=L\left(\frac{1}{2}-\varepsilon\right)$ into Eq. \eqref{eq:propag_1_T_APP}, we obtain
\begin{equation}
p^{(B)}_0\!\left(a_N(T),T\right)
=
\frac{1}{L}
+
\frac{2}{L}\sum_{n=1}^{\infty}
\cos\left(2\pi n\left(\frac{1}{2}-\varepsilon\right)\right)e^{-\pi^2 n^2 T}.
\end{equation}
We then rewrite $\cos\left(2\pi n z_N(T)\right)=\cos(\pi n-2\pi n\varepsilon)=(-1)^n\cos(2\pi n\varepsilon)$, and use $\cos(2\pi n\varepsilon)\approx 1$ for small $\varepsilon$. This gives
\begin{equation}
p^{(B)}_0\!\left(a_N(T),T\right)\approx
\frac{1}{L}+\frac{2}{L}\sum_{n=1}^{\infty}(-1)^n e^{-\pi^2 n^2 T}.
\end{equation}
We now recognize here the function $\vartheta(T)$ introduced in Eq. \eqref{eq:ThetaTfirstdef}. Therefore,
\begin{equation}
p^{(B)}_0\!\left(a_N(T),T\right)\approx \frac{\vartheta(T)}{L}.
\end{equation}

\paragraph{Crossover regime $(T\sim T_c(N))$.}

In Appendix \ref{sec:aNderivation_APP}, we established that the crossover regime emerges in a narrow region around
\begin{equation}
T_c(N)=\frac{1}{4u_N^2}\approx\frac{1}{4\ln N},
\end{equation}
where $u_N=\operatorname{erfc}^{-1}(2/N)$. More precisely, we consider
\begin{equation}
T=T_c(N)-\frac{s}{4u_N^4}\approx T_c(N)-\frac{s}{(2\ln N)^2},
\end{equation}
which means that we are probing a window of width $O\!\left((\ln N)^{-2}\right)$ around $T_c(N)$.
In this region, for large $N$, Eq. \eqref{eq:aNintermregimefinal_APP} gives
\begin{equation}
\frac{a_N(T)}{L}
=
\frac{1}{2}-\varepsilon
\approx
\frac{1}{2}-\frac{1}{4\ln N}\operatorname{arsinh}\!\left(\frac{e^s}{2}\right).
\end{equation}
Thus, $a_N(T)$ lies at a distance $O\!\left(1/\ln N\right)$ from the right boundary $L/2$.
As in Appendix \ref{sec:aNderivation_APP}, in this regime it is sufficient to retain only the $k=0$ and $k=1$ terms in the representation \eqref{eq:propag_3_T_APP} of the propagator. This gives
\begin{equation}
p^{(B)}_0\!\left(a_N(T),T\right)\approx
\frac{1}{L\sqrt{\pi T}}
\left[
e^{-(1/2-\varepsilon)^2/T}
+
e^{-(1/2+\varepsilon)^2/T}
\right],
\end{equation}
where we used $z=\frac{a_N(T)}{L}=\frac{1}{2}-\varepsilon$.
Factoring out the common exponential term, we obtain
\begin{equation}
p^{(B)}_0\!\left(a_N(T),T\right)\approx
\frac{2}{L\sqrt{\pi T}}
e^{-1/(4T)}
e^{-\varepsilon^2/T}
\cosh\!\left(\frac{\varepsilon}{T}\right).
\end{equation}
Since here $\varepsilon=O\!\left(1/\ln N\right)$ and $T=O\!\left(1/\ln N\right)$, we have
\begin{equation}
\frac{\varepsilon^2}{T}=O\!\left(\frac{1}{\ln N}\right)\to 0,
\end{equation}
so that, at leading order, $e^{-\varepsilon^2/T}\approx 1$.
Therefore,
\begin{equation}
p^{(B)}_0\!\left(a_N(T),T\right)\approx
\frac{2}{L\sqrt{\pi T}}
e^{-1/(4T)}
\cosh\!\left(\frac{\varepsilon}{T}\right).
\end{equation}
We can now use Eq. \eqref{eq:IDK_APP}, which is valid in this regime, and obtain
\begin{equation}
p^{(B)}_0\!\left(a_N(T),T\right)\approx
\frac{1}{LNT}
\coth\!\left(\frac{\varepsilon}{T}\right).
\end{equation}
Moreover, from Eq. \eqref{eq:aNintermregimefinal_APP}, we know that
\begin{equation}
\frac{\varepsilon}{T}\approx 
\operatorname{arsinh}\left(\frac{e^{s}}{2}\right).
\end{equation}
Hence
\begin{equation}
\coth\!\left(\frac{\varepsilon}{T}\right)=
\coth\!\left[\operatorname{arsinh}\left(\frac{e^{s}}{2}\right)\right].
\end{equation}
Using the identity $\cosh y=\sqrt{1+\sinh^2 y}$, this can be rewritten as
\begin{equation}
\coth\!\left[\operatorname{arsinh}\left(\frac{e^{s}}{2}\right)\right]=
\sqrt{1+4e^{-2s}}.
\end{equation}
Therefore,
\begin{equation}
p^{(B)}_0\!\left(a_N(T),T\right)\approx
\frac{1}{LNT}\sqrt{1+4e^{-2s}}.
\end{equation}
Finally, using $T\approx T_c(N)\approx \frac{1}{4\ln N}$, we arrive at
\begin{equation}
p^{(B)}_0\!\left(a_N(T),T\right)\approx
\frac{4\ln N}{LN}\sqrt{1+4e^{-2s}}.
\end{equation}

\section{Intermediate regime for the gap statistics in the BC}\label{sec:intermregimeGAPBC_APP}

In this appendix we derive the intermediate scaling regime for the gap
statistics in the BC geometry, given in Eqs. \eqref{eq:recapGAPBC_begin} and \eqref{eq:h2maintextGAP_begin}. This regime connects the typical sector,
where the first gap is of order $O(1/N)$, to the large-deviation sector,
where it is of order $O(1/\ln N)$. We start from Eq.~\eqref{eq:GAP_general_base},
which we report here for convenience:
\begin{equation}\label{eq:gapgeneral_APP}
    \operatorname{Prob}\{d_1=g\}=
    \nu_B^2\int_0^\infty dT e^{-\nu_B^2 T}\,
    \varrho_B(T)\,e^{-\varrho_B(T)g},
\end{equation}
where $\varrho_B(T)=N p_0^{(B)}(a_N(T),T)$ is the local density at the typical position of the maximum at fixed reset age $T$.
For compactness, we denote here by $P_{<}(g)$ and $P_{>}(g)$ the contributions
to Eq.~\eqref{eq:gapgeneral_APP} coming from the intervals $0<T<T_c(N)$ and
$T>T_c(N)$, respectively. Thus,
\begin{equation}
\begin{aligned}
\operatorname{Prob}\{d_1=g\}
&=
\nu_B^2\int_0^{T_c(N)} dT e^{-\nu_B^2T}
\varrho_B(T)e^{-\varrho_B(T)g}
+
\nu_B^2\int_{T_c(N)}^{\infty} dT e^{-\nu_B^2T}
\varrho_B(T)e^{-\varrho_B(T)g}  \\
&\equiv P_{<}(g)+P_{>}(g).
\end{aligned}
\end{equation}
As discussed in the main text and derived in Appendix~\ref{sec:aNderivation_APP}, 
the crossover between the Gaussian regime and the boundary-dominated regime occurs around
\begin{equation}
    T_c(N)=\frac{1}{4u_N^2}
    \approx \frac{1}{4\ln N},
    \qquad
    u_N=\operatorname{erfc}^{-1}\!\left(\frac{2}{N}\right).
\end{equation}
In the two outer regimes, the local edge density behaves as
\begin{equation}
    \varrho_B(T)\approx
    \begin{cases}
    \dfrac{2\sqrt{\ln N}}{L\sqrt{T}},
    & T<T_c(N), \\[8pt]
    \dfrac{N}{L}\vartheta(T),
    & T>T_c(N),
    \end{cases}
\end{equation}
where
\begin{equation}
    \vartheta(T)=1+2\sum_{n=1}^{\infty}(-1)^n e^{-\pi^2 n^2 T}.
\end{equation}
The matching between these two behaviors takes place in a narrow window
$|T-T_c(N)|=O((\ln N)^{-2})$. In this window, writing
\begin{equation}
    T=T_c(N)-\frac{s}{4(\ln N)^2},
    \qquad s=O(1),
\end{equation}
one finds
\begin{equation}\label{eq:intermregimeforrho_APP}
    \varrho_B(T)\approx
    \frac{4\ln N}{L}\sqrt{1+4e^{-2s}}.
\end{equation}
The intermediate regime for the gap statistics is obtained by rescaling the first gap as
\begin{equation}
    g=\frac{L y}{(\ln N)^2},
    \qquad y=O(1).
\end{equation}
We now evaluate Eq.~\eqref{eq:gapgeneral_APP} in this scaling limit. We first consider the contribution from the interval $[0,T_c(N)]$. Inside this
interval, the narrow crossover region close to $T_c(N)$ has width
$O((\ln N)^{-2})$. Since in that region $\varrho_B(T)=O(\ln N/L)$, its total contribution is of order $O(1/(L\ln N))$ (see \eqref{eq:intermregimeforrho_APP}) and it is therefore subleading. The leading contribution instead comes from the Gaussian part of the interval, where
\begin{equation}
    \varrho_B(T)\approx\frac{2\sqrt{\ln N}}{L\sqrt{T}}.
\end{equation}
Thus
\begin{equation}
    P_<(g)
    \approx
    \nu_B^2\int_0^{T_c(N)} dT
    \frac{2\sqrt{\ln N}}{L\sqrt{T}}
    \exp\!\left[
    -\frac{2y}{(\ln N)^{3/2}\sqrt{T}}
    \right],
\end{equation}
where we have used $e^{-\nu_B^2 T}\approx 1$ in this range. Setting
$T=\frac{v}{\ln N}$, and using $T_c(N)\approx 1/(4\ln N)$, we obtain
\begin{equation}
    P_<(g)
    \approx
    \frac{2\nu_B^2}{L}
    \int_0^{1/4}\frac{dv}{\sqrt{v}}\,
    \exp\!\left[
    -\frac{2y}{\ln N\sqrt{v}}
    \right].
\end{equation}
Taking the large $N$ limit at fixed $y=O(1)$ gives
\begin{equation}\label{eq:contrib1_APP}
    P_<(g)
    \approx
    \frac{2\nu_B^2}{L}
    \int_0^{1/4}\frac{dv}{\sqrt{v}}
    =
    \frac{2\nu_B^2}{L}.
\end{equation}
We now consider the contribution from the interval $[T_c(N),\infty)$. For times $T=O(1)$ one has $\vartheta(T)=O(1)$, and therefore
\begin{equation}
    \varrho_B(T)g
    \approx
    \frac{N}{L}\vartheta(T)\frac{Ly}{(\ln N)^2}
    =
    \frac{Ny}{(\ln N)^2}\vartheta(T)
    \to \infty .
\end{equation}
Thus the factor $e^{-\varrho_B(T)g}$ suppresses the contribution of times
$T=O(1)$. The leading contribution from $T>T_c(N)$ therefore comes from a region just above $T_c(N)$.
We introduce again the crossover variable
\begin{equation}
    T=T_c(N)-\frac{s}{4(\ln N)^2}.
\end{equation}
The interval $T>T_c(N)$ corresponds to $s<0$. In this regime,
Eq.~\eqref{eq:intermregimeforrho_APP} gives
\begin{equation}
    \varrho_B(T)\approx
    \frac{4\ln N}{L}\sqrt{1+4e^{-2s}},
\end{equation}
and
\begin{equation}
    dT\approx -\frac{ds}{4(\ln N)^2}.
\end{equation}
Therefore the contribution from the right of $T_c(N)$ reads, at leading order,
\begin{equation}
P_>(g)
\approx
\frac{\nu_B^2}{L\ln N}
\int_{-\ln N}^{0} ds\,
\sqrt{1+4e^{-2s}}\,
\exp\left[
-\frac{4y}{\ln N}\sqrt{1+4e^{-2s}}
\right].
\end{equation}
The exponential factor selects values of $s$ such that $-s$ is large. In
this range, $\sqrt{1+4e^{-2s}}\approx 2e^{-s}$,
and therefore
\begin{equation}
    P_>(g)
    \approx
    \frac{2\nu_B^2}{L\ln N}
    \int_{-\ln N}^{0} ds\, e^{-s}
    \exp\!\left[
    -\frac{8y}{\ln N}e^{-s}
    \right].
\end{equation}
We now perform the change of variable $u=\frac{8y}{\ln N}e^{-s}$. Since
\begin{equation}
    e^{-s}ds=-\frac{\ln N}{8y}\,du,
\end{equation}
the integration bounds become
\begin{equation}
    s=0 \quad \Rightarrow \quad u=\frac{8y}{\ln N},
    \qquad
    s=-\ln N \quad \Rightarrow \quad u=\frac{8yN}{\ln N}.
\end{equation}
Hence
\begin{equation}
    P_>(g)
    \approx
    \frac{2\nu_B^2}{L\ln N}
    \frac{\ln N}{8y}
    \int_{8y/\ln N}^{8yN/\ln N}du\,e^{-u}.
\end{equation}
Taking the large $N$ limit at fixed $y=O(1)$ gives
\begin{equation}\label{eq:contrib2_APP}
    P_>(g)
    \approx
    \frac{\nu_B^2}{4Ly}
    \int_0^\infty du\,e^{-u}
    =
    \frac{\nu_B^2}{4Ly}.
\end{equation}
Combining the two contributions (Eq. \eqref{eq:contrib1_APP} and Eq. \eqref{eq:contrib2_APP}), we finally obtain the intermediate scaling form
\begin{equation}
    \operatorname{Prob}\{d_1=g\}
    \approx
    \frac{1}{L}\,
    h_2\!\left(\frac{(\ln N)^2 g}{L}\right),
    \qquad
    g=O\!\left(\frac{L}{(\ln N)^2}\right),
\end{equation}
with
\begin{equation}
    h_2(y)=2\nu_B^2+\frac{\nu_B^2}{4y}.
\end{equation}

\section{EVS for $V(x)=\kappa |x|$}\label{sec:alphaequal1_APP}

In this section, we study our resetting Brownian gas of $N$ particles confined by the linear potential $V(x)=\kappa |x|$. This corresponds to the case $\alpha=1$ of the family
$V(x)=\kappa |x|^\alpha$ considered in Sec. \ref{sec:GeneralPotential}. Our goal is to determine the EVS of the rightmost particle $M_1$. As discussed in Sec.~\ref{sec:EVS}, the key
ingredient is the reset-free single-particle propagator.
For a particle starting at the origin at time $t=0$, the propagator is (see Ref.~\cite{Touchette2010})
\begin{equation}\label{eq:propag_alpha1_APP}
p(x,t)=
\frac{1}{\sqrt{4\pi Dt}}
\exp\!\left[-\frac{(|x|+\kappa t)^2}{4Dt}\right]+
\frac{\kappa}{4D}e^{-\kappa |x|/D}
\operatorname{erfc}\!\left(\frac{|x|-\kappa t}{2\sqrt{Dt}}\right).
\end{equation}
At fixed reset age $\tau$, the particles are IID with common density
$p(x,\tau)$. We denote by $a_N(\tau)$ the typical position of the maximum at fixed
$\tau$, defined by
\begin{equation}\label{eq:defa_Nalpha1_APP}
\int_{a_N(\tau)}^{\infty} dx\, p(x,\tau)=\frac{1}{N}.
\end{equation}
Since the maximum probes the large-$x$ tail of the propagator, we use the
asymptotic form, valid for fixed $t=O(1)$ and $x\to+\infty$,
\begin{equation}\label{eq:tailpropagalpha1_APP}
p(x,t)\sim
\frac{1}{\sqrt{4\pi Dt}}
\exp\!\left[-\frac{(x+\kappa t)^2}{4Dt}\right].
\end{equation}
Substituting this expression into Eq.~\eqref{eq:defa_Nalpha1_APP}, we obtain
\begin{equation}\label{eq:aNalpha1noappr_APP}
a_N(\tau)\approx \sqrt{4D\tau}\,u_N-\kappa\tau,
\end{equation}
where $u_N=\operatorname{erfc}^{-1}\!\left(\frac{2}{N}\right)$ and $u_N\sim \sqrt{\ln N}$ for large $N$.
As in other cases considered Sec. \ref{sec:EVS}, the dominant source of fluctuations of
$M_1$ comes from the fluctuations of the reset age $\tau$. Therefore, at leading
order, we can approximate the conditional distribution of $M_1$ at fixed $\tau$ by a delta peak centered at $a_N(\tau)$. This gives
\begin{equation}\label{eq:alpha1EVSapprox_APP}
\operatorname{Prob}\{M_1=m\}
=r\int_0^\infty d\tau\, e^{-r\tau}\,
\delta\!\left(m-a_N(\tau)\right).
\end{equation}
At leading order for $N\to\infty$, the term $-\kappa\tau$ in
Eq.~\eqref{eq:aNalpha1noappr_APP} is subleading compared to
$\sqrt{4D\tau}\,u_N$. We may therefore write
\begin{equation}\label{eq:alpha1aNasfree_APP}
a_N(\tau)\approx \sqrt{4D\tau\ln N}.
\end{equation}
This is exactly the same expression for $a_N(T)$ as in the unconfined resetting gas \cite{Biroli2023}. Hence,
inserting Eq.~\eqref{eq:alpha1aNasfree_APP} into
Eq.~\eqref{eq:alpha1EVSapprox_APP}, one recovers the unconfined scaling form
\begin{equation}\label{eq:maxalpha1interm_APP}
M_1=L_N Z,
\qquad
L_N=\sqrt{\frac{4D}{r}\ln N},
\end{equation}
where the rescaled variable $Z$ has density $S(z)=2z e^{-z^2}$.
This shows that the potential $V(x)=\kappa |x|$ belongs, at leading order, to the same EVS universality class as the free case.
However, the convergence to this limiting form is very slow. For the values of $N$ accessible in simulations, typically $N\sim 10^6$, the subleading term $-\kappa\tau$ in Eq.~\eqref{eq:aNalpha1noappr_APP} can still produce visible finite-$N$ corrections. To account for these corrections, we keep the full expression
\begin{equation}
a_N(\tau)\approx \sqrt{4D\tau}\,u_N-\kappa\tau.
\end{equation}
Substituting it into Eq.~\eqref{eq:alpha1EVSapprox_APP} gives
\begin{equation}
\operatorname{Prob}\{M_1=m\}=
r\int_0^\infty d\tau\, e^{-r\tau}\,
\delta\!\left(m-\sqrt{4D\tau}\,u_N+\kappa\tau\right).
\end{equation}
Equivalently, this means that the random variable $M_1$ can be represented as
\begin{equation}
M_1=\sqrt{4D}\,u_N\sqrt{\tau}-\kappa\tau,
\end{equation}
where $\tau$ is exponentially distributed with density $r e^{-r\tau}$.

We now introduce the change of variable $\tau=x^2/r$, with $x\geq 0$.
Since $\tau$ is exponentially distributed with density $r e^{-r\tau}$, the variable $x$ has density
\begin{equation}
P_x(x)=2x e^{-x^2},\qquad x\geq 0.
\end{equation}
In terms of $x$, the maximum can therefore be written as
\begin{equation}\label{eq:maxalpha1xsquared_APP}
M_1=\widetilde L_N x-c_2x^2,
\end{equation}
where
\begin{equation}\label{eq:Ltildec2alpha1_APP}
\widetilde L_N=2\sqrt{\frac{D}{r}}\,u_N,
\qquad
c_2=\frac{\kappa}{r},
\end{equation}
and $u_N=\operatorname{erfc}^{-1}\!\left(2/N\right)$ behaves as $u_N\approx \sqrt{\ln N}$ for large $N$.
Here $\widetilde L_N$ is the finite-$N$ version of the asymptotic scale
\begin{equation}
L_N=\sqrt{\frac{4D}{r}\ln N},
\end{equation}
since $\widetilde L_N\sim L_N$ as $N\to\infty$. Since $u_N\approx\sqrt{\ln N}$, the first term in Eq.~\eqref{eq:maxalpha1xsquared_APP} grows as $\sqrt{\ln N}$, while the quadratic correction remains of order $O(1)$ for typical $x=O(1)$. Therefore, the leading large-$N$ limit is again the unconfined result in Eq.~\eqref{eq:maxalpha1interm_APP}. Nevertheless, at finite $N$ the correction proportional to $c_2$ is not negligible and must be kept in order to describe the numerical data accurately.
We can now obtain a finite-$N$ corrected approximation by performing a change of variables from $x$ to $m$. The equation
\begin{equation}
m=\widetilde L_N x-c_2x^2
\end{equation}
has two solutions,
\begin{equation}
x_{\pm}(m)=
\frac{\widetilde L_N\pm\sqrt{\widetilde L_N^2-4c_2m}}{2c_2}.
\end{equation}
The exact change of variables would contain the sum of the contributions from both roots. However, for large $N$, the root $x_+(m)$ is typically of order $\widetilde L_N/c_2$, and is therefore large. Since $P_x(x)\propto e^{-x^2}$, its contribution is exponentially suppressed. We thus retain only the dominant branch $x_-(m)$.
Using
\begin{equation}
\operatorname{Prob}\{M_1=m\}\approx
P_x(x_-(m))
\left|\frac{dx_-(m)}{dm}\right|,
\end{equation}
we obtain, for $0<m<\widetilde L_N^2/(4c_2)$,
\begin{equation}\label{eq:finiteNalpha1dist_APP}
\operatorname{Prob}\{M_1=m\}\approx
\frac{
\widetilde L_N-\sqrt{\widetilde L_N^2-4c_2m}
}{
c_2\sqrt{\widetilde L_N^2-4c_2m}}
\exp\!\left[
-\left(
\frac{
\widetilde L_N-\sqrt{\widetilde L_N^2-4c_2m}}{
2c_2}
\right)^2
\right].
\end{equation}
This expression still depends explicitly on $N$ through $\widetilde L_N$.
It should therefore be interpreted as a finite-$N$ correction to the
asymptotic scaling form, obtained by retaining the dominant branch in the
change of variables. In the limit $N\to\infty$, one has
$\widetilde L_N\sim L_N$ and $c_2/L_N\to0$. Thus
Eq.~\eqref{eq:finiteNalpha1dist_APP} reduces to the universal unconfined result
\begin{equation}
    S(z)=2z e^{-z^2},
    \qquad
    M_1=L_N z .
\end{equation}
For the finite system sizes used in our simulations, this correction
significantly improves the comparison with the numerical data, as shown by the
dotted lines in Fig.~\ref{fig:generalpotential}.

\section{EVS for $V(x)=\kappa \sqrt{|x|}$}\label{sec:potenzsqrt_APP}

In this section, we expand upon the discussion of the case $\alpha=0.5$ introduced in Sec. \ref{sec:GeneralPotential}. We first note that the potential 
\begin{equation}\label{eq:potentialalpha0dot5_APP}
    V(x)=\kappa \sqrt{|x|}
\end{equation}
corresponds to a confining force $F(x)=-V'(x)=-\frac{\kappa}{2\sqrt{|x|}}\operatorname{sign}(x)$, which diverges as $x \to 0$. 
To avoid numerical instabilities during the simulations, we introduced a regularized potential of the form:
\begin{equation}\label{eq:potentialregularized_APP}
    V_{\epsilon}(x) = \kappa \left[ \left( x^2 + \epsilon^2 \right)^{1/4} - \sqrt{\epsilon} \right]
\end{equation}
with $\epsilon \ll 1$. This regularization yields the associated force
\begin{equation}
    F_\epsilon(x)=-\frac{\kappa}{2}x(x^2+\epsilon^2)^{-3/4}.
\end{equation}
which remains finite and regular at the origin ($x=0$). For large distances with respect to $\epsilon$, this regularized potential correctly recovers the original exact force. In our numerical simulations, we set $\epsilon=10^{-2}$. Consequently, the deviation from the exact potential is strictly localized near the origin. Because EVS are governed by the large-distance behavior of the system, this regularization should not affect our asymptotic results.
\begin{figure}[h!]
    \centering
    \includegraphics[width=0.6\textwidth]{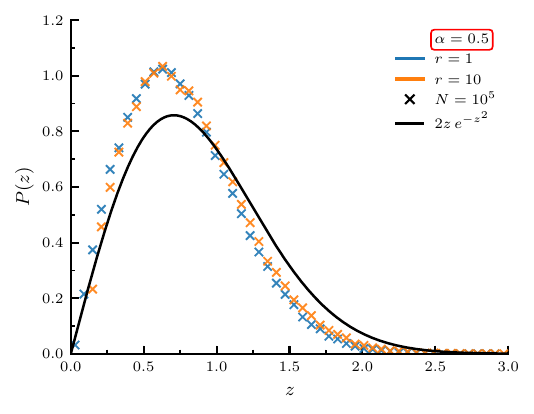}
   \caption{EVS for the potential $V(x)=\kappa \sqrt{|x|}$. Crosses show numerical simulations for $\kappa=1$, $N=10^5$ and two values of the resetting rate, $r=1$ and $r=10$. The horizontal axis is the rescaled maximum $z=M_1/\sqrt{(4D/r)\ln N}$, while the solid black line corresponds to the scaling function $S(z)=2z e^{-z^2}$. The two empirical distributions collapse onto each other when expressed in terms of $z$, even though $r$ is varied by one order of magnitude. This contrasts with the behavior observed for $\alpha>1$ in Fig.~\ref{fig:generalpotential}, where changing $r$ modifies the shape of the distribution.}
    \label{fig:alpha0dot5_rdiffer_APP}
\end{figure}

Furthermore, we present additional numerical simulations to support the claim that the case $\alpha=0.5$ belongs to the unconfined EVS universality class defined in the main text. As previously discussed, standard numerical simulations for these systems often exhibit prominent finite-size effects. However, a strong piece of evidence supporting our conclusion is the invariant behavior of the rescaled maximum
\begin{equation}
    z = \frac{M_1}{\sqrt{\frac{4D}{r}\ln N}},
\end{equation}
when we vary $r$.
As shown in Fig. \ref{fig:alpha0dot5_rdiffer_APP}, the empirical distribution of $z$ does not change even when the resetting rate $r$ is varied by an order of magnitude. This robustness strongly contrasts with the behavior observed for $\alpha > 1$ (e.g., $\alpha=1.5, 2, 3$), where the rescaled distribution changes drastically upon varying $r$, as previously illustrated in Fig. \ref{fig:generalpotential}. We believe that this is another point suggesting that our claim that the unconfined universality class in Eq. \eqref{eq:resultEVSunconfined} is valid for all $0\leq\alpha\leq 1$.

\end{appendix}

\end{document}